\documentclass[pra,onecolumn, showpacs,showkeys,secnumarabic,amssymb,amsmath,superscriptaddress,floatfix]{revtex4-1}
\usepackage{amssymb,amsmath}
\usepackage{color}
\usepackage{tikz}
\usepackage{mathtools}
\usepackage{graphics}
\usepackage{graphicx}
\usepackage{enumitem}
\usepackage{bm}
\usepackage{standalone}
\usepackage{color}
\begin{document}

\title{Plateau dynamics with quantized oscillations of a strongly driven qubit}

\author{ Yejia Chen$^{1,2}$, Zhiguo L\"{u}$^{1,4}$\footnote{Electronic address:~\url{zglv@sjtu.edu.cn}}, Yiying Yan$^3$, Hang Zheng$^1,^4$\footnote{Electronic address:~\url{hzheng@sjtu.edu.cn}}}

\affiliation{$^1$Key Laboratory of Artificial Structures and Quantum Control (Ministry of Education), School of Physics and Astronomy,Shanghai Jiao Tong University, Shanghai 200240, China\\
$^2$Zhiyuan College, Shanghai Jiao Tong University, Shanghai 200240, China \\
$^3$Department of Physics, School of Science, Zhejiang University of Science and Technology, Hangzhou 310023, China\\
$^4$Collaborative Innovation Center of Advanced Microstructures, Nanjing University, Nanjing 210093, China
}
\date{\today}

\begin{abstract}

We present an interesting dynamical temporal localization of a strongly driven two-level system (TLS), a plateau with quantized oscillation, by an analytical and transparent method, the counter-rotating-hybridized rotating-wave (CHRW) method. This approach, which is based on unitary transformations with a single parameter, treats the rotating and counter-rotating terms on equal footing. In the unitarily transformed representation, we find that it is the multiple-harmonic terms shown in the transformed Hamiltonian that make a crucial contribution to the generation of the exotic plateau phenomenon. By comparing the results of the numerically exact calculation and several other methods, we show that the CHRW results obtained by analytical formalism involving the collective effects of multiple harmonics are in good agreement with the numerical results, which illustrates not only the general tendency of the dynamical evolution of strongly driven TLS, but also the interesting phenomena of plateaus. The developed CHRW method reveals two kinds of plateau patterns: zigzag plateau and armchair plateau, and quantitative analyses are given to comprehensively describe the features of the plateau phenomenon. The plateau phenomenon has a periodical pattern whose frequency is double the driving frequency, and possesses quantized oscillations the number of which has a certain, precise value. Besides, fast oscillation is produced on every plateau which is determined by the relevant driving parameters of the TLS. Our main results are as follows:
(i) in the large-amplitude oscillatory case, it turns out that the collective effects of all even harmonics contribute to the generation of zigzag plateau with quantized oscillation, and the general tendency of evolution coincides with the result of the original CHRW method because of a linear trend of the phase function; (ii) in the small-amplitude oscillatory case, the dynamics of the coherent destruction of tunneling under strong driving is exactly exhibited by including the odd-harmonic effect, which offers a novel dynamical pattern, namely, armchair plateau possessing a two-stair structure rather than the complete destruction. Besides, the characteristics of the plateau (position, frequency, the envelope and number of quantized oscillations) are revealed by our analytical formalism. Both of the dynamical patterns are of great interest to strongly driven physics in the ongoing research on driven TLS systems.
\end{abstract}
\pacs{42.50.Ct 42.50.Pq 03.67.-a 03.65.-w}
\maketitle

\section{introduction}
The dynamics of driven quantum systems is an intriguing component of quantum physics which enjoys a long history of both experimental and theoretical researches \cite{Hanggi,Leggett-book_of_dissipative_TLS,Kockum}. Among these numerous investigations, two-level systems (TLSs), which attract the researchers in nearly every field for its simplicity, have been widely used to explore and demonstrate the fantastic phenomena of quantum physics. With the progress of experimental technology, many TLSs and some variants have received much attention in recent decades, such as TLSs with relaxation \cite{TLS_relax:origin,TLS_relax:David_PRL,TLS_relax:Gorlicki_adiabatic,TLS_relax:Karabanov_math,TLS_relax:Lapert,TLS_relax:Naum,TLS_relax:Ofek_disorder}, TLSs coupled to bosons \cite{Leggett-book_of_dissipative_TLS,TLS_boson:Castella_coherent_ctrl,TLS_boson:Chenu,TLS_boson:Plohn_path-integral,TLS_boson:Raedt_thermo,TLS_boson:Shen_QPT}, driven dissipative TLSs\cite{EF:Dakhnovskii,EF:Wang}, periodically driven qutrit\cite{Qutrit:Grimaudo_twoQ,Qutrit:Han,Qutrit:QED,Qutrit:Vitanov_Arbitrary_SU3}, etc. Of these systems, the simplest and widely investigated one is the Rabi model \cite{Rabi:origin}, which in the localized representation is written as
\begin{eqnarray}\label{origin}
H(t) & = & -\frac{\Delta}{2}\sigma_{x}-\frac{A}{2}\cos(\omega t)\sigma_{z}   \nonumber\\
     & = & -\frac{\Delta}{2}\sigma_{x}-\frac{A}{4}(e^{-i\omega t}\sigma_{+}+e^{i\omega t}\sigma_{-})-\frac{A}{4}(e^{i\omega t}\sigma_{+}+e^{-i\omega t}\sigma_{-}),
\end{eqnarray}
where $\sigma_{x,(y,z)}$ is Pauli matrix, $\sigma_{\pm}=(\sigma_z\pm i\sigma_y)/2$,  $\Delta$ is tunneling frequency independent of time and $A\cos(\omega t)$ is a harmonic driving field with amplitude $A$ and frequency $\omega$. The last two terms in Eq. (\ref{origin}) represent the counter-rotating (CR) couplings. On the other hand, the Rabi model has been utilized and illustrated by experiments \cite{EXP:Berezovsky,EXP:Deblock,EXP:Nakamura}. These experimental progresses substantiate the effectiveness of the TLS models, and inspire the further development of theoretical techniques to give reasonable exposition for novel dynamics in a broader parameter regime. Especially, when the driving amplitude $A$ is comparable to or even larger than the tunneling frequency $\Delta$, i.e., the driving parameter is in strongly or ultrastrongly driving regimes, the driven systems that have been realized in flux and spin qubits, generate novel dynamical phenomena\cite{EXP:spin-qubit,fluxstrongdriving,SCstrongdriving,Yu-strongdriving}.

A lot of methods have been introduced to study the dynamics of the Rabi model and improved for the past eighty years \cite{Rabi:origin,Rabi:analytic_solution1,Rabi:coupled_qubit,Rabi:CR_term,Rabi:integrability,Rabi:untrastrong_regime}, aiming to give a valid approximate analytical solution to depict major dynamical characteristics and phenomena of interest. The Rabi rotating wave approximation (RWA) method \cite{Rabi:origin}, which is regarded as the initial method to analytically solve the dynamical evolution of the TLS, has been applied and developed for many years. However, it is a basic fact that the Rabi-RWA method breaks down in the strong driving regime or off-resonant situation since the CR terms can not be neglected in comparison with the rotating one. Actually, many interesting phenomena are attributed to the contribution of CR terms, such as coherent destruction of tunneling (CDT) \cite{CDT:origin,CDT:stabilizing_CDT,CDT:experiment}, Bloch-Siegert(BS) shift \cite{BS:Bloch,BS:Yan,BS:Zhang}, quantum Zeno effect \cite{QZE}, etc. Therefore, various methods sprang up in order to improve the analytical results in some parameter regime and offer perspectives to discover the insight of these significant phenomena, such as the RWA in a rotating frame (RWA-RF) \cite{RWA-RF}, Floquet theory \cite{Shirley,Yi}, transfer-matrix approach \cite{RWA-RF,Garraway}, etc.

Over all of the advanced approaches, it is worth emphasizing the efficacy of the counter-rotating-hybridized rotating-wave method (CHRW method) \cite{CHRW:origin}, which has been extensively utilized in many researches \cite{CHRW:bias,CHRW:battery,CHRW:Effects_of_CR,CHRW:2photon-Rabimodel,citeCHRW:BS_shift_q_to_c_transition,citeCHRW:controlled_state_transfer,citeCHRW:LFM,citeCHRW:maximal_quantum_Fisher_information_semi_Rabi_model}. In comparison with the classical Rabi-RWA method, it takes the effects of the CR terms into consideration by a unitary transformation, which results in the RWA mathematical formalism applicable to both the far off-resonant situation ($\Delta \gg \omega$ or $\Delta \ll \omega $) and the intermediate-strong driving strength case. Although, in the strong driving case, the original CHRW method demonstrates the accurate frequency of the overall oscillatory evolution, it cannot give a detailed local pattern of the dynamical evolution, such as a dynamical phenomenon, termed as ``plateau with quantized oscillation''.

Plateaus with quantized oscillation, firstly named in this work, is an intriguing dynamical pattern and its intrinsic mechanism is not clearly demonstrated until now though it is noticed in previous works\cite{plateau:magnetic_flux, plateau:dc-ac-LiCN}. This stunning phenomenon happens in the strong driving condition. The particular dynamical pattern consists of plateaus periodically occurring with the double frequency of the driving field, and fast oscillation with a certain period number in each plateau. Although the original CHRW method could predict the accurate frequency of the overall oscillatory evolution, it ignores some local oscillatory property that strongly driven TLSs possess. A possible reason for the failure to account for this phenomenon is that the original CHRW Hamiltonian doesnot take into account higher order harmonics, as shown in the harmonic expansion of Hamiltonian in the interaction picture. Therefore, an improved method needs to be introduced to serve as a bridge between the multiple-harmonic effect and the intriguing plateau phenomenon.

The aim of our work is to reveal the relation of this plateau phenomenon to the multiple harmonics. Based on the simplest form of the TLSs shown in Eq. (\ref{origin}), we develop the CHRW method, which is realized by applying another unitary transformation to the modified Hamiltonian of the original CHRW method in Ref. \cite{CHRW:origin}. The improved method, named as the CHRW2 method in this work, has an extraordinary performance in a much wider parameter regime than before, especially on the strong driving condition. From the viewpoint of the formalism of the CHRW2 method, the plateau phenomenon of large-amplitude oscillatory dynamics is determined by the even-harmonic effect naturally. Besides, the concise form of the analytical solution provides further quantitative discussion about dynamical structures of the plateau, and emphasizes the important collective role of all the even-harmonic terms on the evolution of strongly driven TLS.

On the other hand, the CDT phenomenon, which points out the magical destruction of the TLS dynamics under the control of specific driving parameters, is presented by a novel explanation in this work. Instead of the traditional recognition, completely frozen dynamics without any deviation from the initial state, the CDT exhibits actually a complex structure. The occupation probability oscillates periodically between two kinds of plateaus close to each other, and  quantized oscillation with small amplitude is also shown on each plateau. From the CHRW method, it would be clear for us to illustrate that the even-harmonic effect diminishes under the CDT condition and the odd-harmonic effect establishes dominance over the dynamics. Additionally, analogous analysis offers the quantitative features of the plateau structure, which presents the advantage of this analytical method.

The structure of the paper is organized as follows: In Sec. II, we improve the original CHRW method in Ref. \cite{CHRW:origin} by applying two unitary transformations successively to the Hamiltonian of the Rabi model, and derive analytical results of $\langle \sigma_x(t) \rangle$, $\langle \sigma_y(t) \rangle$ and $\langle \sigma_z(t) \rangle$ which exhibit the even-harmonic effect in the strong driving regime. Then in Sec. III, we demonstrate the effectiveness of the CHRW2 method by comparing with the results of the numerically exact method, the original CHRW and Rabi-RWA methods, and illustrate the intriguing phenomenon of plateau with quantized oscillation in the time evolution. Finally, we give quantitative results of some physical quantities describing the plateau dynamics. In Sec. IV, we obtain an effective Hamiltonian by utilizing the CDT condition under strong driving condition . Then we present another kind of plateau phenomenon which exhibits the odd-harmonic effect and provide detailed quantitative analysis of the plateau. In Sec. V we give the conclusion of this paper.

\section{Counterrotating- hybridized rotating-wave method} \label{CHRW2 method}
To give a more complete physics on strong driving, in this section we attempt to improve the CHRW method with unitary transformations. We illustrate an important dynamical phenomenon, i.e., a plateau structure with quantized oscillation. Meanwhile, we demonstrate the significant underlying reasons, the effect of even-harmonic processes, on this kind of dynamical pattern.

\subsection{CHRW Hamiltonian} \label{CHRWHamilton}
In order to investigate the effect of the multi-harmonic process on the dynamics, we apply the unitary transformations to the Rabi model and obtain an effective Hamiltonian. More importantly, the even-harmonic processes are taken into account to analytically resolve the dynamics under strong driving condition. $\Phi(t)$ is defined as the corresponding wave function of the TLS satisfying the Schr\"odinger equation
$
    i\frac{d}{dt}\Phi(t)=H\Phi(t).
$
By the unitary transformation with a generator $S$ \cite{CHRW:origin}, the Schr\"odinger equation becomes
$
i\frac{d}{dt}\Phi^{\prime}(t)=H^{\prime}\Phi^{\prime}(t)
$
, where
$
\Phi^{\prime}(t)=e^{S}\Phi(t)
$
and
\begin{eqnarray}\label{unitran}
H^{\prime}(t)=e^{S}H(t)e^{-S}-ie^{S}\frac{de^{-S}}{dt}.
\end{eqnarray}
Now we apply two unitary transformations to the Rabi model in succession, and the generators are
\begin{eqnarray}
  S_{1}&=&-i\frac{A}{2\omega}\xi\sin(\omega t)\sigma_{z}, \label{S1}\\
  S_{2}&=&-i \sum_{k=1}^{\infty}\frac{\Delta J_{2k} \left(\frac{A}{\omega}\xi \right)} {2k\omega} \sin(2k\omega t)\sigma_{x}, \label{S2}
\end{eqnarray}
where $\xi$ is a parameter determined later and $J_{2k}\left(\frac{A}{\omega}\xi \right)$ is the $2k$-order Bessel function of the first kind. Then we can get the transformed Hamiltonian as follows,
\begin{eqnarray}
  H_{1}(t) &=& e^{S_{1}}H(t)e^{-S_{1}}-ie^{S_{1}}\frac{de^{-S_{1}}}{dt} \nonumber \\
     &=& -\frac{\Delta}{2}J_{0}\left(\Xi\right)\sigma_{x} -\Delta \sum_{n=1}^{\infty} J_{2n}\left(\Xi\right)\cos(2n\omega t)\sigma_x   \nonumber\\
     &&  -\Delta \sum_{n=1}^{\infty} J_{2n-1}\left(\Xi\right)\sin\left[(2n-1)\omega t\right]\sigma_y-\frac{A}{2}\left(1-\xi\right)\cos(\omega t)\sigma_{z}, \label{trans_step1}
\end{eqnarray}
\begin{eqnarray}
  H_{2}(t) &=& e^{S_{2}}H_{1}(t)e^{-S_{2}}-ie^{S_{2}}\frac{de^{-S_{2}}}{dt} \nonumber \\
     &= &  -\frac{\Delta}{2}J_{0}\left(\Xi\right)\sigma_{x}-\Delta \sum_{n=1}^{\infty} J_{2n-1}\left(\Xi\right)\sin\left[(2n-1)\omega t\right]\times \nonumber \\
      & &  \left\{\cos\left[\sum_{k=1}^{\infty} X_{k}\sin(2k\omega t)\right]\sigma_{y}+\sin\left[\sum_{k=1}^{\infty} X_{k}\sin(2k\omega t)\right]\sigma_{z} \right\}     \nonumber\\
       & &-\frac{A}{2}\left(1-\xi\right)\cos(\omega t)\left\{\cos\left[\sum_{k=1}^{\infty} X_{k}\sin(2k\omega t)\right]\sigma_{z}\right.  \nonumber\\
       &&-\left.\sin\left[\sum_{k=1}^{\infty} X_{k}\sin(2k\omega t)\right]\sigma_{y}\right\}, \label{trans_step2}
\end{eqnarray}
where $X_{k}=\Delta J_{2k}\left(\Xi\right)/k\omega$ and $\Xi=A\xi/\omega$. It is obvious to see the expansion of Eq. (5) according to the orders of the harmonics: zeroth harmonic(zero-$\omega$ term), even harmonics ($2n\omega$ terms, $n=1,2,\cdot\cdot\cdot$) and odd harmonics ($(2n-1)\omega$ terms, $n=1,2,\cdot\cdot\cdot$). In Eq. (\ref{trans_step1}), the second term consists of all even harmonics, while the last two terms includes all odd harmonics. Since the product of odd harmonics and even harmonics of the $\cos$ or $\sin$ factor happens in the last two terms of Eq. (6), it leads to odd harmonics in $H_2$.  Thus, after the successive two unitary transformation, $H_2(t)$ is composed of zeroth and all odd harmonics. In other words, we have removed all even harmonics ($n\geq 2$). This is a key point of this work. Until now, no approximation has been made.


In the following treatment, we divide the Hamiltonian $H_2$ into two parts $H_2=H_{\rm CHRW2}+H'$, $H_{\rm CHRW2}$ keeps the zeroth and first harmonics of the transformed Hamiltonian ($n\omega, n=0, 1$).
\begin{eqnarray} \label{CHRW2}
  H_{\rm CHRW2}& =& -\frac{\Delta}{2}J_{0}\left(\Xi\right)\sigma_{x}+\Delta J_{1}\left(\Xi\right)\prod_{k=1}^{\infty}J_{0}(X_{k})\sin(\omega t)\sigma_{y} \nonumber \\
   &-& \frac{A}{2}[1-\xi]\prod_{k=1}^{\infty}J_{0}(X_{k})\cos(\omega t)\sigma_{z}
\end{eqnarray}
and $H^{\prime}=H_{2}-H_{\rm CHRW2}$. Apparently, $H'$ includes all higher-order odd harmonics ($3 \omega, 5\omega,...$) generally accompanying the higher-order Bessel functions.

All higher-order odd harmonic terms in $H^{\prime}$ which are corresponding to multi-harmonic processes in the unitarily transformed presentation are neglected since they cause little variation during the integral with regard to a long period which is numerically proved in the next section. Then we choose a proper parameter $\xi$ so that the coefficients of the counter-rotating terms $\sigma_{+}\exp(i\omega t)$ and $\sigma_{-}\exp(-i\omega t)$ in Eq. (\ref{CHRW2}) become zero, which gives
\begin{equation}\label{parameter-determined}
  \frac{A}{2}(1-\xi)-\Delta J_{1}(\Xi)=0.
\end{equation}
Thus we obtain the reformulated rotating wave Hamiltonian,
\begin{equation}\label{Heff_final}
  H_{\rm CHRW2} = - \frac{\tilde{\Delta}}{2}\sigma_{x}-\frac{\tilde{A}}{4}\left(e^{-i\omega t}\sigma_{+}+e^{i\omega t}\sigma_{-}\right).
\end{equation}
Here, $\tilde{\Delta}$ and $\tilde{A}$ are renormalized tunneling strength and renormalized driving strength, which read
\begin{eqnarray}
  \tilde{\Delta} &=& \Delta J_{0}\left( \frac{A}{\omega}\xi\right),  \label{deltatilde} \\
  \tilde{A} &=& 2A(1-\xi)\prod_{k=1}^{\infty} J_{0}(X_{k}),  \label{Atilde}
\end{eqnarray}
respectively. This is called the counter-rotating-hybridized rotating-wave method. It is obvious that the CHRW Hamiltonian possesses the same mathematical structure as the RWA one except for those modified physical quantities. In order to calculate dynamical quantities of the CHRW Hamiltonian shown in Eq. (\ref{Heff_final}), we acquire the well-known RWA results. To distinguish the present CHRW Hamiltonian in Eq. (\ref{Heff_final}) from that introduced in the CHRW method \cite{CHRW:origin}, we denote the original CHRW Hamiltonian in Ref. \cite{CHRW:origin} as `CHRW1', and the the present Hamiltonian in Eq. (\ref{Heff_final}) as `CHRW2' in the following. Actually, when $X_k=0$, the CHRW2 immediately returns to the CHRW1, which indicates that the CHRW2 method works well in the valid parameter region of the CHRW1 \cite{CHRW:origin}. More importantly, by applying the second transformation to $H_1$ in Eq. (\ref{trans_step1}), we have taken into account the effect of all even-harmonic terms. Therefore, the CHRW2 method can be used to investigate the physics of strong driving and may reveal the ``even-harmonic'' collective effect.

\begin{figure}
  \centering
  \includegraphics[width=8cm]{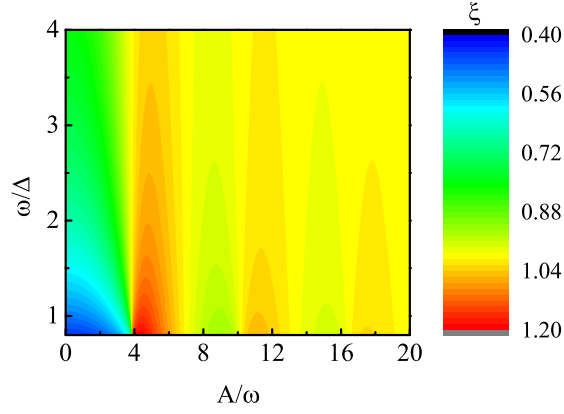}\\
  \caption{(Color online) The value of $\xi$ as functions of $\omega/\Delta$ and $A/\omega$, which is determined by Eq. (8).}\label{Figreplyxi}
\end{figure}

For the the parameter regime of interest $A \gg \omega$ and $\Delta \sim \omega$, one immediately obtains $\xi \sim 1$ which is illustrated in Fig. \ref{Figreplyxi}. Then,
one gets
 \begin{equation}
  X_{k}= \frac{\Delta J_{2k}\left(\Xi\right)}{k\omega}{\longrightarrow}
  \frac{\Delta}{k\omega}\sqrt{\frac{2\omega}{\pi A \xi}}\cos\left(\frac{A}{\omega}\xi-k\pi-\frac{\pi}{4}\right)\ll 1.
\end{equation}
Thus, $J_{n}(X_{k}) \ll 1~(n \neq 0)$. Consequently, it is reasonable to simplify Eq. (\ref{trans_step2}) with the drop of $H'$ to Eq. (\ref{CHRW2}). We ignore all higher harmonics in the sums of coefficients $\sigma_z$ and $\sigma_y$ involved in the product of multiple higher order Bessel functions, because the value of Bessel function decreases rapidly when its order increases.

\subsection{Physical quantities}

Similar to the mathematical formalism of the RWA method, one can find an exact solution of the Schr\"odinger equation with the Hamiltonian in Eq.(\ref{Heff_final}). Supposing $\Phi(t=0)=\left(\begin{array}{c} 1\\ 0 \end{array}\right)$ without loss of generality, one obtains $\Phi(t=0)=\Phi^{\prime}(t=0)$ beacuase of the generator $S(t)$ proportional to sinusoidal function $\sin(\omega t)$, and

\begin{eqnarray}
  \Phi^{\prime}(t) &=& \frac{1}{\sqrt{2}} c_{1}(t)\left(\begin{array}{c}
                                                   1 \\
                                                   1
                                                 \end{array}\right) + \frac{1}{\sqrt{2}} c_{2}(t) \left(\begin{array}{c}
                                                                                            1 \\
                                                                                            -1
                                                                                          \end{array}\right),\\
  c_{1}(t) &=& e^{\frac{i\omega t}{2}}\left[\frac{1}{\sqrt{2}}\cos\left(\frac{\tilde{\Omega}_{R}t}{2}\right)+i\frac{\frac{\tilde{A}}{2}-\tilde{\delta}}{\sqrt{2}\tilde{\Omega}_{R}}\sin\left(\frac{\tilde{\Omega}_{R}t}{2}\right)\right] ,\\
 c_{2}(t) &=& e^{-\frac{i\omega t}{2}}\left[\frac{1}{\sqrt{2}}\cos\left(\frac{\tilde{\Omega}_{R}t}{2}\right)+i\frac{\frac{\tilde{A}}{2}+\tilde{\delta}}{\sqrt{2}\tilde{\Omega}_{R}}\sin\left(\frac{\tilde{\Omega}_{R}t}{2}\right)\right],
\end{eqnarray}
where $\tilde{\delta}=\omega-\tilde{\Delta}, \tilde{\Omega}_{R}=\sqrt{\tilde{\delta}^{2}+\tilde{A}^{2}/4}$.
In this representation, one can easily calculate the average values of $\sigma_x, \sigma_y $, and $\sigma_z $,
\begin{eqnarray}
  X^{\prime} &= \langle \Phi^{\prime}(t)|\sigma_{x}|\Phi^{\prime}(t) \rangle = & -\frac{\tilde{A}\tilde{\delta}}{\tilde{\Omega}_{R}^{2}}\sin^{2}\left(\frac{\tilde{\Omega}_{R}t}{2}\right),\label{x_org}\\
  Y^{\prime} &= \langle \phi^{\prime}(t)|\sigma_{y}|\Phi^{\prime}(t) \rangle = & \sin(\omega t)\left[\cos^{2}\left(\frac{\tilde{\Omega}_{R}t}{2}\right)+\frac{\left(\frac{\tilde{A}}{2}\right)^{2}-\tilde{\delta}^{2}}{\tilde{\Omega}_{R}^{2}}\sin^{2}\left(\frac{\tilde{\Omega}_{R}t}{2}\right)\right]\nonumber\\
  &&-\frac{\tilde{\delta}}{\tilde{\Omega}_{R}}\cos(\omega t)\sin\left(\tilde{\Omega}_{R}t\right),\label{y_org}\\
  Z^{\prime} &= \langle \Phi^{\prime}(t)|\sigma_{z}|\Phi^{\prime}(t) \rangle = & 1-2\left\{\left[\frac{\tilde{A}}{2\tilde{\Omega}_{R}}\sin\left(\frac{\tilde{\Omega}_{R}t}{2}\right)\sin\left(\frac{\omega t}{2}\right)\right]^{2} \right.\nonumber \\
  &&+\left.\left[\cos\left(\frac{{\tilde\Omega_{R}}t}{2}\right)\sin\left(\frac{\omega t}{2}\right)-\frac{\tilde{\delta}}{{\tilde\Omega}_{R}}\sin\left(\frac{{\tilde\Omega_{R}}t}{2}\right)\cos\left(\frac{\omega t}{2}\right)\right]^{2}\right\},\label{z_org}
\end{eqnarray}
respectively.
To acquire the average values of the physical quantities in the laboratory representation, one obtains
\begin{eqnarray}
  X &=& \langle \Phi(t)|\sigma_{x}|\Phi(t) \rangle = \langle \Phi^{\prime}(t)|e^{S_{2}}e^{S_{1}}\sigma_{x}e^{-S_{1}}e^{-S_{2}}|\Phi^{\prime}(t) \rangle ,\\
  \label{xwave}
  Y &=& \langle \Phi(t)|\sigma_{y}|\Phi(t) \rangle = \langle \Phi^{\prime}(t)|e^{S_{2}}e^{S_{1}}\sigma_{y}e^{-S_{1}}e^{-S_{2}}|\Phi^{\prime}(t) \rangle ,\\ \label{ywave}
  Z &=& \langle \Phi(t)|\sigma_{z}|\Phi(t) \rangle = \langle \Phi^{\prime}(t)|e^{S_{2}}e^{S_{1}}\sigma_{z}e^{-S_{1}}e^{-S_{2}}|\Phi^{\prime}(t) \rangle .  \label{zwave}
\end{eqnarray}
One can see that the average values in the laboratory representation can be expressed by those in the unitarily transformed representations under a linear transformation,
\begin{eqnarray}
  \left(\begin{array}{c}
    X \\
    Y \\
    Z
  \end{array}
  \right)
   &=& T^{\prime} \left(\begin{array}{c}
                          X^{\prime} \\
                          Y^{\prime} \\
                          Z^{\prime}
                        \end{array}\right),
 \end{eqnarray}
 \begin{eqnarray}
  T^{\prime} &=& \left(\begin{array}{ccc}
                         \cos\left(\Xi\sin(\omega t)\right) & \sin\left(\Xi\sin(\omega t)\right)\cos\left(S_{je}(\omega t)\right) & \sin\left(\Xi\sin(\omega t)\right)\sin\left(S_{je}(\omega t)\right)  \\
                         -\sin\left(\Xi\sin(\omega t)\right) & \cos\left(\Xi\sin(\omega t)\right)\cos\left(S_{je}(\omega t)\right) & \cos\left(\Xi\sin(\omega t)\right)\sin\left(S_{je}(\omega t)\right) \\
                         0 & -\sin\left(S_{je}(\omega t)\right) & \cos\left(S_{je}(\omega t)\right)
                       \end{array}\right), \label{transform_matrix_for_even}
\end{eqnarray}
where
\begin{equation}\label{Sje}
  S_{je}(\omega t)=\sum_{k=1}^{\infty}\frac{\Delta J_{2k}\left(\Xi\right)}{k\omega}\sin(2k\omega t).
\end{equation}
Assuming $S_{je}\rightarrow 0$, one obtains the transformation matrix in the CHRW1 method,
\begin{equation}\label{transform_matrix_simple}
T=\left( \begin{array}{ccc}
           \cos(\Xi\sin(\omega t)) & \sin(\Xi\sin(\omega t)) & 0 \\
          -\sin(\Xi\sin(\omega t)) & \cos(\Xi\sin(\omega t)) & 0 \\
           0 & 0 & 1
         \end{array}\right).
\end{equation}

The physical quantities of the three methods (RWA, CHRW1, and CHRW2) are shown in the Table \ref{tab:compare}. It is obvious to see that both the CHRW1 and CHRW2 methods have modified the bare physical quantities into effective ones because of the unitary transformations. In comparison with the CHRW1 method, the CHRW2 method, which takes the even-harmonic effect into consideration, has different effective physical quantities because of the unitary transformation Eq. (\ref{trans_step2}).  Moreover, the dynamical quantities of interest are subject to linear transformations in both the CHRW methods, and exhibits the sophisticated physical picture of the driven TLS. Combining modified physical parameters with transformations for the observable physical quantities, we turn out that the CHRW methods are superior to the RWA one because of their accurate dynamics in a much broader parameter regime \cite{CHRW:origin}.

\begin{table}
\caption{The comparison of quantities for different methods}
\begin{tabular}{c c c c}
  \hline\hline
  Method & RWA & CHRW1 & CHRW2 \\
  \hline
  diving strength & $A$ & ~~~$2A(1-\xi)$ &~~ $2A(1-\xi)\prod_{k}J_{0}\left(X_{k}\right)$ \\
  Tunneling frequency & $\Delta$ &~~~ $~\Delta J_{0}\left(\frac{A}{\omega}\xi\right)$ & $\Delta J_{0}\left(\frac{A}{\omega}\xi\right)$ \\
  Transformation matrix & $I$ &~~~ $T$ &~~ $T^{\prime}$\\
  \hline\hline
\end{tabular} \\
Note: $I$ denotes the identity matrix and $T$ defined in Eq. (\ref{transform_matrix_simple}) represents the transformation matrix derived from the original CHRW method. \label{tab:compare}
\end{table}

Although the CHRW2 method has the same parameter equation as that of the CHRW1, it has an important reduced factor $\prod_{k=1}^{\infty} J_{0}\left(X_{k}\right)$ in $\tilde{A}$, which makes adjustments towards the general tendency in the time evolution. Furthermore, the CHRW2 method, whose transformation matrix is equipped with dressed factors resulting from the unitary transformation $S_{2}$, exhibits the even-harmonic effect that the CHRW1 does not show. We will investigate this effect in the next section. It is just the transformation of the CHRW2 method that reveals an interesting plateau structure in the evolution of $\langle \sigma_z(t) \rangle$ under strong driving.

From Eqs. ({\ref{zwave}})-(\ref{transform_matrix_for_even}), one obtains the population dynamics $Z=\langle\sigma_z(t)\rangle$,
\begin{eqnarray}
  Z &=& \cos\left[S_{je}(\omega t)\right]\left\{  1-2\left\{\left[\frac{\tilde{A}}{2\tilde{\Omega}_{R}}\sin\left(\frac{\tilde{\Omega}_{R}t}{2}\right)\sin\left(\frac{\omega t}{2}\right)\right]^{2}\right.\right. \nonumber \\
  && +\left.\left.\left[\cos\left(\frac{\tilde{\Omega}_{R}t}{2}\right)\sin\left(\frac{\omega t}{2}\right)-\frac{\tilde{\delta}}{\tilde{\Omega}_{R}}\sin\left(\frac{\tilde{\Omega}_{R}t}{2}\right)\cos\left(\frac{\omega t}{2}\right)\right]^{2}\right\}\right\} \nonumber \\
  && -\sin\left[S_{je}(\omega t)\right]\left\{ \sin(\omega t)\left[\cos^{2}\left(\frac{\tilde{\Omega}_{R}t}{2}\right)+\frac{\left(\frac{\tilde{A}}{2}\right)^{2}-\tilde{\delta}^{2}}{\tilde{\Omega}_{R}^{2}}\sin^{2}\left(\frac{\tilde{\Omega}_{R}t}{2}\right)\right]\right.\nonumber\\
  && -\left. \frac{\tilde{\delta}}{\tilde{\Omega}_{R}}\cos(\omega t)\sin\left(\tilde{\Omega}_{R}t\right)\right\}. \label{z_complete}
\end{eqnarray}
When $S_{je}(\omega t)\rightarrow 0$, one gets the $\langle \sigma_z (t)\rangle_{\rm CHRW1}$ of the CHRW1 method.
The CHRW1 result shares the same form as that of the RWA method, except for some renormalized physical quantities. In contrast, the explicit expression of the CHRW2 method has taken into account of multiple even harmonics with the key dressed factors $\cos(S_{je})$ and $\sin(S_{je})$. As a result, it leads to some interesting characters of driving dynamics. The precise condition for the validity of the CHRW1 method has been reported in Ref. [46]. Moreover, the CHRW2 method can work in the parameter regime of the strong driving ($A \gg \omega$) as well as in the valid parameters of the CHRW1 method which is demonstrated in the next section.

\section{Even-harmonic processes under strong driving} \label{eveneff}

In this section, the strongly driven dynamics of the CHRW2 method by Eq.(26) are given in comparison with those of other methods (the CHRW1, RWA, numerically exact methods) to demonstrate the even-harmonics effects on dynamics. Meanwhile, we will explain the interesting characters of the population dynamics $\langle \sigma_z(t) \rangle$.

\subsection{Population dynamics} \label{{calc_even}}
In strong driving cases an interesting dynamical pattern, i.e., plateau dynamics with quantized oscillations happens. We show the time evolution $\langle \sigma_z(t) \rangle$ of the CHRW2 method in the resonance case ($\Delta=\omega$) for different strong driving strengths in Fig. \ref{fig1}, and also give the results of other methods for comparison. It is obvious to see that the CHRW2 method considering the even harmonic effect gives a zigzag type plateau with small-amplitude quantized oscillation on short time scale which is a pronounced character of strong driven dynamics. It not only shows a slow large-amplitude oscillation with the same frequency of the numerical one, but also illustrates periodic plateaus with an oscillatory frequency 2$\omega$ which is exactly the same as that of the numerical calculation. In the comparison of the CHRW1 result, we demonstrate that the novel phenomenon `periodic plateau' of the TLS results from even-harmonic effect. It is clear to see that the RWA method breaks down, of which the result deviates largely from the numerically exact one. Actually, the dynamics of the RWA always oscillates with an amplitude $\langle\sigma_z\rangle_{\rm Max}=1$ and a constant frequency nearly equal to $\omega$. In contrast to the numerically exact result, the CHRW1 method shows a correct general tendency of the $\langle \sigma_z(t) \rangle$ on a longer time scale. However, it does not exhibit the exact details of the dynamical structure during the time evolution. Besides, quantized oscillations mean that the number of the fast periodic oscillations on each plateau is a certain, precise integer value. This kind of quantization is characterized by an integer number corresponding to the fast periodic oscillation on each plateau which is discussed in the next subsection.

Moreover, we show the off-resonance case of $\langle \sigma_z(t) \rangle$ with different tunneling frequencies for $A/\omega=10$ in Fig. \ref{fig2}. When $\Delta < \omega$ shown in Figs. \ref{fig2}(a)-\ref{fig2}(c), the results of the CHRW2 demonstrate that $\langle \sigma_z(t) \rangle$ exhibits slower oscillation as $\Delta$ decreases which agrees well with the numerically exact result. The RWA method fails to give the correct evolution while both of the two CHRW methods do provide the general tendency of the dynamics. Besides, the plateaus illustrated by the CHRW2 result possess a certain periodical pattern which coincides with the numerical result for the off-resonance condition. Hence, the dynamical pattern of the plateaus turns out the improvement of the CHRW2 beyond the CHRW1 method in the strong driving case. Likewise, when $\Delta$ is a bit larger than $\omega$ ($\Delta=1.2\omega$ in Fig.\ref{fig2}(d)), it is obvious to see that the CHRW2 method still performs well, demonstrating the profound efficiency of the CHRW2 method in much wider parameter region.

Further, in order to examine the validity of the CHRW2 method, we also show the Fourier transform (FT) of the $\langle \sigma_z(t) \rangle$ of Figs. \ref{fig1} and \ref{fig2},
\begin{equation}\label{fft}
  F(\nu)=\left|\int_{-\infty}^{\infty} dt \langle\sigma_z(t)\rangle e^{i\nu t}\right|,
\end{equation}
which are shown in Figs. \ref{FigFFT1} and \ref{FigFFT2}, respectively. From the FT results, it is clear that the values of discrete frequencies obtained by the CHRW2 method are in good agreement with the exactly numerical results for all the peaks in each case. As the ratio of $\Delta/\omega$ increases, the frequency of the highest peak, which is corresponding to the frequency $|\tilde{\Omega}_R-\omega|$, increases, too. On the other hand, as the driving strength increases, the frequency of the main peak decreases. In order to explicate the information given in Fourier space, let us take Fig. \ref{FigFFT1}(d) ($A=19\omega,\Delta=\omega$) as an example. As expected, the RWA method shows only a single peak at $\nu=\omega$, in other words, the resonance frequency results from the rotating wave interactions. In contrast, the CHRW1 method accurately shows a main peak at $\nu=|\tilde{\Omega}_R-\omega|$ since it considers the effect of counter-rotating coupling. Nevertheless, the CHRW2 method exhibits all the peaks perfectly in accordance with the exact solution. Moreover, the frequency component corresponding to $\langle \sigma_z(t) \rangle$  occurs at $\nu=2k \omega \pm \left(\omega-\tilde{\Omega}_R\right)$, or approximately $\nu=2k \omega \pm \tilde{\Delta}$ for $A \gg\omega$, which can be easily obtained in Eq. (\ref{z_complete}) by expanding the multi-harmonic factor, $S_{je}$, in Taylor series and retaining sum frequency terms. Furthermore, the multiple even harmonic factors $\cos(S_{je})$ and $\sin(S_{je})$ determines the split pairs of peaks at $\nu=2k \omega \pm \left(\omega-\tilde{\Omega}_R\right)$, which is an important character of the qubit response away from the CDT condition. As a result, the even-harmonic effect is presented in a more explicit way in the Fourier space, where we can see the split pairs of the peaks resulting in the important plateau phenomenon in the dynamical picture. The splitting between each pair of peaks is described by $2\left|\omega - \tilde{\Omega}_R \right|$. From Fig. \ref{FigFFT2}, we find that the splitting becomes larger as the tunneling frequency $\Delta$ goes up. It also manifests the validity of the CHRW method in extracting major frequency information of the population dynamics.

In summary, the dynamics $\langle \sigma_z(t) \rangle$ under strong driving exhibits an intriguing mode of time evolution, which generally oscillates harmonically with a large amplitude near to $1$ in a long-time evolution and locally presents a plateau structure during a certain interval. Furthermore, it is worth noting that these plateaus consist of fast oscillations with a much smaller amplitude, of which the relevant properties are dependent on the relative values of $\Delta/\omega$ and $A/\omega$. In the next subsection we will analyze how the even-harmonic terms have effects on the occurrence of this phenomenon and derive concise expressions of some quantities comprehensively characterizing the fast oscillations of the plateau with a small amplitude.

\subsection{Sudden jumps and plateaus phenomena}\label{calc_even}
We would explain the interesting feature of the time evolution $\langle \sigma_{z}(t) \rangle$ consisting of zigzag plateaus: sudden jump and plateau. To validate the efficiency and accuracy of the CHRW2 method introduced in Sec. \ref{CHRW2 method}, we would give a simple approximate expression of Eq. (\ref{z_complete}) for certain parameters. On the physical condition of interest, where $A\gg \omega$ and $\Delta$ is comparable to $\omega$, it is easy to check that the parameter $\xi$ is close to 1. Therefore, we can explicitly verify that $\tilde{A} \ll \omega$ and $\tilde{\Omega}_{R} \approx \tilde{\delta}$ by Eqs. (\ref{deltatilde}) and (\ref{Atilde}). Then, we rewrite Eqs. (\ref{y_org}) and (\ref{z_org}) as
\begin{eqnarray}
  Y^{\prime} &\approx& \sin\left[(\omega-\tilde{\Omega}_{R})t\right], \label{y_org_simple}\\
  Z^{\prime} &\approx& \cos\left[(\omega-\tilde{\Omega}_{R})t\right], \label{z_org_simple}
\end{eqnarray}
respectively.
By Eq. (\ref{z_complete}), we obtain
\begin{equation}\label{Z_middle}
  Z=\cos[(\omega-\tilde{\Omega}_{R})t+S_{je}(\omega t))].
\end{equation}
Since $\tilde{A}$ and $\tilde{\Delta}$ are quite smaller compared with $\omega$, thus we get $\omega-\tilde{\Omega}_R \approx \tilde{\Delta}$. By means of a mathematical equality
\begin{equation}
  \cos(x\sin \alpha)= \sum_{n=-\infty}^{\infty} J_n(x)\cos n\alpha,
\end{equation}
we readily obtain
\begin{equation}\label{z_simplest}
  Z=\cos \left[\Delta \int_{0}^{t} d \tau \cos\left(\Xi \sin(\omega \tau)\right) \right].
\end{equation}

\begin{figure}
  \includegraphics[width=8cm]{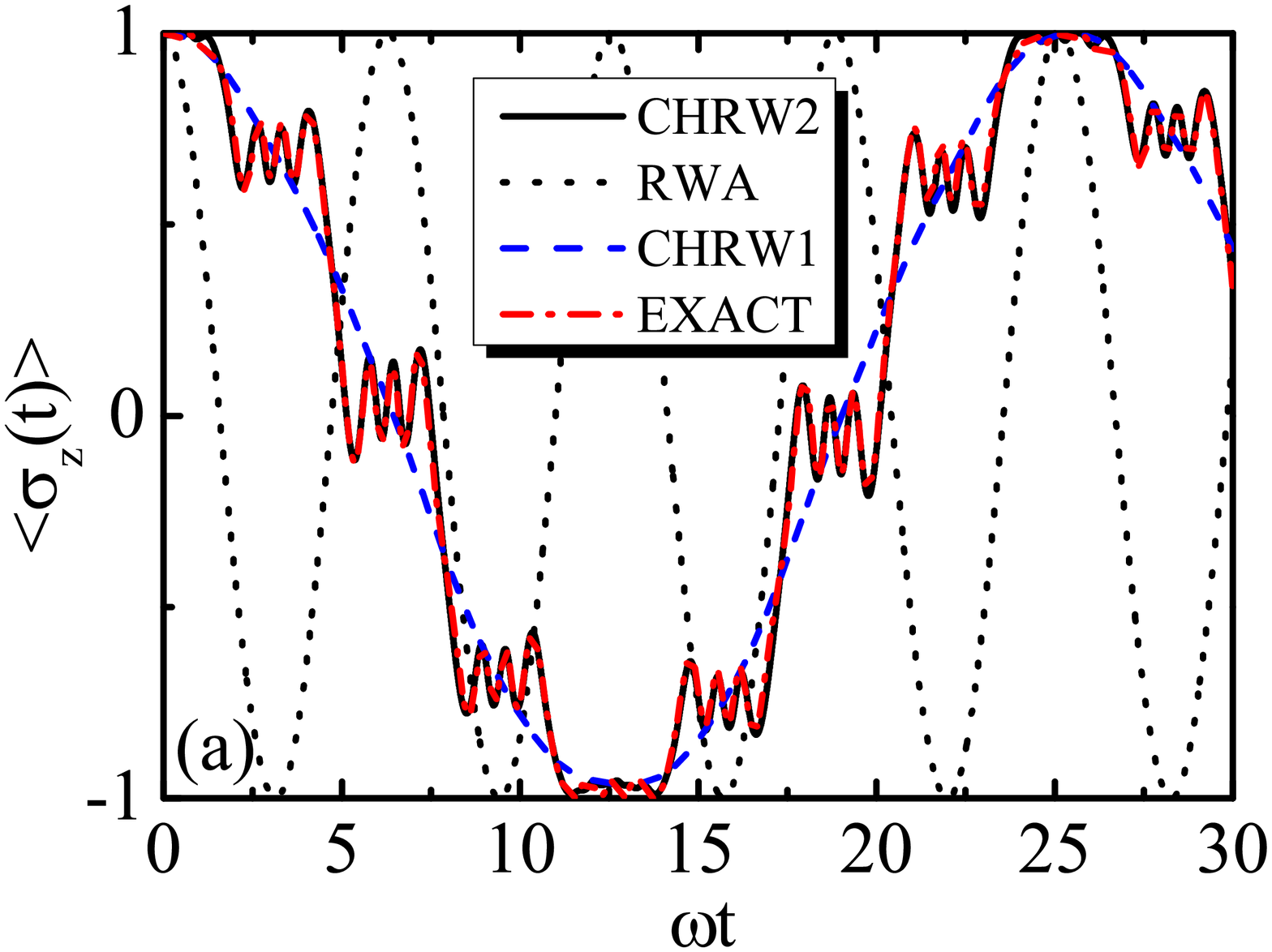}
  \includegraphics[width=8cm]{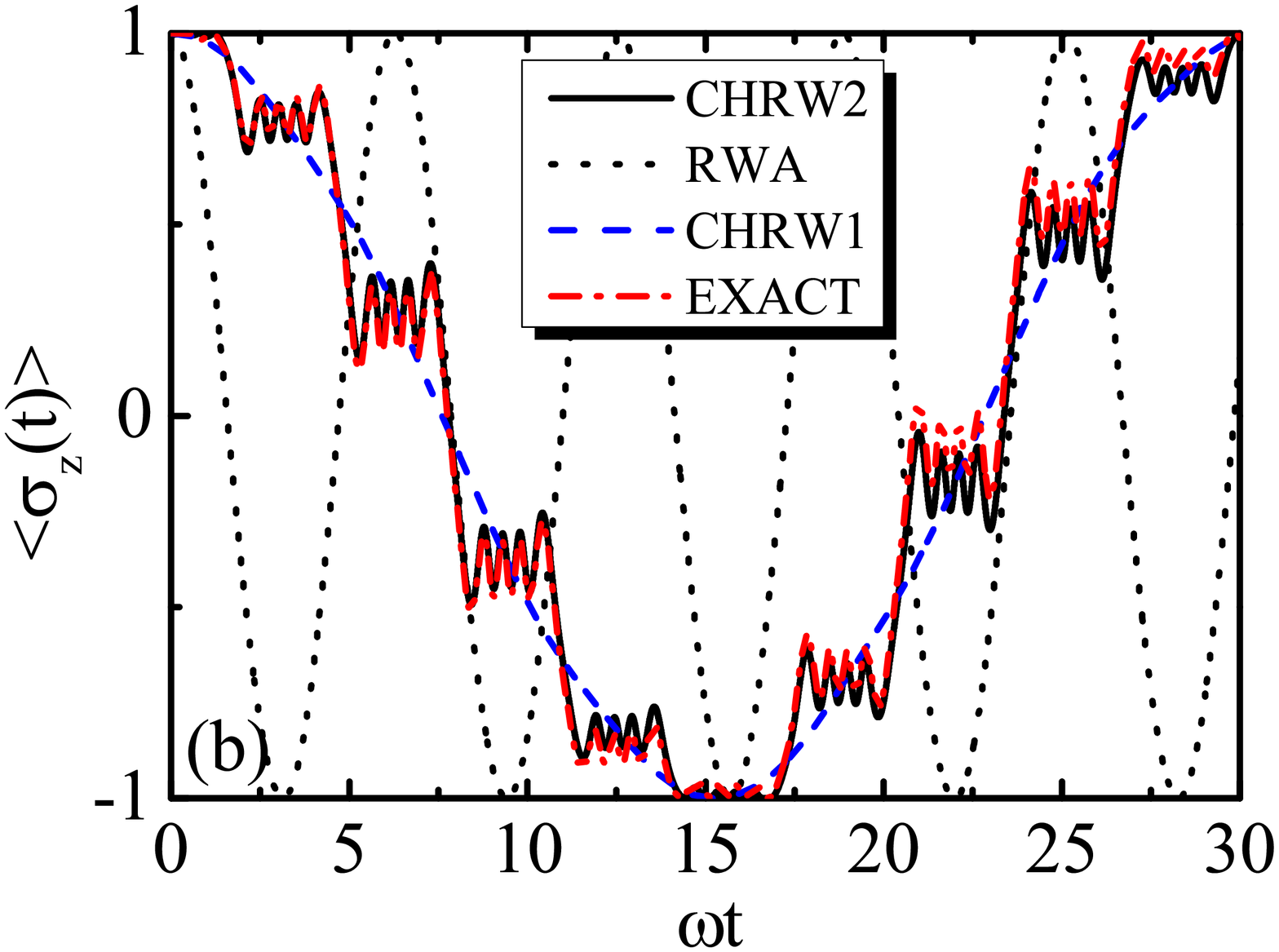}\\
  \includegraphics[width=8cm]{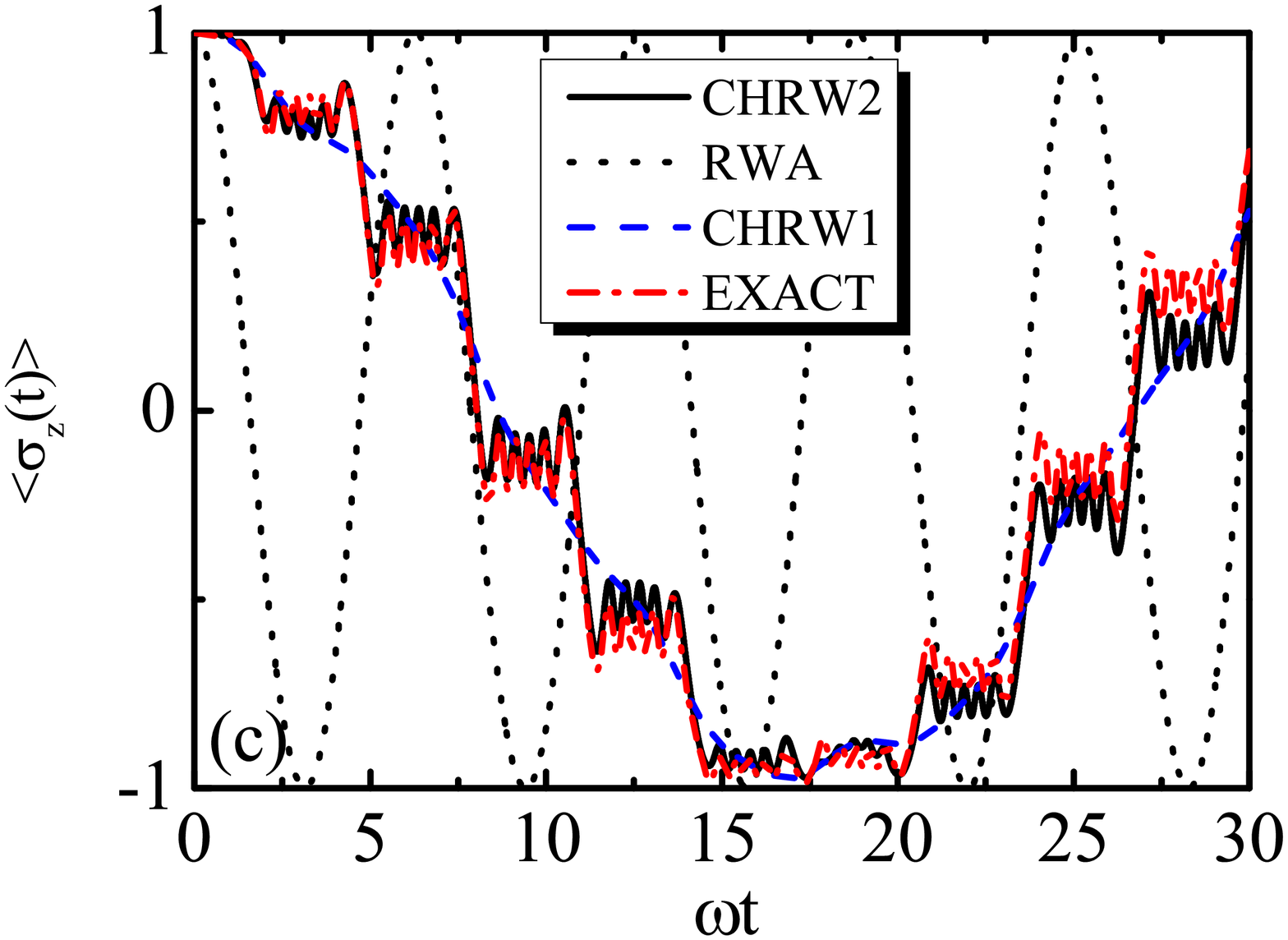}
  \includegraphics[width=8cm]{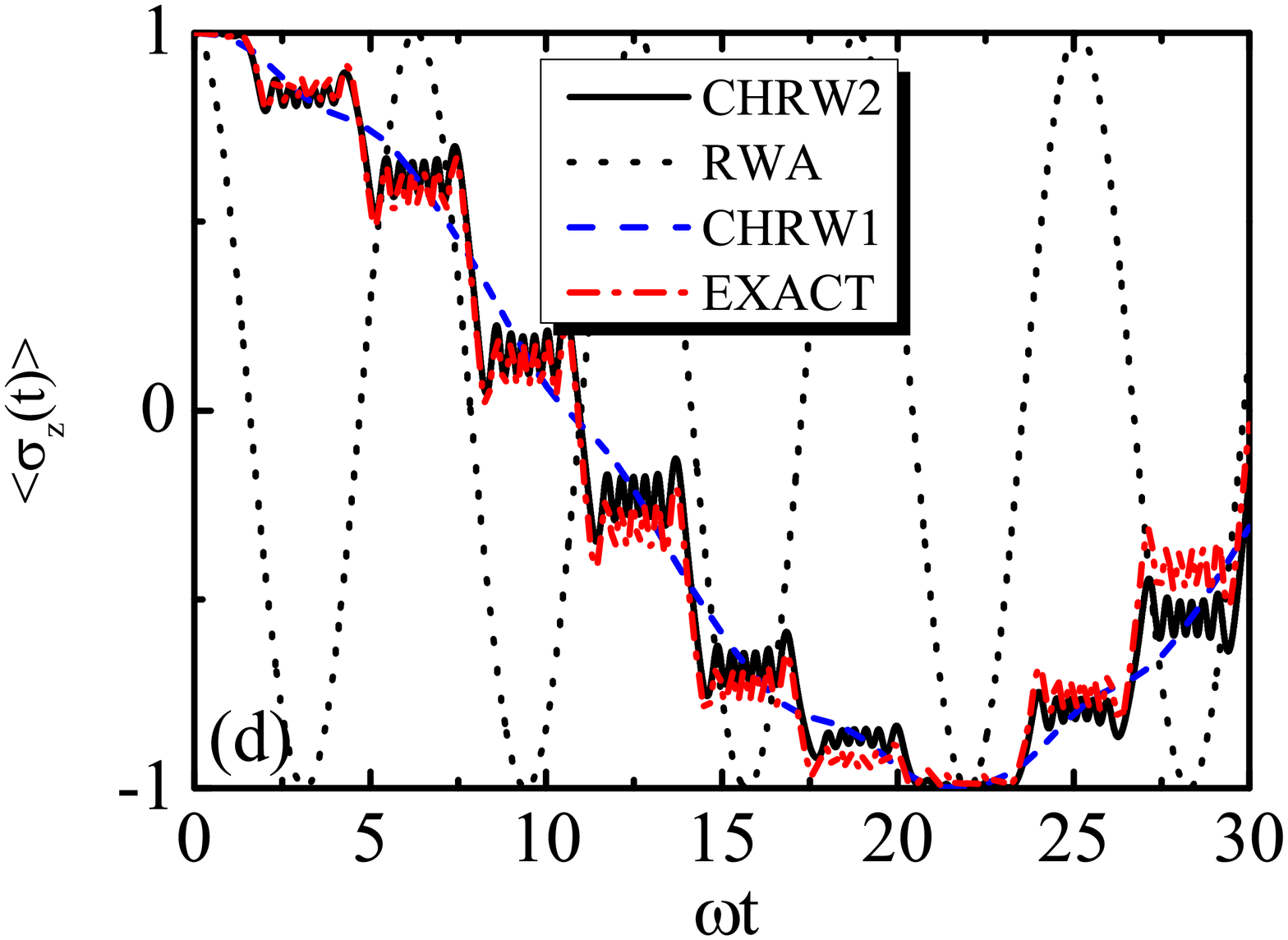}\\
  \caption{(Color online) $\langle \sigma_z(t) \rangle$ as a function of dimensionless time $\omega t$ for $A/\omega$=10, 13, 16 and 19 in the on-resonance case ($\Delta=\omega$), shown in (a), (b), (c) and (d), respectively. In each graph, the black line is the result of the CHRW2 method obtained by Eq. (\ref{z_complete}). }\label{fig1}
\end{figure}

\begin{figure}
  \includegraphics[width=8cm]{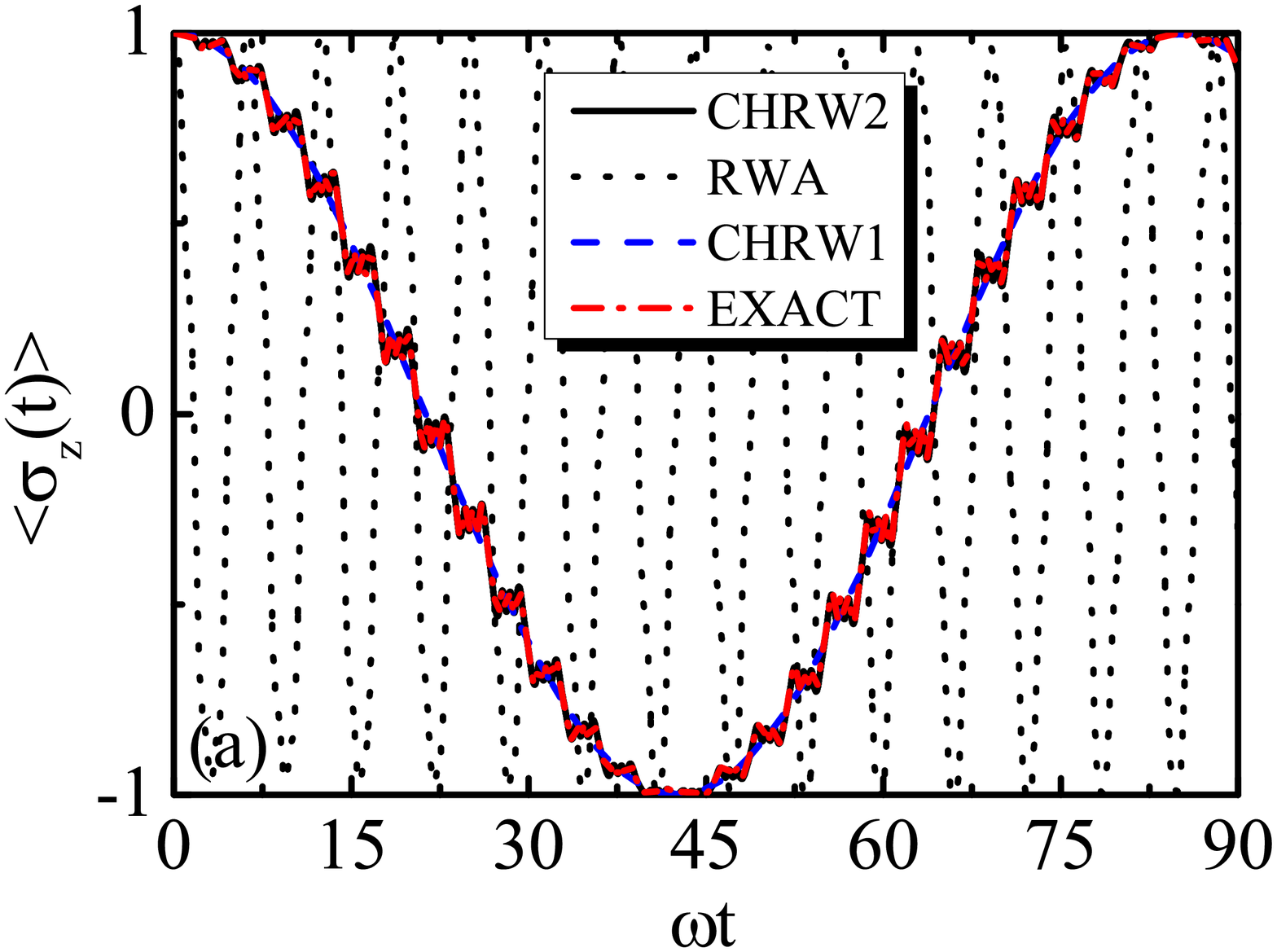}
  \includegraphics[width=8cm]{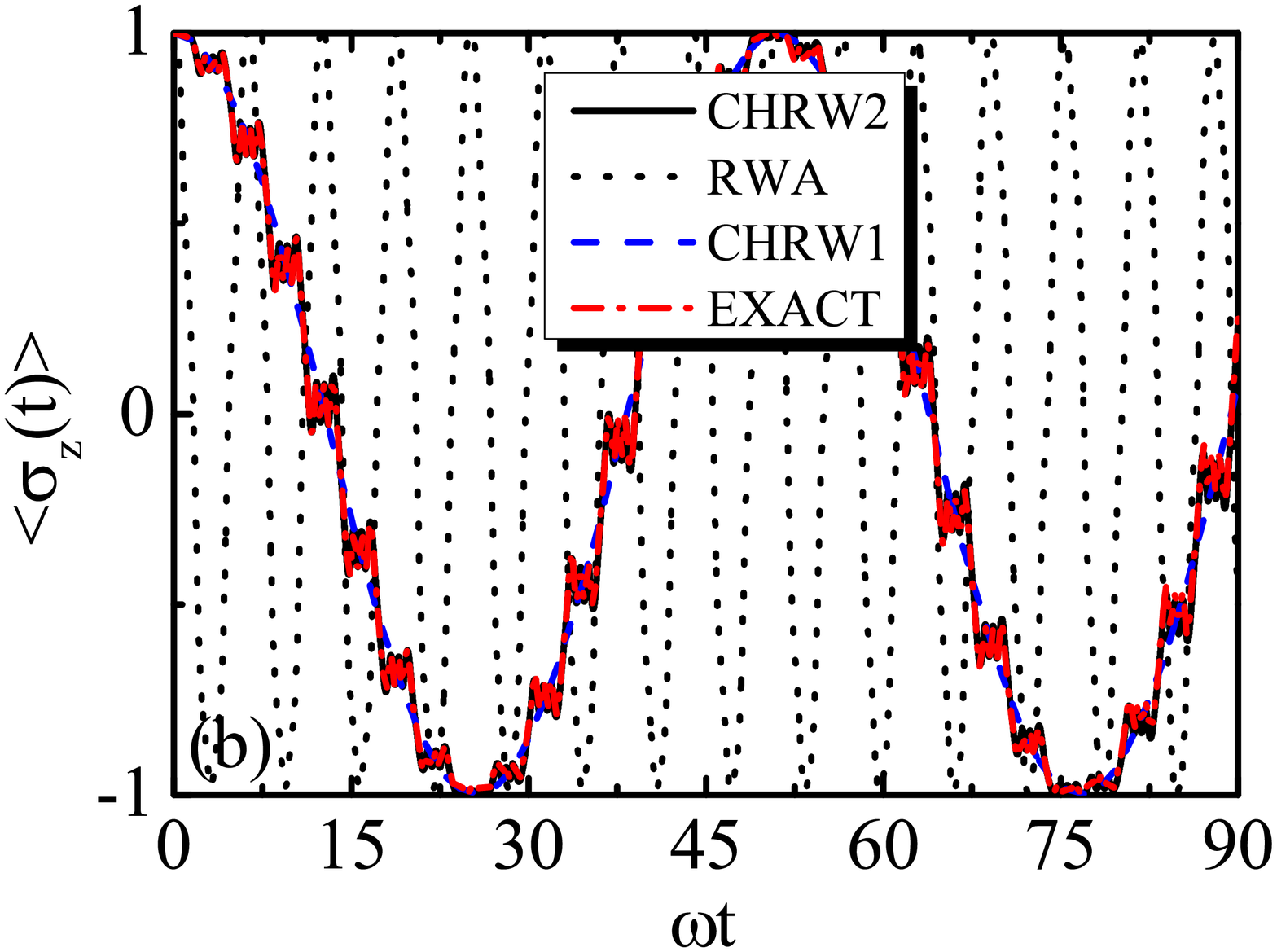}\\
  \includegraphics[width=8cm]{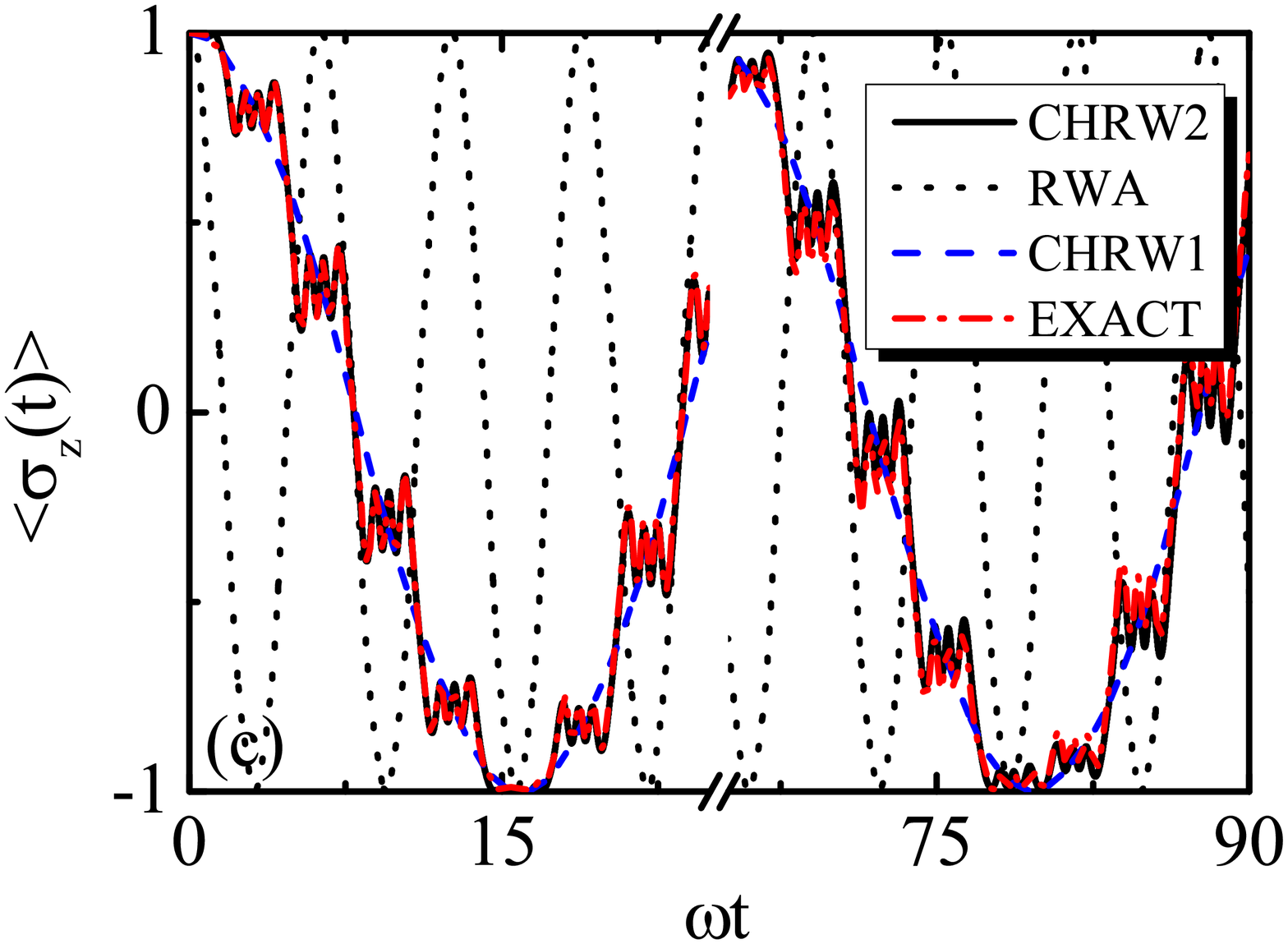}
  \includegraphics[width=8cm]{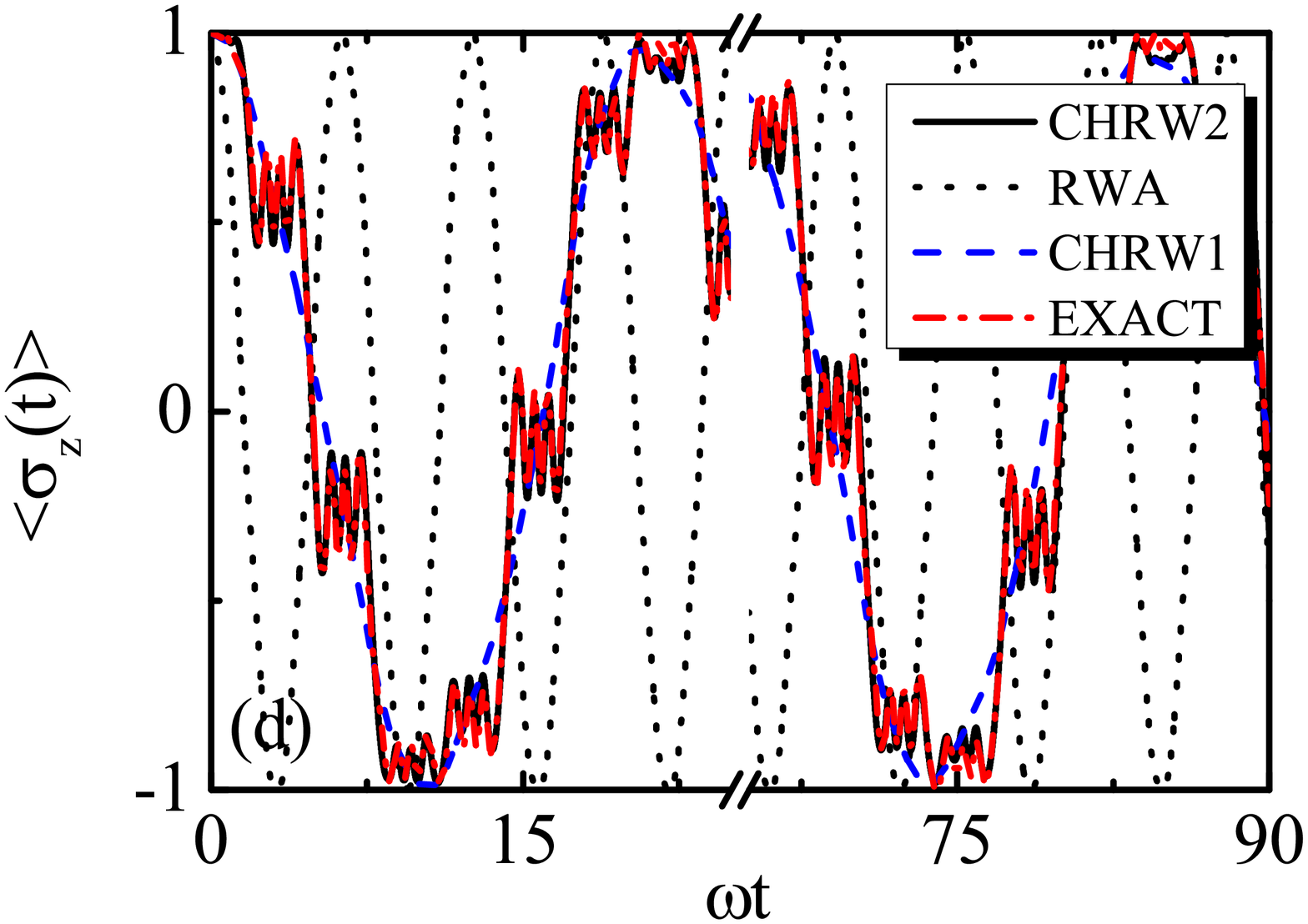}\\
  \caption{(Color online) $\langle \sigma_z(t) \rangle$ as a function of dimensionless time $\omega t$, under the off-resonance case with $A/\omega=10$ for $\Delta/\omega=0.3, 0.5, 0.8, 1.2$ shown in (a), (b), (c) and (d), respectively.
  In each graph, the black line is the result of the CHRW2 method obtained by Eq. (\ref{z_complete}).}\label{fig2}
\end{figure}

\begin{figure}
  \centering
  \includegraphics[width=8cm]{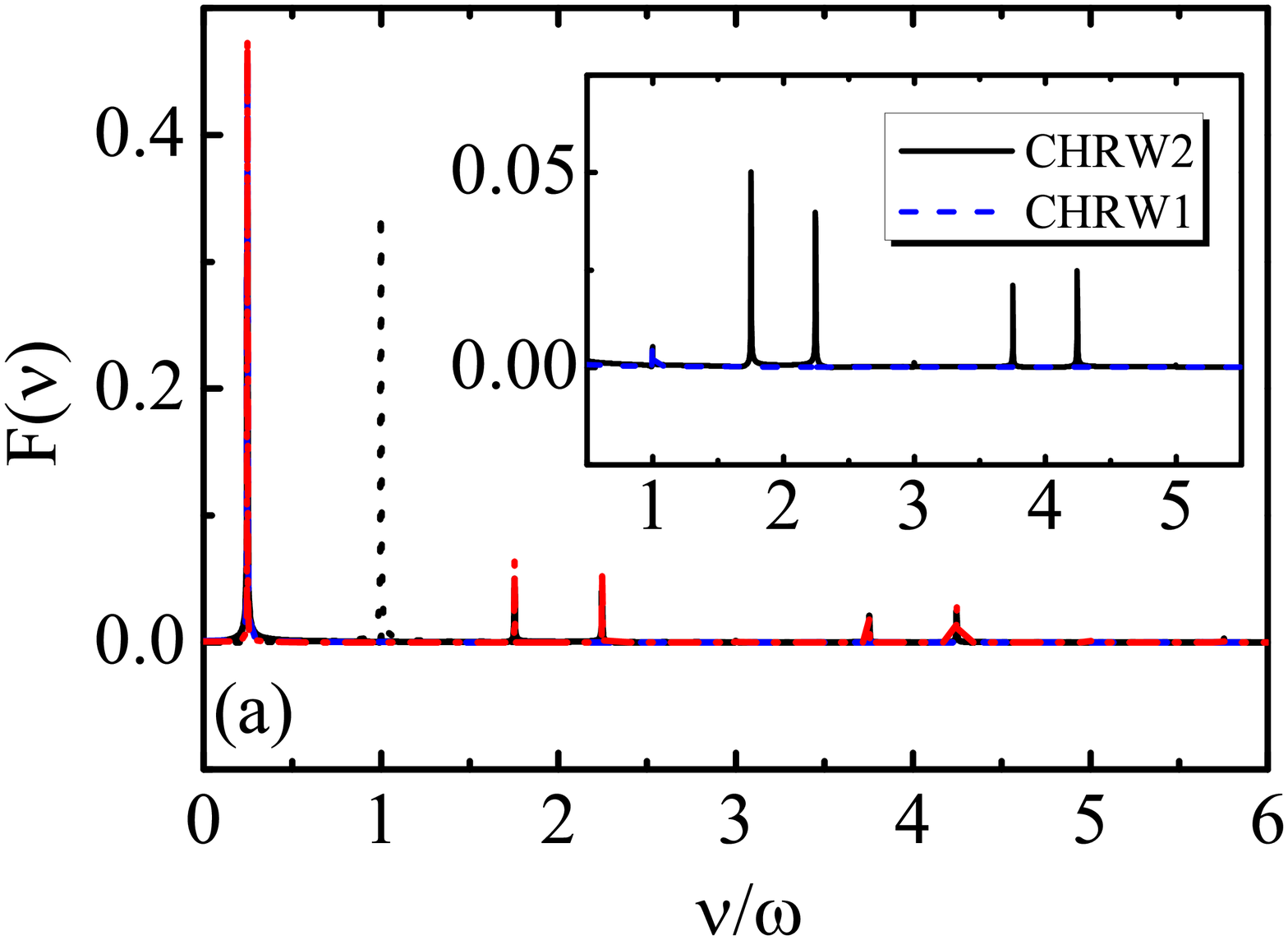}
  \includegraphics[width=8cm]{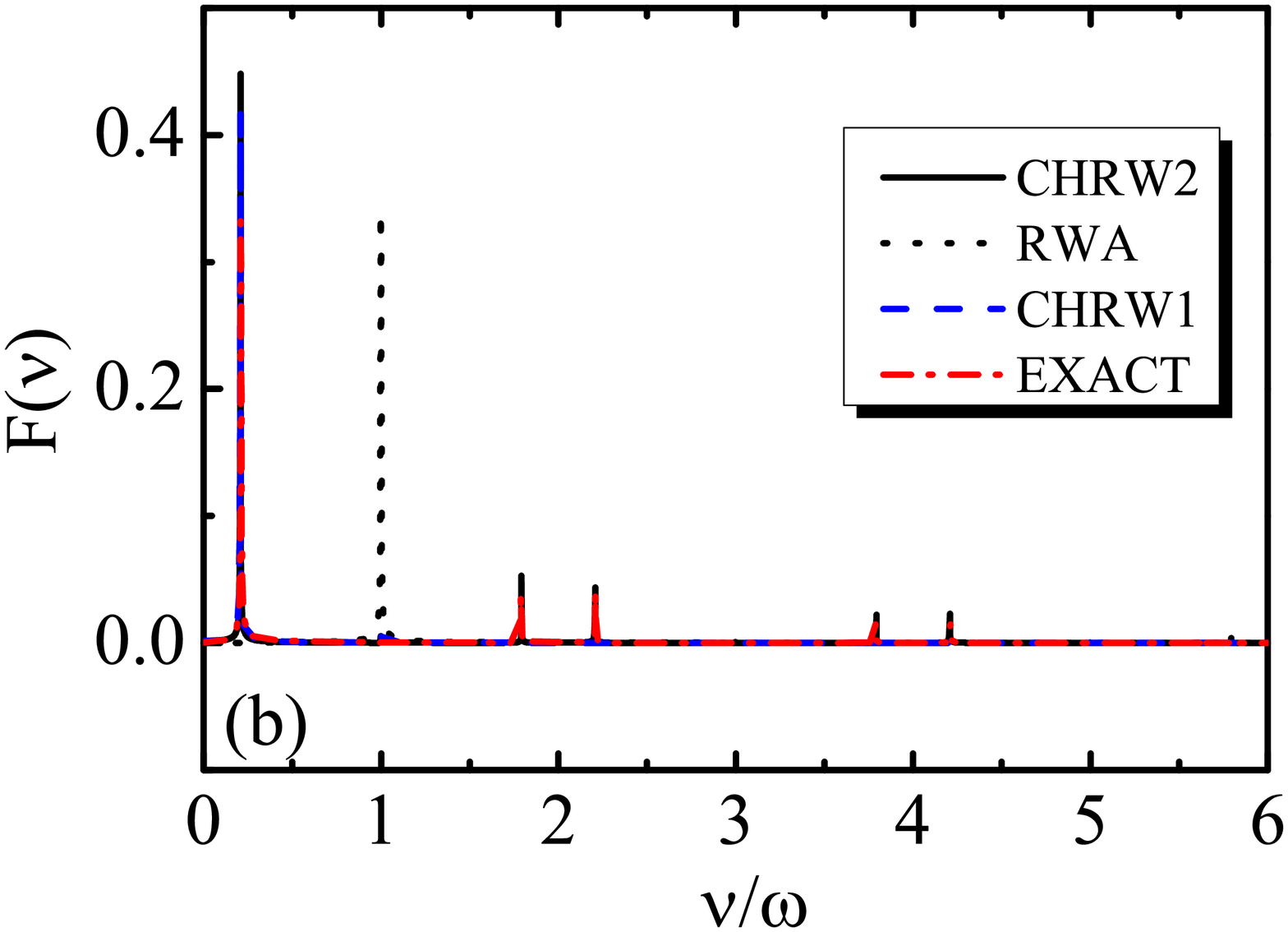}\\
  \includegraphics[width=8cm]{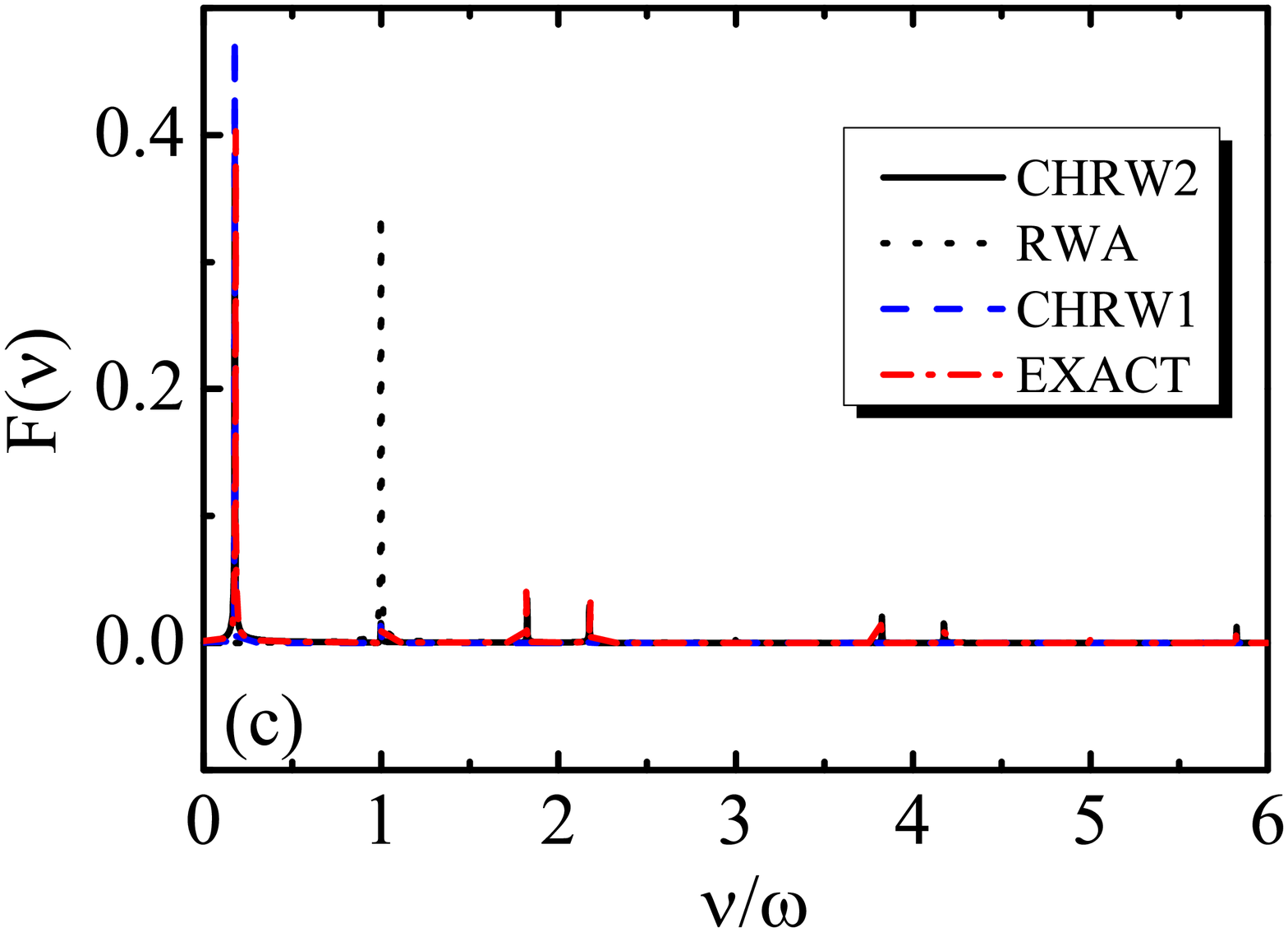}
  \includegraphics[width=8cm]{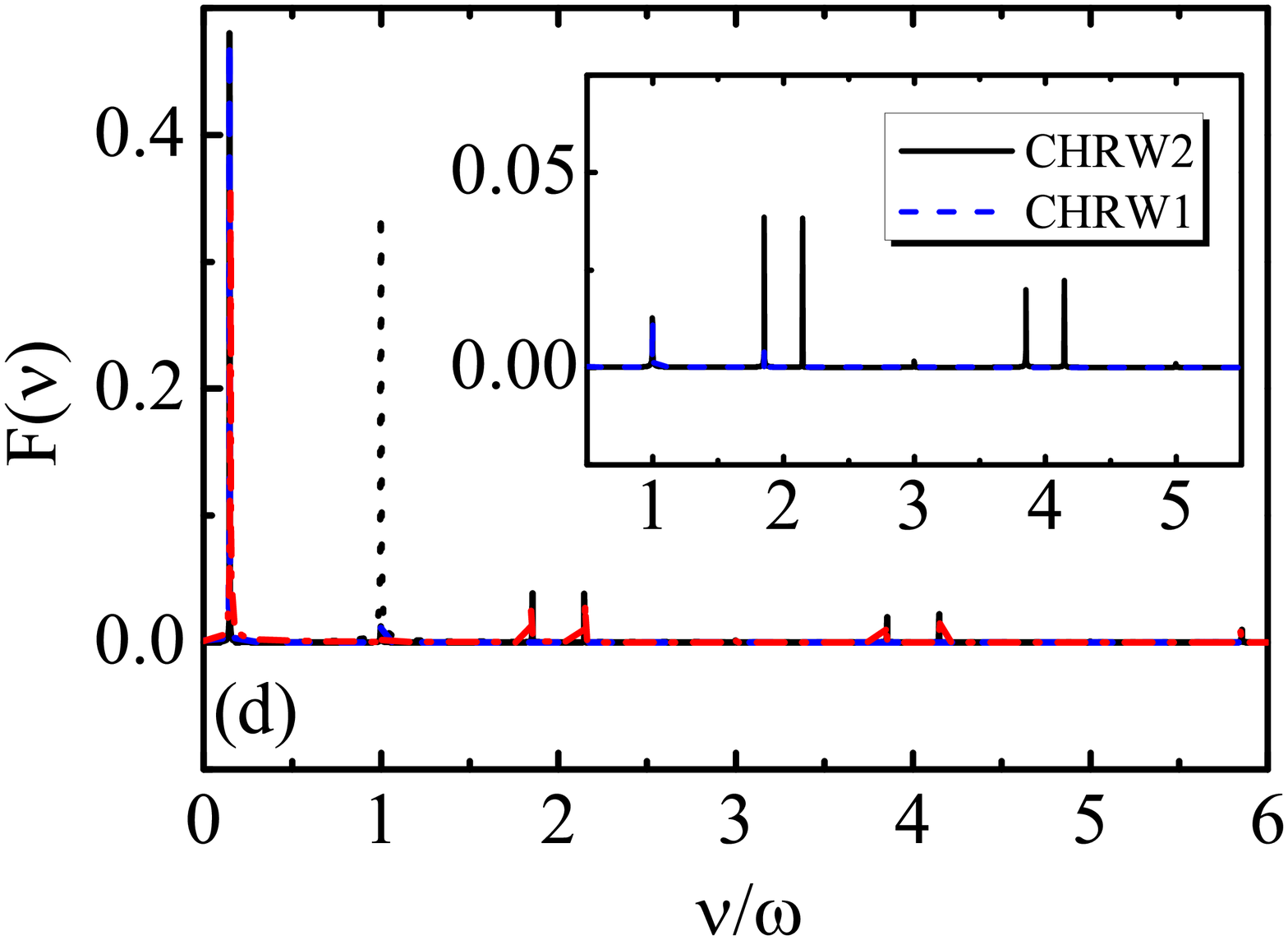}\\
  \caption{(color online) Fourier spectrum of $\langle \sigma_z \rangle$ with the same parameters as in Fig. \ref{fig1}. In each graph, the black line is the result of the CHRW2 method. The black dotted line is the result of the RWA method. The blue dashed line is the result of the CHRW1 method, and the red dash-dotted line is the result of numerically exact method. The insets of (a) and (d) show detailed comparison of the CHRW1 and the CHRW2 methods.} \label{FigFFT1}
\end{figure}

\begin{figure}
  \centering
  \includegraphics[width=8cm]{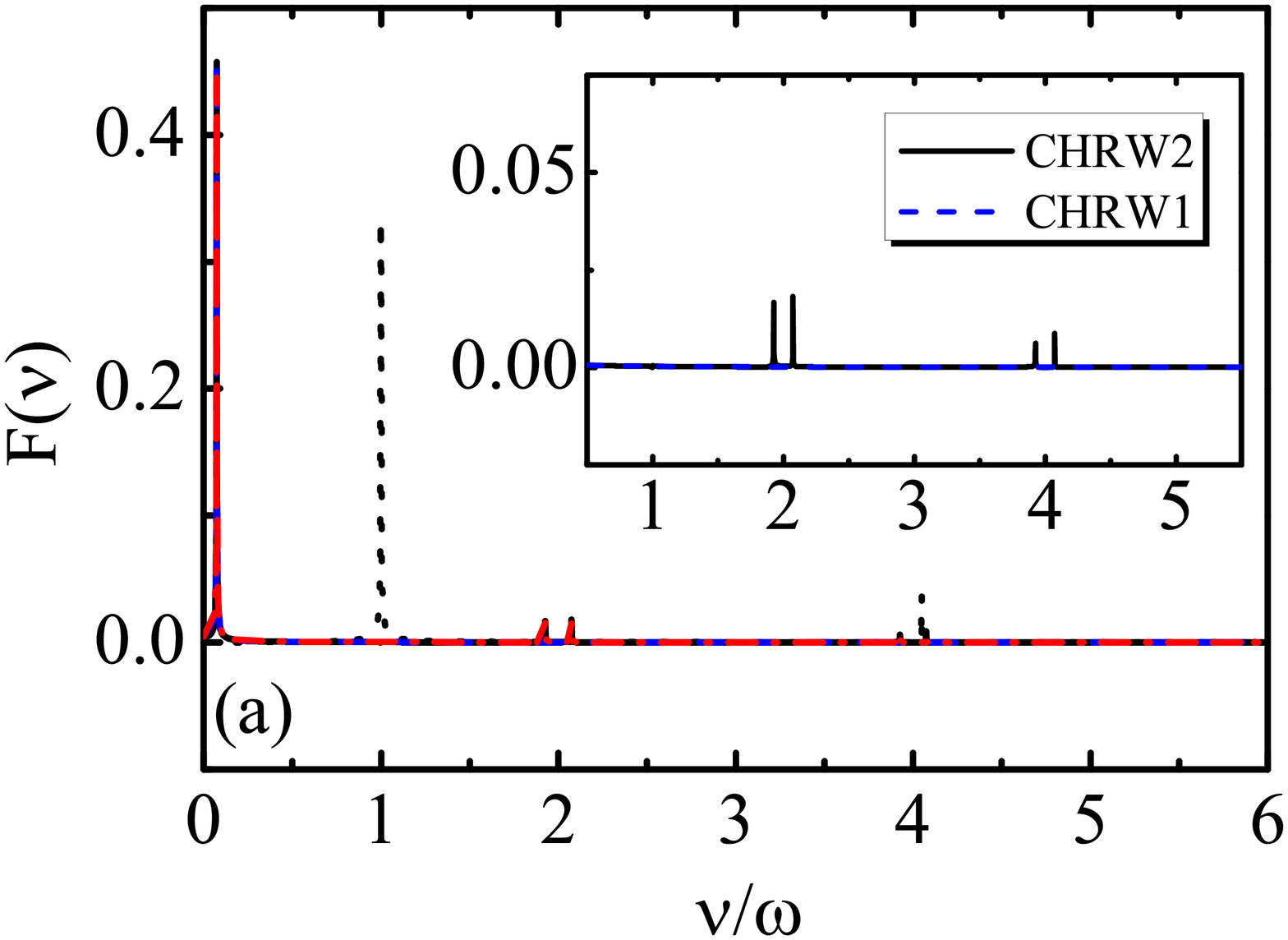}
  \includegraphics[width=8cm]{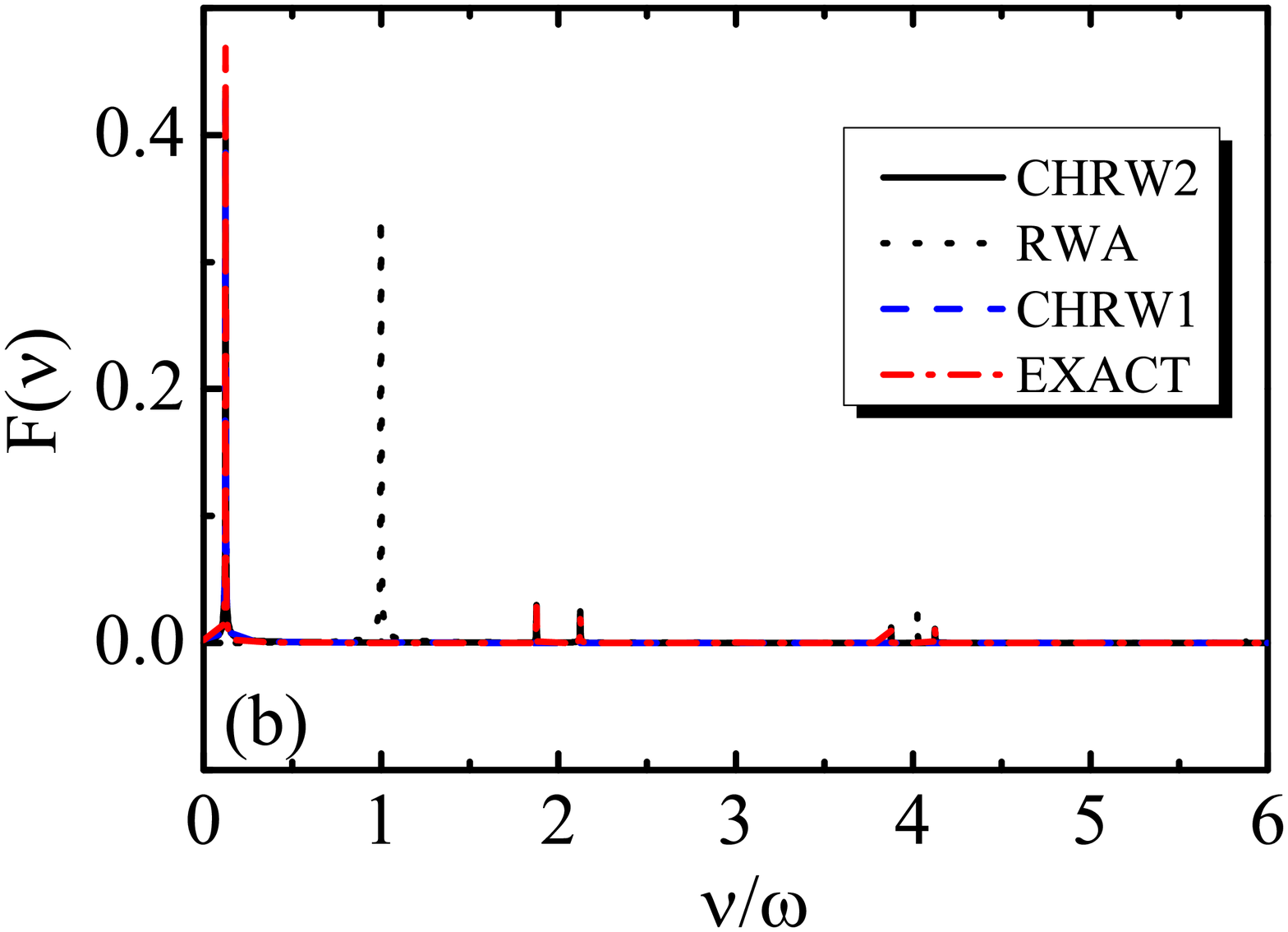}\\
  \includegraphics[width=8cm]{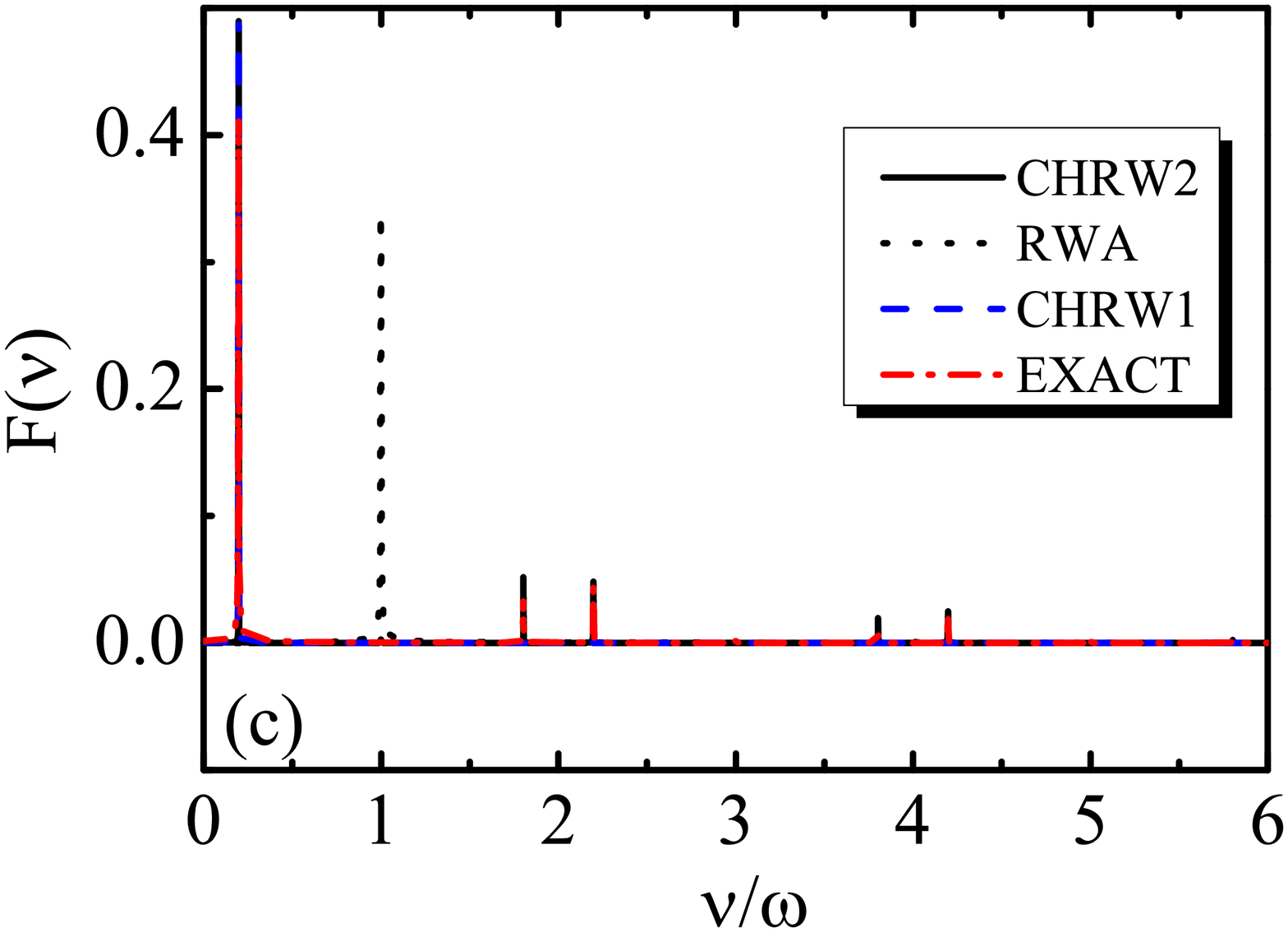}
  \includegraphics[width=8cm]{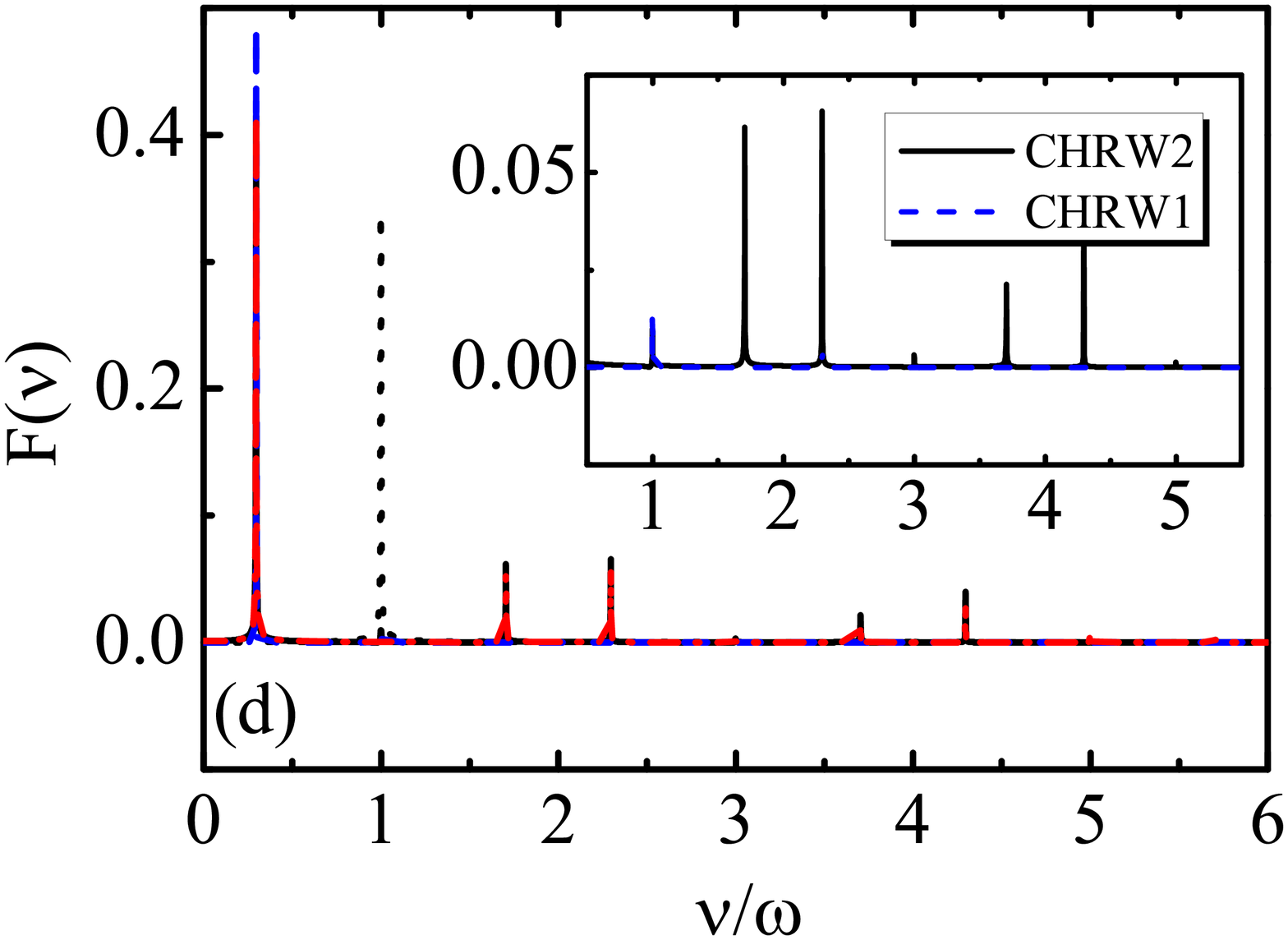}\\
  \caption{(color online) Fourier spectrum of $\langle \sigma_z \rangle$ with the same parameters as in Fig. \ref{fig2}.  The insets of (a) and (d) show detailed comparison of the CHRW1 and the CHRW2 methods.} \label{FigFFT2}
\end{figure}

\begin{figure}
  \centering
  \includegraphics[width=8cm]{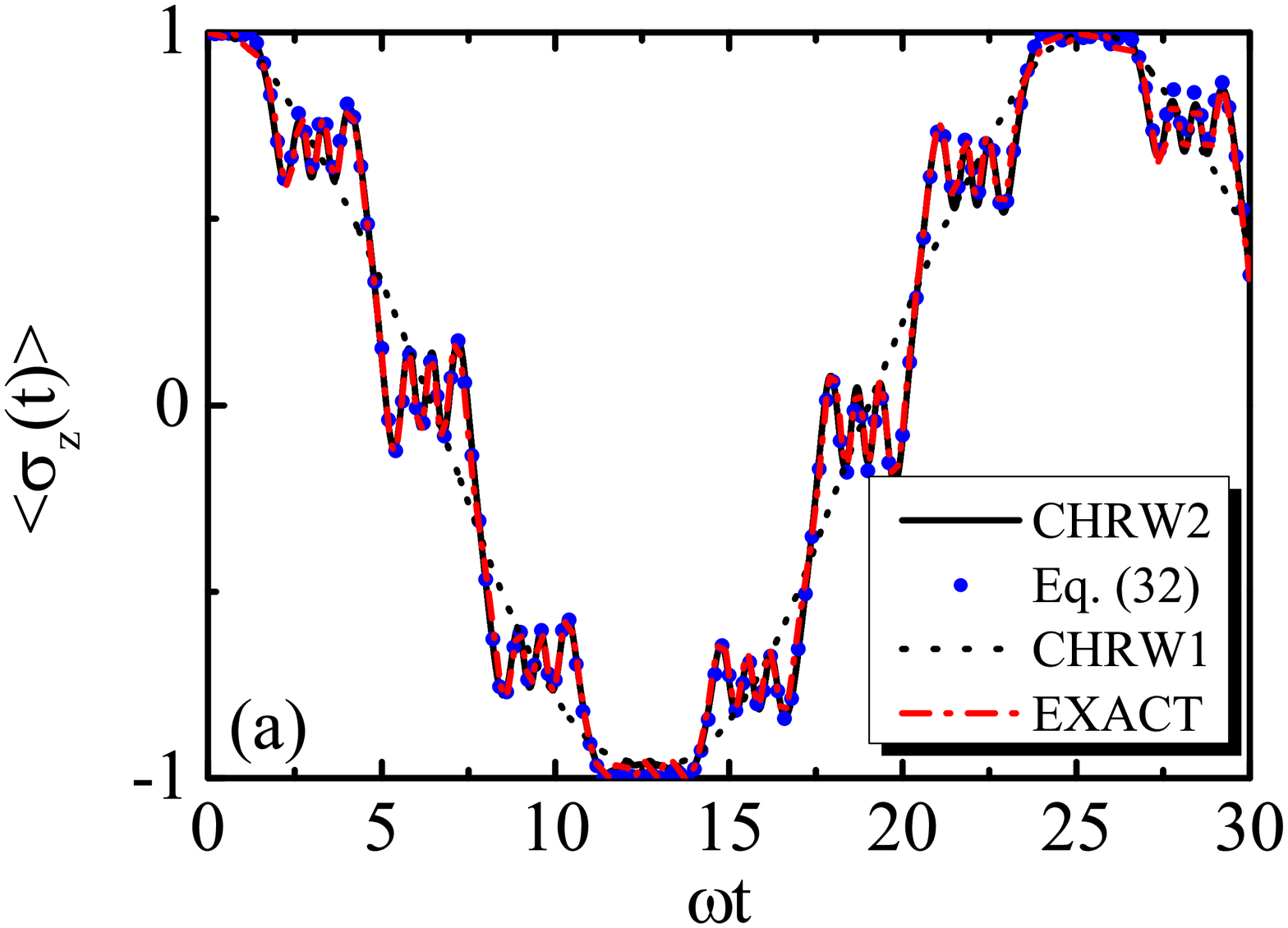}
  \includegraphics[width=8cm]{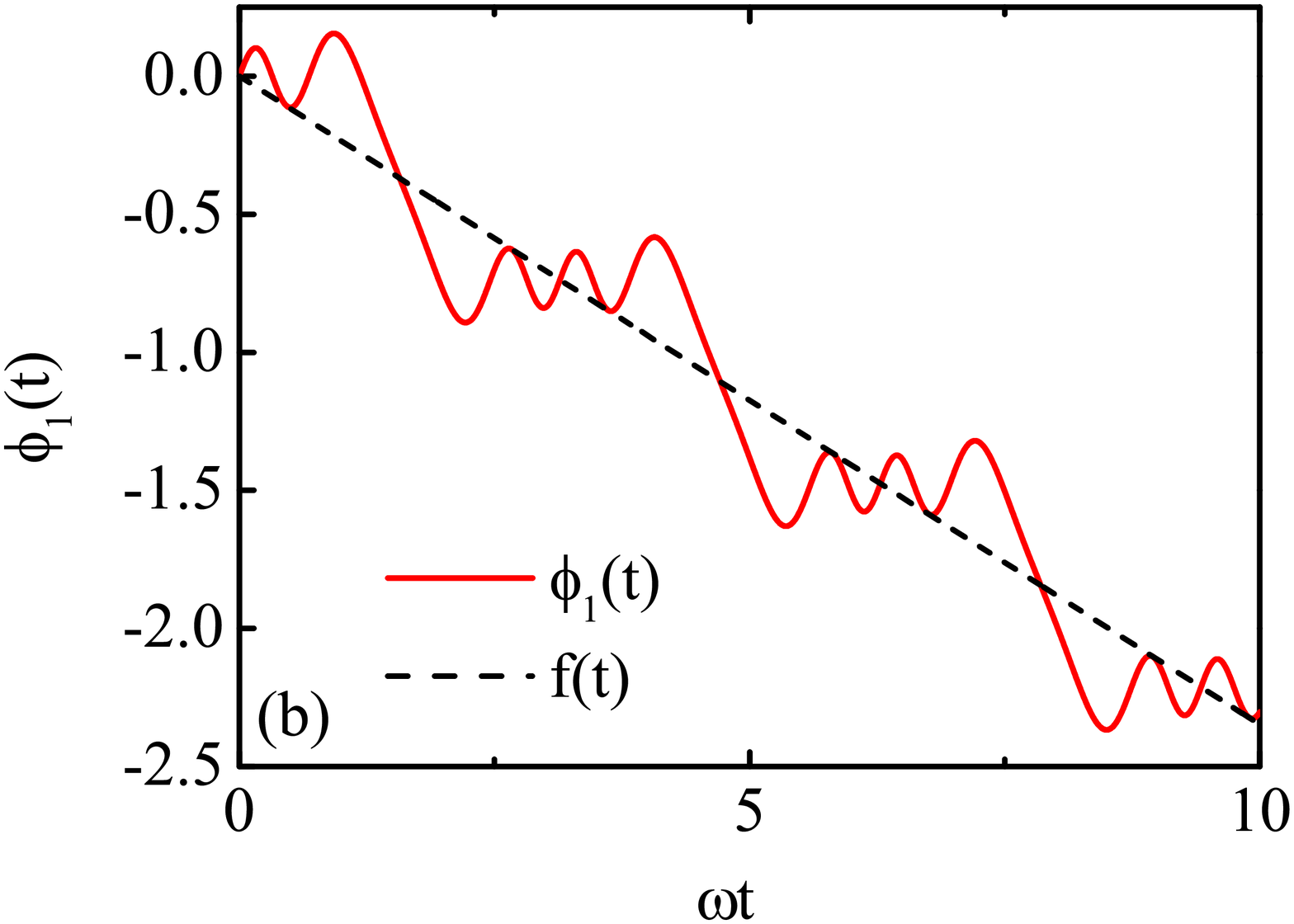}\\
  \caption{(Color online) (a) $\langle \sigma_z(t) \rangle$ as a function of dimensionless time $\omega t$ for $A/\omega=10$ and $\Delta=\omega$.
  The result of the CHRW2 method obtained by Eq. (\ref{z_complete}) is plotted by the black line. The result of Eq. (32) with $\xi=1$ is plotted by the blue dots. The result of the CHRW1 method is plotted by the black dotted line.
  The result of numerically exact method is plotted by the red dash-dotted line. (b) The phase function $\phi_1(t)=\Delta \int_{0}^{t} \cos\left[\frac{A}{\omega} \sin(\omega \tau)\right] d\tau$ shown in  Eq. (\ref{z_simplest}),
  as a function of dimensionless time $\omega t$, for $A=10\omega$ and $\Delta=\omega$.  The linear function $f(t)=\tilde{\Delta}t$ shows the general tendency of the phase function $\phi_1(t)$.}\label{fig2_add}
\end{figure}

This analytical result can give accurate dynamics under the condition above. For example, in comparison with the result in Fig. {\ref{fig1} }(a) for $A/\omega=10$ and $\Delta=\omega$, we show the result of $\langle \sigma_z(t) \rangle$ calculated by Eq. (\ref{z_simplest}) with $\xi=1$ in Fig. {\ref{fig2_add}}. Definitely, the result of Eq. (\ref{z_simplest}) is in perfect agreement with the CHRW2 and numerically exact results. Furthermore, the value of Eq. (\ref{z_simplest}) is determined by its phase function defined as $\phi_1(t)=\Delta \int_{0}^{t} d \tau \cos\left[\Xi \sin(\omega \tau)\right] $ which includes all even harmonics. We illustrate the phase function with $A=19 \omega$ and $\Delta=\omega$ in Fig. \ref{fig2_add}(b).

The key dynamical characters of $\langle \sigma_z(t) \rangle$ are attributed to the properties of the phase function $\phi_1(t)$. When $t=t_n=n\pi/\omega$, it is clear to see that the values of the phase function $\phi_1(t)$ satisfy a linear function $f(t)=\tilde{\Delta}t$ and, hence $Z \approx \cos(\tilde{\Delta}t_n)$ at the same time, which signifies the efficiency of the CHRW1 and CHRW2 methods by correctly illustrating the general tendency of time evolution. Moreover, the phase function $\phi_1(t)$ involving all even harmonics presents fast oscillation with a small amplitude around the average value $\tilde{\Delta}n\pi/\omega$ when $t$ is during the interval $\left(\frac{n\pi}{\omega}-\frac{\pi}{2\omega}, \frac{n\pi}{\omega}+\frac{\pi}{2\omega}\right)$. What is more, it also exhibits a sharp variation only when $t$ is very close to $\frac{(2n-1)\pi}{2\omega}$. The characters of $\phi_1(t)$ determine the dynamical features of $\langle\sigma_z(t)\rangle$.  That is why the dynamical pattern of $\langle\sigma_z\rangle$ is called the plateau phenomenon with sudden jump, i.e., zigzag plateau. Furthermore, the zigzag plateau comes from the collective effect of all even harmonics because of the explicit phase function $\phi_1(t)$. From detailed analysis in Appendix. \ref{appen1}, we can see that the oscillations of these plateaus satisfy a quasi-periodic structure illustrated by $\Xi\sin(\omega t)=m\pi+\frac{\pi}{2}$, and accordingly the period number of any plateau is obtained as $N=\lfloor\frac{2A}{\omega\pi}\rfloor-\lfloor\frac{A}{\omega\pi}\rfloor$, where $\lfloor \bullet \rfloor$ is a floor function. It is really notable that the period number is independent of $\Delta$ in the parameter region of interest. Table 2 shows the corresponding precise values of the period number $N$ in Figs. \ref{fig1} and \ref{fig2}, in good agreement with the rule above. The periodic oscillation of each plateau is called quantized oscillation. It is clear to see in Fig. \ref{fig1} that the number $N$ increases as the driving strength $A$ increases. For example, $N$ increases from $3$ to $6$ when $A/\omega$ raises from 10 to 19. However, Table II shows the number $N$ of Fig. \ref{fig2} for the fixed $A/\omega=10$ does not change with the increase of $\Delta$.

\begin{table}
\caption{ The number of period for each zigzag plateau in Figs. \ref{fig1} and  \ref{fig2}.}
\centering
\begin{tabular}{c c c c c}
  \hline\hline
  Fig.\ref{fig1} & (a) & (b) & (c) & (d)\\
  \hline
  $N$  & 3 & 4 & 5 & 6\\
  \hline
  Fig.\ref{fig2} & (a) & (b) & (c) & (d)\\
  \hline
  $N$ & 3 & 3 & 3 & 3\\
  \hline\hline
\end{tabular} \\
Here $N$ is the number of period of small-amplitude oscillation for each plateau.
\end{table}

Another remarkable feature of the plateau structure is the amplitude of the oscillation, i.e., the deviation from the average value of a plateau. The envelope of the amplitude of the plateau satisfies
\begin{equation}
  |g_1(t)|<q_1(t)=\frac{2\Delta}{A\cos(\omega t)}.
\end{equation}
which is derived in the Appendix (see  Eq. (\ref{inequality3})).
It is obvious to see that the amplitude of this quantized oscillation is proportional to $\Delta/A$. Subsequently, the oscillation is much suppressed as $A$ increases or $\Delta$ decreases which is clearly seen in Figs. \ref{fig1} and \ref{fig2}.

\section{Odd-harmonic processes under strong driving} \label{oddeff}
The CDT is a fascinating phenomenon in the tunneling dynamics of a driven quantum system, which is corresponding to the situation that tunneling is much suppressed and even frozen for certain driving parameters. From the viewpoint of $\langle \sigma_z(t) \rangle$,  its average value is almost constant and its amplitude is extremely small, namely the system stays in the initial state.

It is known that the CDT phenomenon occurs when $\omega \gg \Delta$ and $A/\omega$ is one of the zeros of $J_0(x)$, the zero-order Bessel function of the first kind \cite{CHRW:origin}. Note that the shift of the CDT condition due to the key parameter $\xi$ in Eq.(\ref{deltatilde}) has been reported in Ref. \cite{CHRW:origin}. It seems that the traditional CDT condition which is corresponding to zeros of $J_0(A/\omega)$ has no relation with $\omega/\Delta$. However, in the discussion of ~Ref. \cite{CHRW:origin}, as $\omega/\Delta$ increases from $2$ to $20$, the driving-induced suppression of tunneling emerges clear only for $\omega/\Delta \geq 6$. For an intermediate large driving frequency, for example, $\omega/\Delta=6$ and $A/\omega=2.3976$,  $\langle\sigma_z(t)\rangle$ exhibits an oscillation with small amplitude. In contrast, the amplitude of $\langle \sigma_z(t) \rangle$ for $A/\omega=2.4048$, which is a zero of $J_0(x)$, comes to increase. This shift of the CDT condition becomes vanishing for $\omega/\Delta \gg 1$ because of $\xi \rightarrow 1$. In this section we investigate the dynamics with a small amplitude in the regime of $A \gg \omega$ and $\Delta \geq \omega$. By the unitary transformation similar to the CHRW method, we will predict an interesting dynamics with peculiar plateau structures under the condition above, which is called the odd-harmonic effect.  We explain the quantized oscillation by the odd-harmonic effect.

\subsection{The odd-harmonic effect}

To begin with, let us come back to Eq. (\ref{origin}) and take the same unitary transformation as Eq. (\ref{S1}) with $\xi=1$. Therefore, we get the Hamiltonian in the unitarily transformed picture as
\begin{equation}\label{odd_1}
    H_{3}(t) = -\frac{\Delta}{2} \sum_{n=-\infty}^{\infty}J_{n}\left(\frac{A}{\omega}\right)\left[\cos(n\omega t)\sigma_x+\sin(n\omega t)\sigma_y\right]. \\
\end{equation}
We denote the $m$th zero of $J_0$ by $\alpha_m $. According to asymptotic behavior of the Bessel function
\begin{equation}\label{approaching}
  J_n\left(\frac{A}{\omega}\right) \longrightarrow \sqrt{\frac{2\omega}{\pi A}}\cos\left(\frac{A}{\omega}-\frac{n\pi}{2}-\frac{\pi}{4}\right),
\end{equation}
(where $A \gg \omega$), we obtain an approximate expression of $\alpha_m$ as
\begin{equation}\label{alpha}
  \alpha_m=\frac{A}{\omega}=m\pi-\frac{\pi}{4},~~~(m \in Z^{+}).
\end{equation}
For $\frac{A}{\omega}=\alpha_m$, i.e., the CDT condition, it is explicit to check that $J_n\left(\frac{A}{\omega}\right) \approx 0$ when $n$ is even and $n \ll A/\omega$ by Eq. (\ref{approaching}). When $n \gg A/\omega$, it is clear that $J_n\left(\frac{A}{\omega}\right) \approx 0$ because of the properties of the Bessel function for a large driving strength $A/\omega \gg 1$. Therefore, it is reasonable to omit all terms with even harmonics and an effective Hamiltonian reads
\begin{eqnarray}
  H_4 &=& -\Delta\sum_{n=1}^{\infty}J_{2n-1}\left(\frac{A}{\omega}\right)\sin[(2n-1)\omega t]\sigma_y,  \label{odd_eff}
\end{eqnarray}
by using $J_{-n}(x)=(-1)^nJ_{n}(x)$ \cite{note1}. Because all the terms remained in Eq. (\ref{odd_eff}) are corresponding to odd harmonics, their effect are so-called ``odd-harmonic'' effect. We obtain an exact wave function by solving Eq. (\ref{odd_eff}),
\begin{equation}\label{odd_solution}
  \Phi^{\prime}(t)=\frac{1}{2}e^{i\frac{\Delta}{2}\int_{0}^{t}\sin\left[\frac{A}{\omega}\sin(\omega \tau)\right]d\tau}\left(\begin{array}{c}
                1 \\
                i
              \end{array}\right)-\frac{i}{2}e^{-{i}\frac{\Delta}{2}\int_{0}^{t}\sin\left[\frac{A}{\omega}\sin(\omega \tau)\right]d\tau}\left(\begin{array}{c}
                i \\
                1
              \end{array}\right),
\end{equation}
where we have supposed $\Phi(0)=\left(\begin{array}{c}
                                      1 \\
                                      0
                                    \end{array}\right)$
without loss of generality and the initial state is the same as that in the transformed frame, i.e. $\Phi(0)=\Phi^{\prime}(0)$ . In comparison with the CHRW1
method, it has a different form of wave function, suggesting the physical quantities, such as $\langle \sigma_z(t) \rangle$, will not be simply frozen at initial values but possess sophisticated dynamics.

In order to calculate some physical quantities of interest, we find that the average values $\langle \sigma_i(t) \rangle$ ($i=x,y,z$), in the unitarily transformed frame are described as
\begin{eqnarray}
  X^{\prime} &=& -\sin\left\{\Delta \int_{0}^{t}d\tau\sin\left[\frac{A}{\omega}\sin(\omega \tau)\right]\right\}, \\
  Y^{\prime} &=& 0, \\
  Z^{\prime} &=& \cos\left\{\Delta \int_{0}^{t}d\tau\sin\left[\frac{A}{\omega}\sin(\omega \tau)\right]\right\}.\label{cdt_z}
\end{eqnarray}
Similar to the CHRW1 method, we acquire the transformation matrix with $\xi=1$,
\begin{equation}\label{odd_transformation}
  T'' = \left( \begin{array}{ccc}
               \cos\left[\frac{A}{\omega}\sin(\omega t)\right] & \sin\left[\frac{A}{\omega}\sin(\omega t)\right]  & 0 \\
               -\sin\left[\frac{A}{\omega}\sin(\omega t)\right]  & \cos\left[\frac{A}{\omega}\sin(\omega t)\right]  & 0 \\
               0 & 0 & 1
             \end{array}\right).
\end{equation}

We investigate the CDT phenomenon of the TLS under strong driving. When the driving strength is extremely larger ($A \gg \omega$), the CDT condition
$J_0\left(\frac{A}{\omega}\right)=0$ leads to an effective Hamiltonian including only odd harmonics. In Ref. \cite{CHRW:origin},  we have demonstrated
$A/\omega=2.4048$ is not a sufficient condition of the CDT, because of the effects of higher-order even harmonics ($n\geq 2$) and odd harmonics ($m\geq 1$).
In fact, for $  A/\omega=m\pi-\pi/4 ~~(m \in Z^{+})$, we can obtain a simpler analytical solution under the CDT condition (in a proper approximation).
In the case $m\gg 1$, the effect of even-harmonic terms, which plays a significant role in the Sec II, diminishes in the Eq. (\ref{cdt_z}). Therefore,
the odd-harmonic effects dominates the CDT dynamics. Furthermore, since the unitary operator $\exp(S_1)$ commutes with $\sigma_z$, we obtain the dynamics of $\langle \sigma_z(t) \rangle$,
\begin{equation} \label{cdt_z2}
  Z = \cos\left\{\Delta \int_{0}^{t}d\tau\sin\left[\frac{A}{\omega}\sin(\omega \tau)\right]\right\},
\end{equation}
which is in contrast with the RWA result in the rotating frame \cite{RWA-RF,Rabi:untrastrong_regime},
\begin{equation} \label{RWA-RF}
  Z_{\rm RWA-RF} = \cos\left[\Delta J_0\left(\frac{A}{\omega}\right) t\right].
\end{equation}

It is obvious that instead of freezing in the initial state, the time evolution of $\langle \sigma_z(t) \rangle$ possessing a time-dependent oscillatory integral similar to Eq. (\ref{z_simplest}), might exhibit an analogous ``plateau'' structure on the CDT condition. In the following subsection we will illustrate the dynamics of $\langle \sigma_z(t) \rangle$ under the CDT condition by several methods, and give analytical results to account for this kind of plateau structure.

\subsection{Population dynamics and armchair plateaus}\label{calc_odd}

\begin{figure}
  \centering
  \includegraphics[width=8cm]{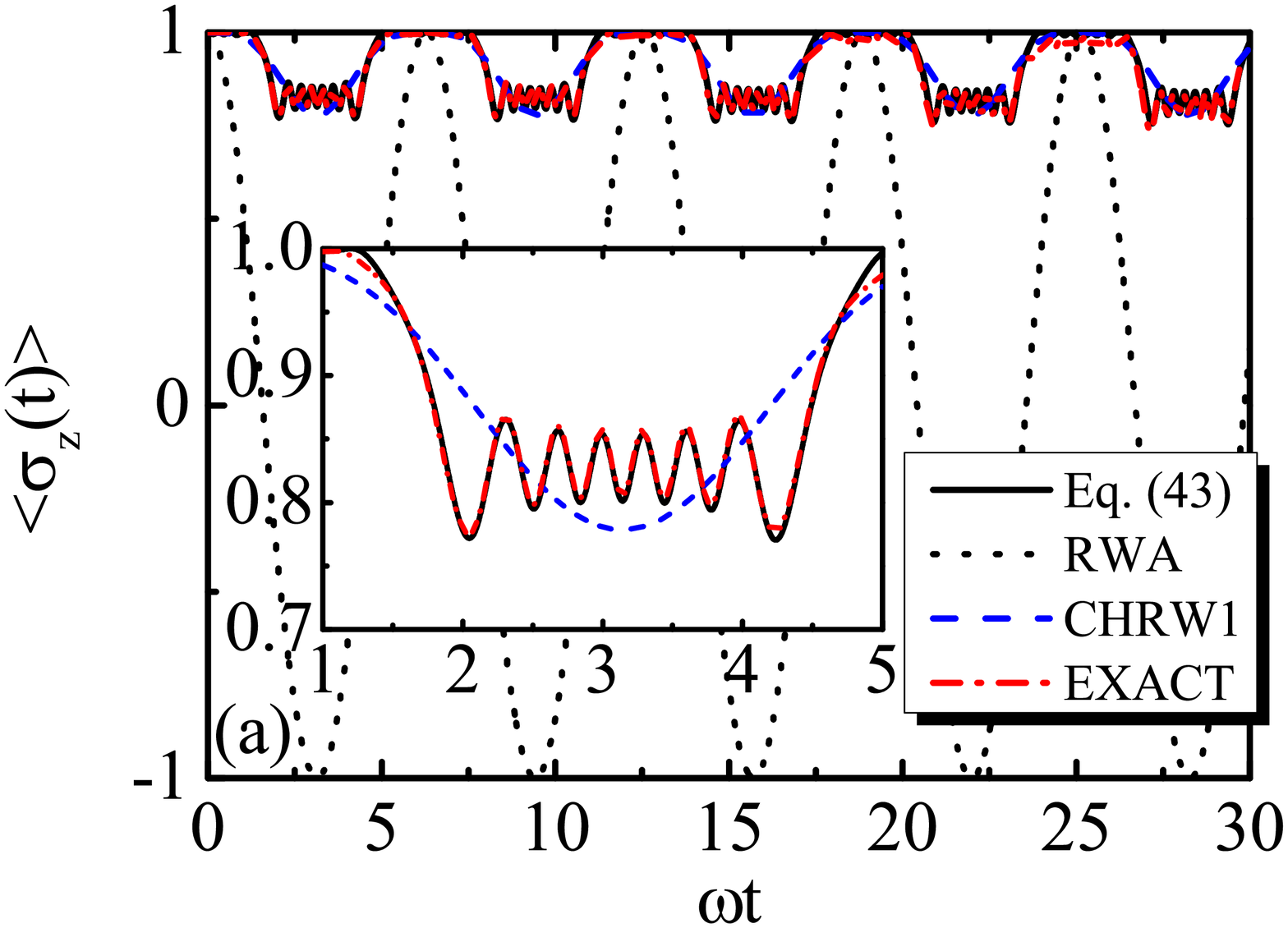}
  \includegraphics[width=8cm]{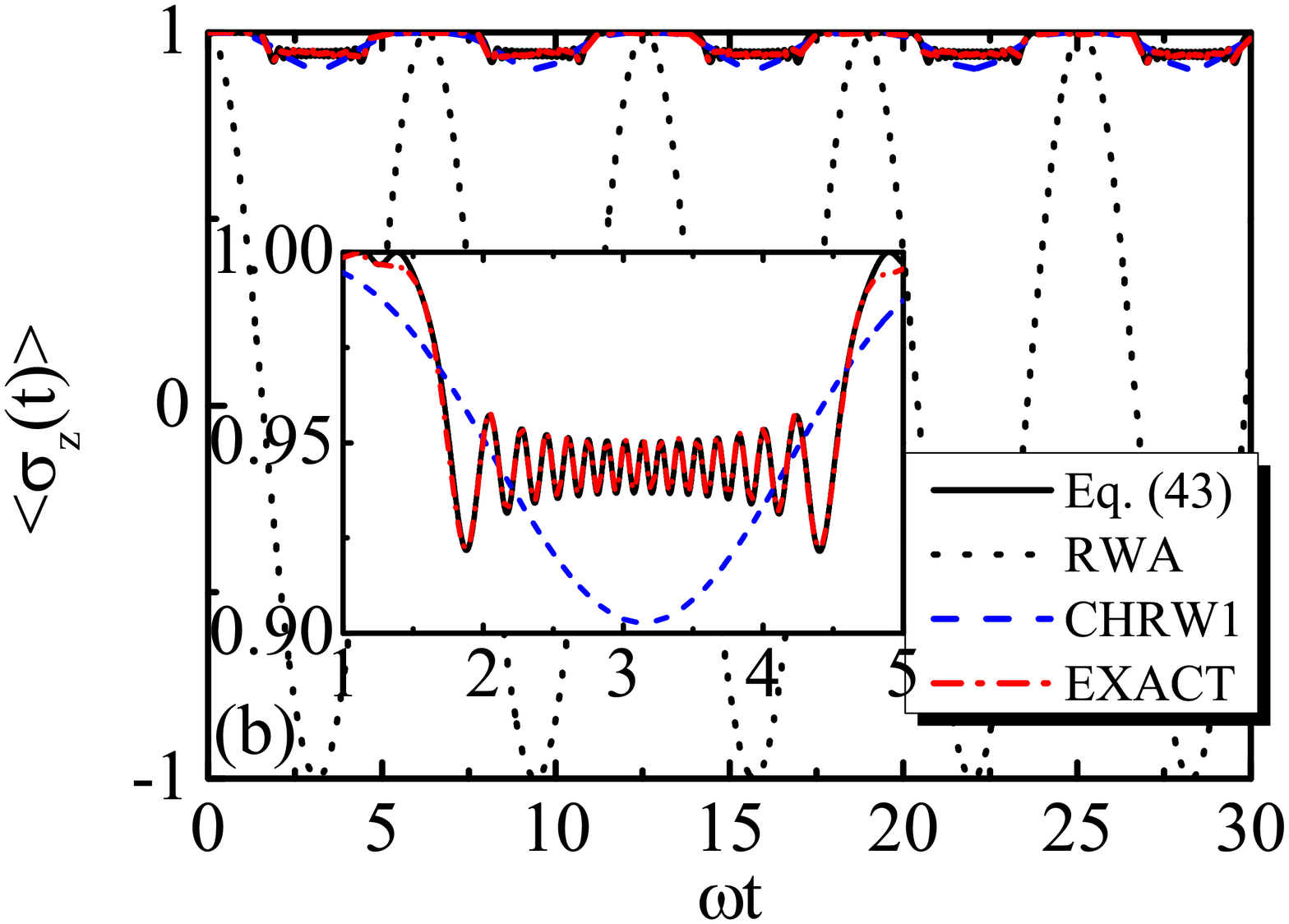}\\
  \caption{(Color online) $\langle \sigma_z(t) \rangle$ as a function of dimensionless time $\omega t$ on CDT conditions, where $A/\omega=6.75\pi$ and $15.75\pi$ in the on-resonance case ($\Delta=\omega$), shown in (a) and (b), respectively.  The black line shows the result of Eq. (\ref{cdt_z2}), the black dotted line shows the RWA results, the blue dashed line shows the CHRW1 results, and the red dash-dotted line shows the numerically exact result. The insets show the detailed comparison of Eq. (\ref{cdt_z2}), the CHRW1 and numerical methods in the aspect of the structure of short-time dynamics.}\label{fig3}
\end{figure}

To illustrate the validity of the simple expression Eq. (\ref{cdt_z2}), we compare its calculated results with those of the previous methods, the RWA, CHRW and
numerically exact calculations. Figure. \ref{fig3} shows the dynamics of $\langle \sigma_z(t) \rangle$ in the resonance case ($\Delta=\omega$) for various driving strengths $A/\omega=6.75\pi$ and $15.75\pi$, respectively.  It is obvious to see that both the CHRW1 method and Eq. (\ref{cdt_z2}) show much suppressed evolutions which are well consistent with the numerically exact result,  in contrast to a periodic oscillation with a large amplitude of the RWA result. However, from the viewpoint of the detailed structure of the dynamics, it is noticeable that only Eq. (\ref{cdt_z2}) gains the exact fast oscillating structure well coinciding with the numerical one, which is considered as the joint contribution of all the odd-harmonic terms, while the CHRW1 shows the main evolution without fast oscillating structure. On the CDT condition, the effects of the even harmonic terms disappears but those of odd harmonics predominate in the time evolution. Since the CHRW1 method have taken into account the zeroth and first harmonics in its Hamiltonian, it could predict the main dynamical tendency corresponding to the first harmonic. In contrast, Eq. (\ref{cdt_z2}) demonstrates the accurate evolution with a fast oscillating structure because it includes all odd harmonics. In other words, the collective effects of all odd harmonics give rise to the pronounced character of the oscillation: armchair plateau.

\begin{figure}
  \centering
  \includegraphics[width=8cm]{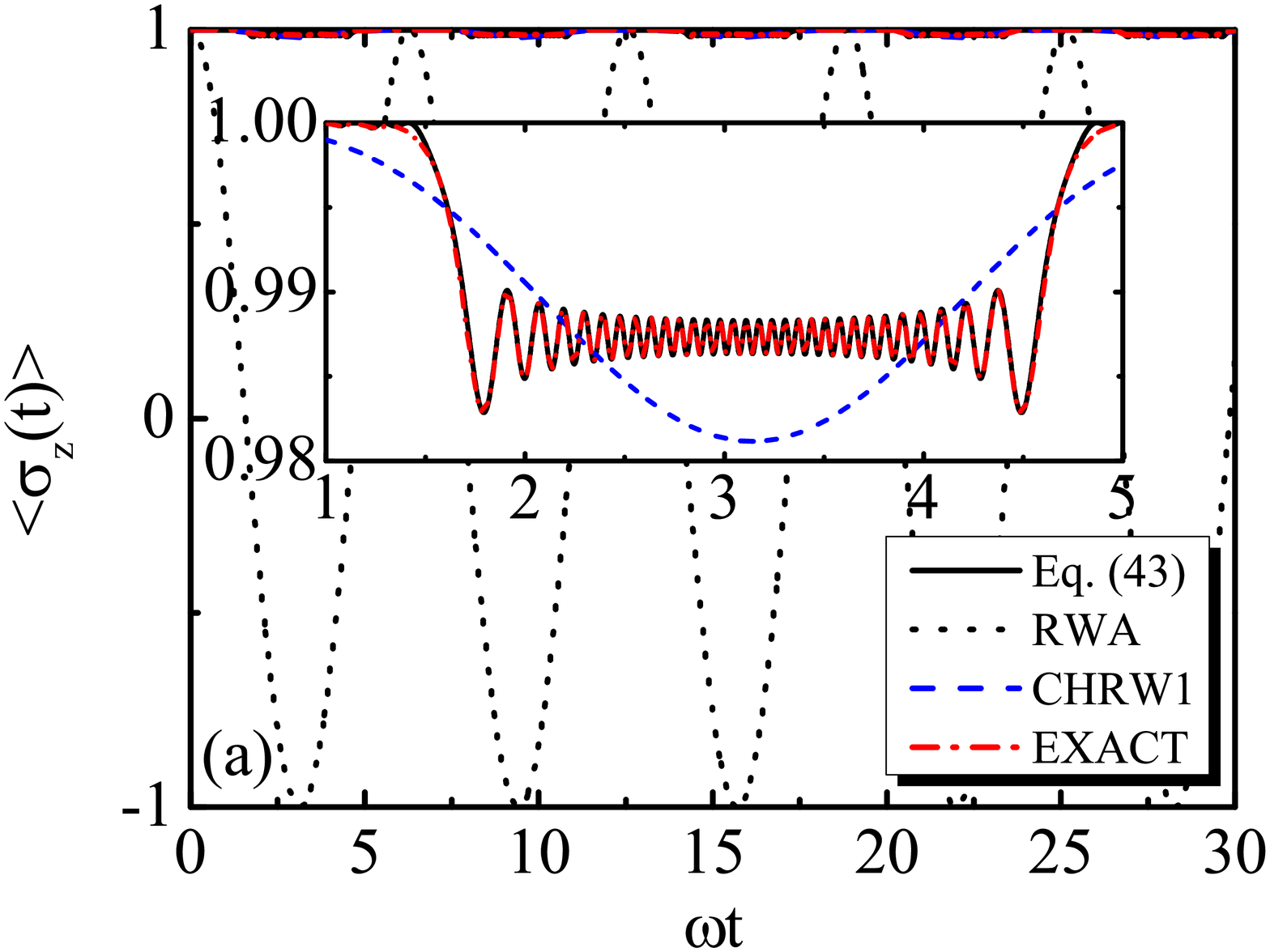}
  \includegraphics[width=8cm]{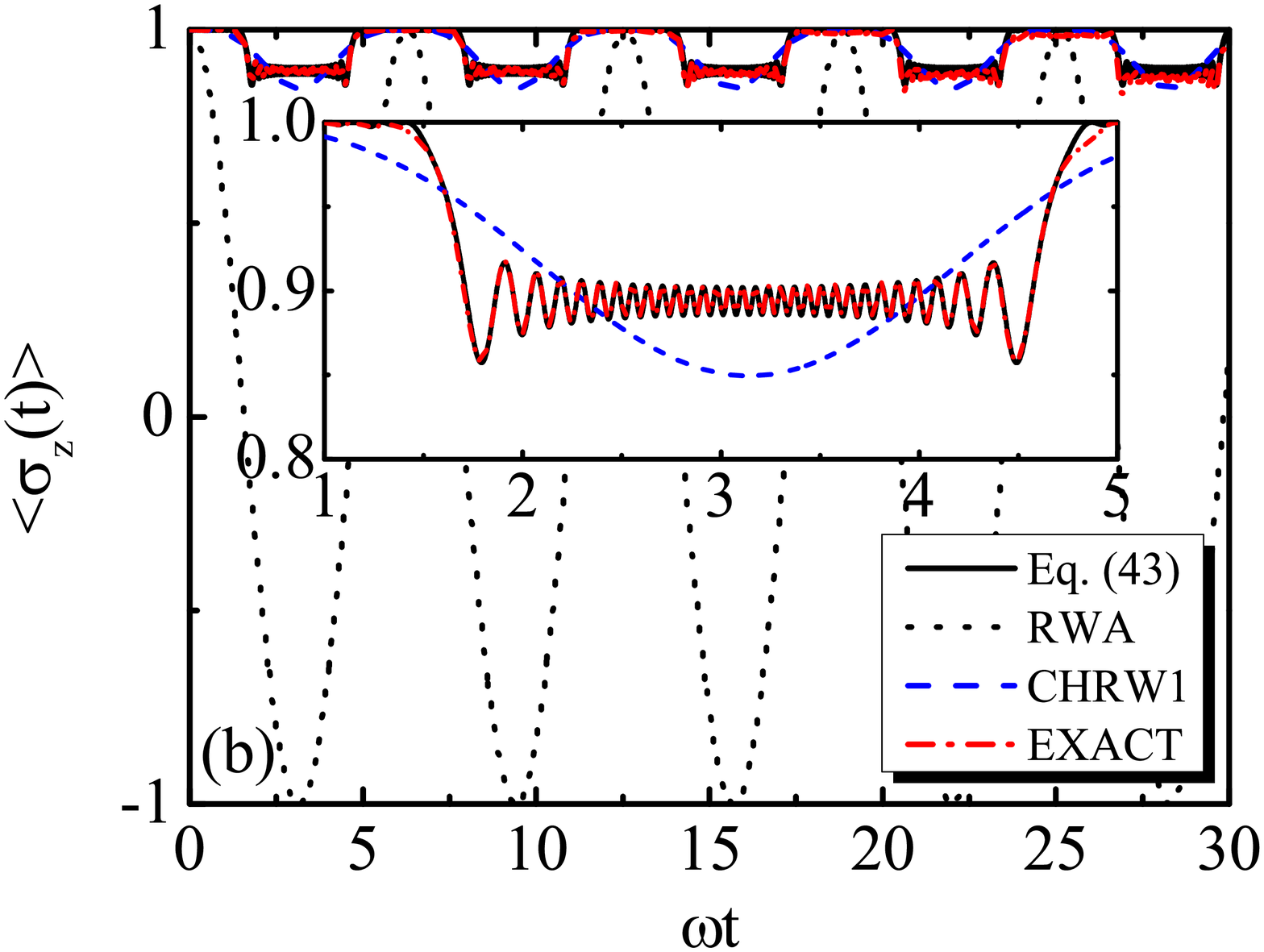}\\
  \caption{(Color online) $\langle \sigma_z(t) \rangle$ as a function of dimensionless time $\omega t$ on CDT conditions with $A/\omega=30.75\pi$ in the off-resonance  cases ($\Delta/\omega=0.6$ and $\Delta/\omega=1.75$), shown in (a) and (b), respectively. The black line shows the result of Eq. (\ref{cdt_z2}), the black dotted line shows the RWA results, the blue dashed line shows the CHRW1 results, and the red dash-dotted line shows the numerically exact result. The insets show the detailed comparison of Eq. (\ref{cdt_z2}), the CHRW1 and numerical methods in the aspect of the structure of short-time dynamics.}\label{fig4}
\end{figure}

In the off-resonance case, the armchair-plateau phenomenon with a fast oscillation structure also happens. In Fig. \ref{fig4}, we show the dynamics of $\langle \sigma_z(t) \rangle$ for $A/\omega=30.75\pi$ with various driving frequencies. When $\omega > \Delta $ (see Fig. \ref{fig4}(a), $\Delta/\omega=0.6$), or  $\omega > \Delta $ (see Fig. \ref{fig4}(b) with $\Delta/\omega=1.75$), the results of  Eq. (\ref{cdt_z2}) exhibit a perfect agreement with the numerical ones from the viewpoint of fast oscillating structure of plateaus. In contrast, the CHRW1 method only shows the general tendency of the evolution but misses the significant plateau structure.

\begin{figure}
  \centering
  \includegraphics[width=8cm]{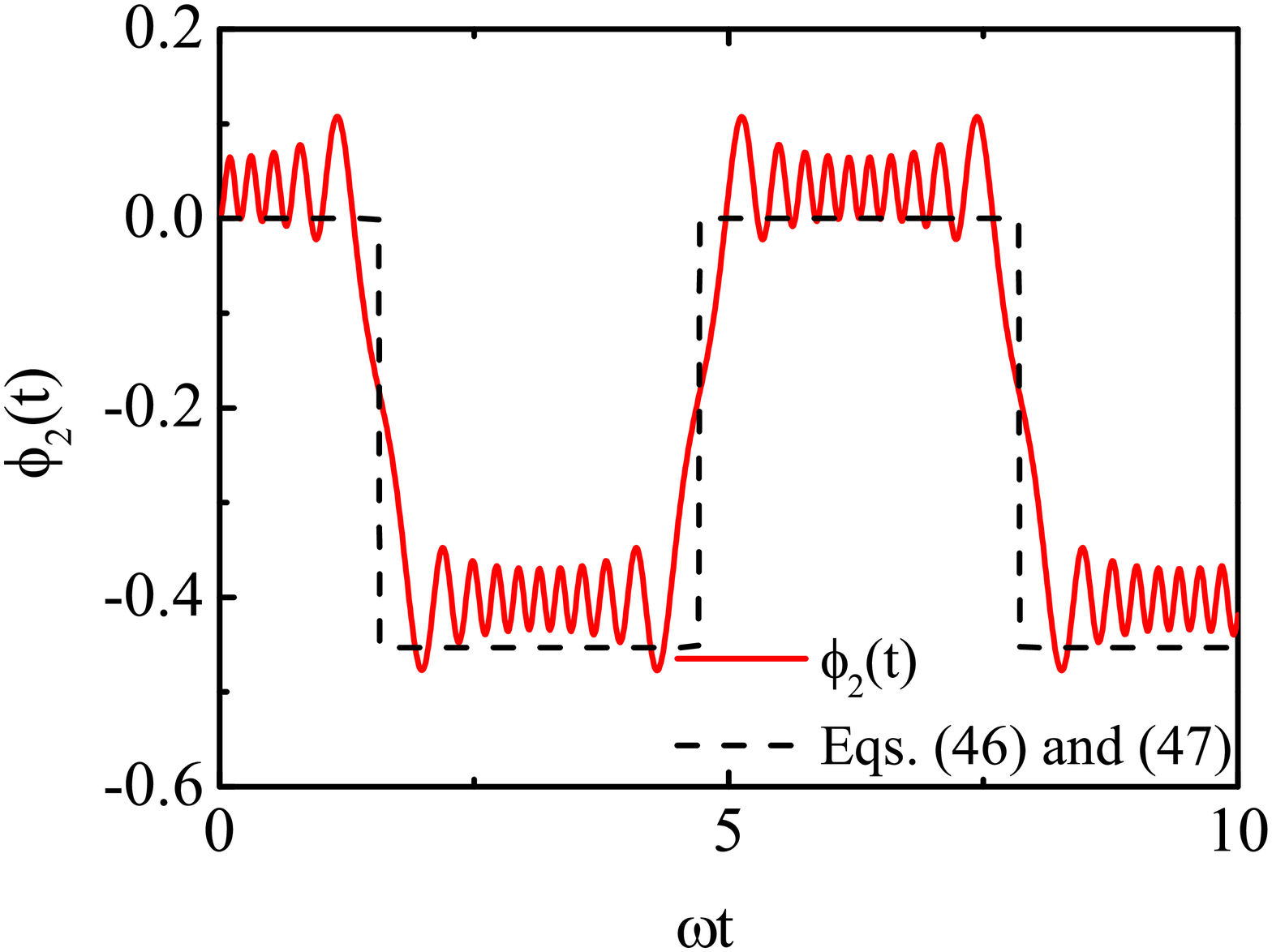}\\
  \caption{(Color online) The phase function, $\phi_2(t)=\Delta \int_{0}^{t} \sin\left[\frac{A}{\omega} \sin(\omega \tau)\right] d\tau$ shown in  Eq. (\ref{cdt_z2}), as a function of dimensionless time $\omega t$, where $A=9.75\pi\omega$ and
  $\Delta=\omega$. The phase function is plotted by red line with a two-stair plateau structure. The mean values of the plateaus are shown as a two-value
  function consisting of $L_1$ and $L_2$ shown by Eq. (\ref{meanvalue1}) and Eq. (\ref{meanvalue3}), respectively, which is plotted by the black dashed line.}\label{fig4_add}
\end{figure}

We will discuss how the odd-harmonic effect leads to the armchair plateau and analyze the feature of the oscillating structure on the CDT condition.
For the sake of further analysis, we denote the phase function $\phi_2(t)=\Delta \int_{0}^{t} \sin\left[\frac{A}{\omega} \sin(\omega \tau)\right] d\tau$, which
is shown in Fig. \ref{fig4_add} with $A=9.75\pi\omega$ and $\Delta=\omega$. In contrast to the phase function $\phi_1(t)$ in Sec. \ref{calc_even}, which has a
linear tendency as a function of time, the function $\phi_2(t)$ consisting purely of odd harmonic terms exhibits a two-stair dynamical structure. In other words, the evolution of $\phi_2(t)$ has two kinds of plateaus alternately occurring and each kind shares the same average value. To verify it, we choose $\phi_2\left(t=\frac{n\pi}{\omega}\right)$ to estimate the average value of every plateau,
\begin{eqnarray}
  \phi_2\left(\frac{n\pi}{\omega}\right) &=& \Delta \int_{0}^{\frac{n\pi}{\omega}} \sin\left[\frac{A}{\omega}\sin(\omega\tau)\right]d\tau \nonumber\\
   &=& 2\Delta \sum_{k=1}^{\infty} \int_{0}^{\frac{n\pi}{\omega}} J_{2k-1}\left(\frac{A}{\omega}\right)\sin[(2k-1)\omega\tau]d\tau \nonumber\\
   &=& 2\Delta [(-1)^{n+1}+1]\sum_{k=1}^{\infty}\frac{1}{(2k-1)\omega}J_{2k-1}\left(\frac{A}{\omega}\right).
\end{eqnarray}
Therefore, it is obvious to classify these values into two groups, each of which has an average value denoted by $L_1$ and $L_2$,
\begin{eqnarray}
\begin{array}{cc}
   L_1 = 0 & (n{~~}{\rm mod}{~~}2=0), \label{meanvalue1}\\
  L_2 = 4\Delta\sum_{k=1}^{\infty}\frac{1}{(2k-1)\omega}J_{2k-1}\left(\frac{A}{\omega}\right) &  (n{~~}{\rm mod} {~~ }2=1)\label{meanvalue2},
\end{array}
\end{eqnarray}
respectively.
In the case of the CDT, where the driving strength can be simplified approximately as $A/\omega=m\pi-\pi/4$, we utilize Eq. (\ref{approaching}) to reduce the sum
in Eq. (\ref{meanvalue2}). As a result, we have
\begin{eqnarray}
  L_2 &\approx& 4\Delta\sum_{k=1}^{\infty} \frac{1}{(2k-1)\omega}\sqrt{\frac{2\omega}{\pi A}}(-1)^{n-k+1} \nonumber\\
   &=& (-1)^{n}\Delta\sqrt{\frac{2\pi}{ A\omega}} \label{meanvalue3}
\end{eqnarray}

It is remarkable that we demonstrate the CDT phenomenon, which is commonly regarded as a complete suppression frozen in the initial state in the previous literatures \cite{CDT:origin,CDT:stabilizing_CDT}, actually processes a stair structure. The general tendency of $\langle \sigma_z(t) \rangle$ changes alternately between the two values shown in Eq. (\ref{meanvalue1}), with the frequency $\omega$, which is clearly seen from Fig. \ref{fig3}. By Eq. (\ref{meanvalue3}), we find that the height of the stair is quantified as $1-\cos(L_2)$, or $1-\cos\left(\Delta\sqrt{\frac{2\pi}{A\omega}}\right)$, which is proportional to $\Delta^2/A\omega$ for strong driving strength. Therefore, Eq. (47) explicitly demonstrates the effect of $A$ and $\Delta$ on the ``freezing'' extent of the initial state. When $\Delta \ll \omega$ and $A \gg\omega$, the relative height of the armchair plateau is so small that the stair structure can not be distinguished. Strictly speaking, the CDT phenomenon should be regarded as a two-stair dynamical structure, and the totally frozen situation is just a limit case. This does contribute to the understanding of the common CDT phenomena under the strong driving condition ($A > \omega$ and $\omega \gg \Delta$).

It is found that the phase function $\phi_2(t)$ determines the dynamics of the plateau. $\phi_2(t)$ not only presents fast oscillations with small amplitude around a certain average value when $t$ is during the interval $\left[\frac{n\pi}{\omega}-\frac{\pi}{2\omega},\frac{n\pi}{\omega}+\frac{\pi}{2\omega}\right]$, but also shows a rapid variation near the boundary of each plateau. According to Appendix \ref{appen1}, the oscillations of these plateaus satisfy a quasi-periodic structure illustrated by $A \sin(\omega t)/\omega =n \pi$, and the number of periods for each plateau is $N=\left\lfloor \frac{2A}{\omega\pi}\right\rfloor -\left\lfloor \frac{A}{\omega\pi}\right\rfloor$, which is notably independent of $\Delta$. Besides, the deviation from the average value of $\phi_2(t)$ corresponding to every plateau satisfies the condition of Eq. (\ref{2inequality3}) (Appendix. \ref{appen1}):
\begin{equation}
  |g_2(t)|<\frac{2\Delta}{A\cos(\omega t)}.
\end{equation}
It indicates that the envelop of the small-amplitude oscillation are tightly dependent on $A$ and $\Delta$. For instance, for $A/\omega=6.75\pi$ in Fig. \ref{fig3}(a), it is readily seen that the bottom plateau of this armchair structure nearly lies at $\langle \sigma_z \rangle =0.82$. As the driving strength increases, such as $A/\omega=15.75\pi$ in Fig. \ref{fig3}(b), the depth of armchair structure is much suppressed, where the bottom plateau is around $\langle \sigma_z \rangle =0.94$. In contrast, when the tunneling frequency $\Delta$ goes up from $\Delta=0.6\omega$ to $\Delta=1.75\omega$ (see Fig. \ref{fig4}), the location of the bottom plateau is varied from the mean value $\langle \sigma_z \rangle=0.997$ to $\langle \sigma_z \rangle =0.9$. It turns clearly out that the depth of armchair structure increases greatly as $\Delta$ increases. Consequently, the quasi-periodicity is determined uniquely by $A/\omega$, $\Delta/A$ controls the amplitude of the oscillation.

\begin{figure}
  \centering
  \includegraphics[width=8cm]{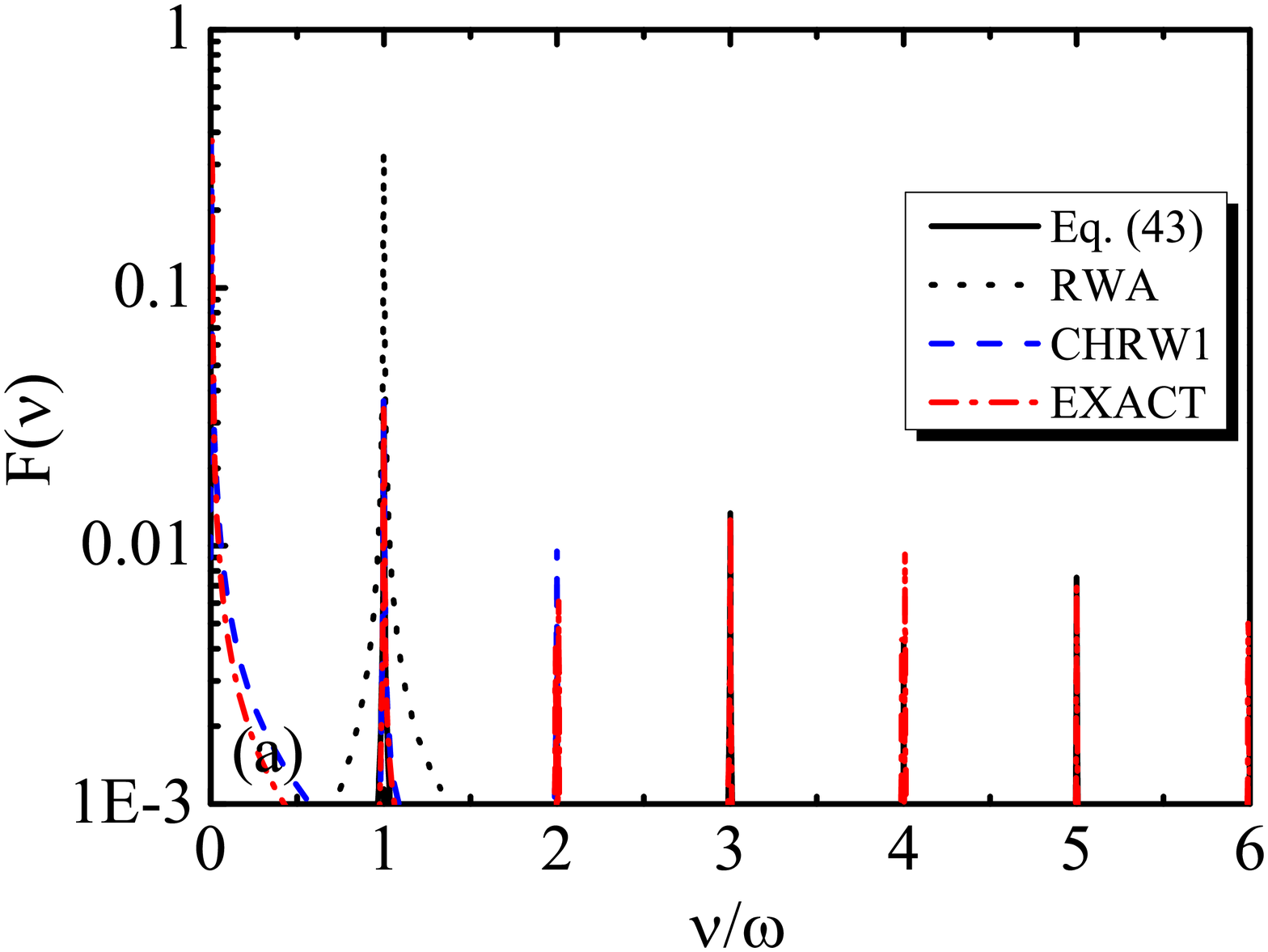}
  \includegraphics[width=8cm]{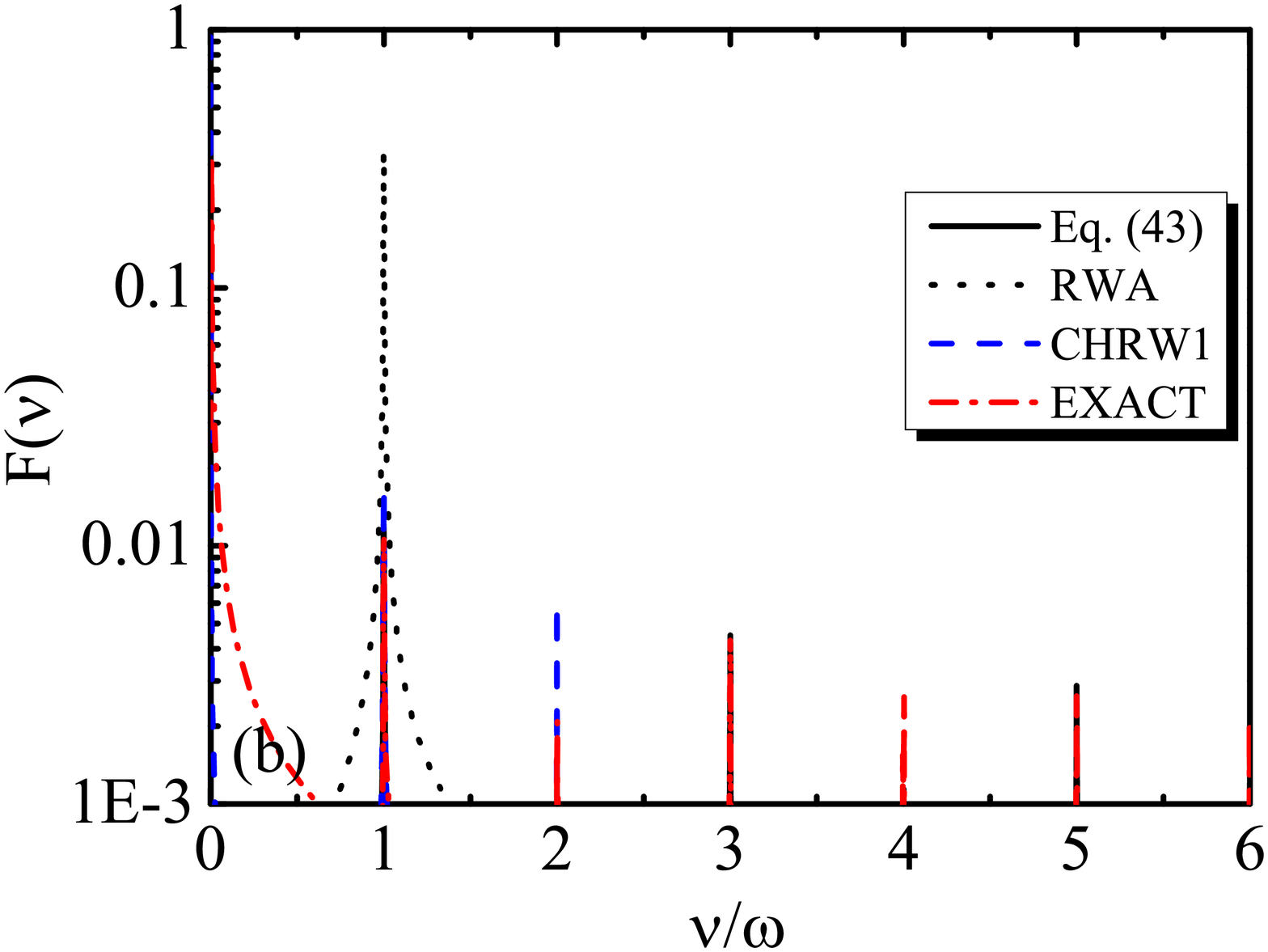}\\
  \caption{(color online) Fourier spectrum of $\langle \sigma_z \rangle$ with the same parameters as in Fig. \ref{fig3}. In each graph, the black line is the result of Eq. (\ref{cdt_z2}).}  \label{FigFFT3}
\end{figure}

\begin{figure}
  \centering
  \includegraphics[width=8cm]{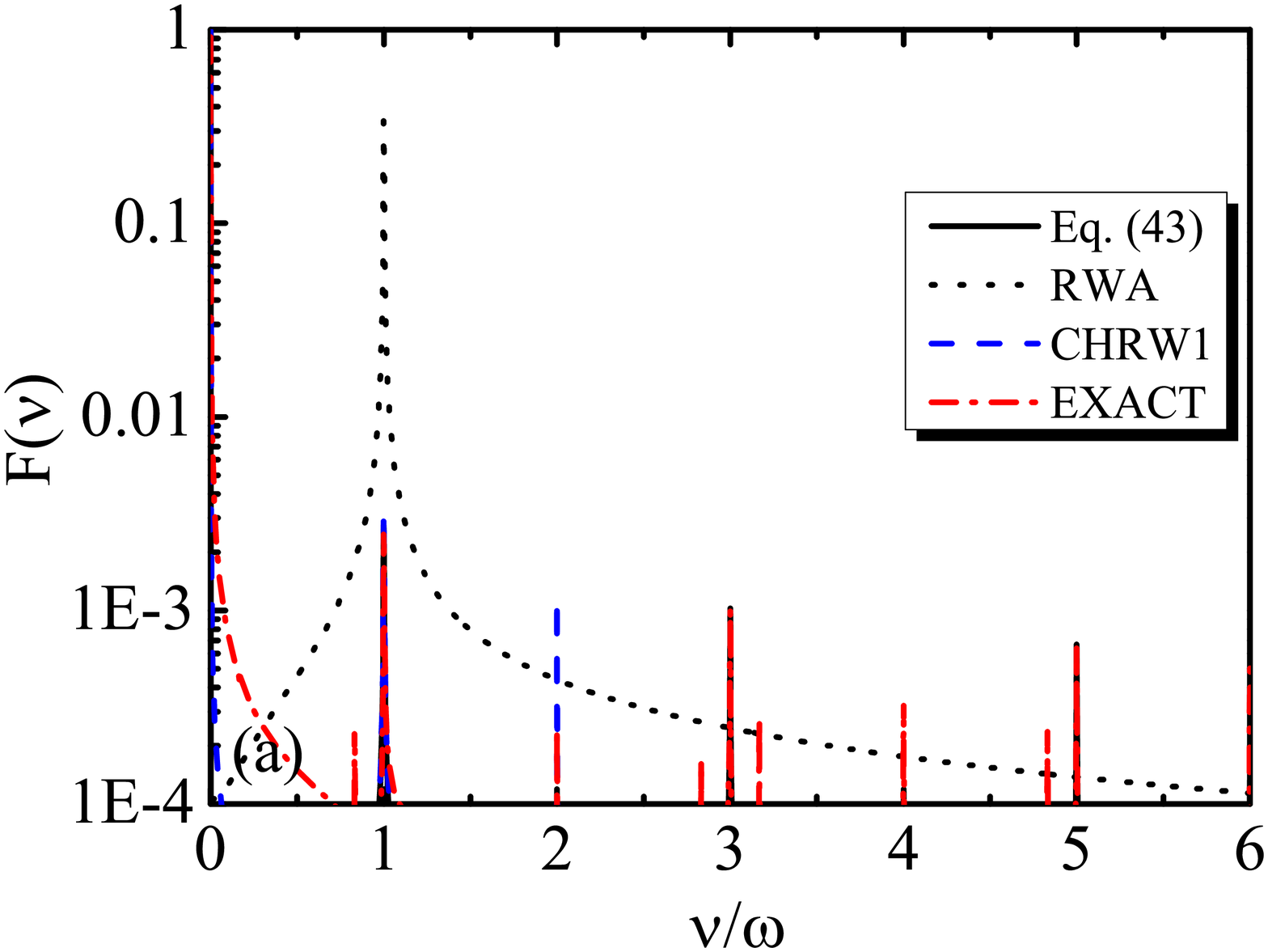}
  \includegraphics[width=8cm]{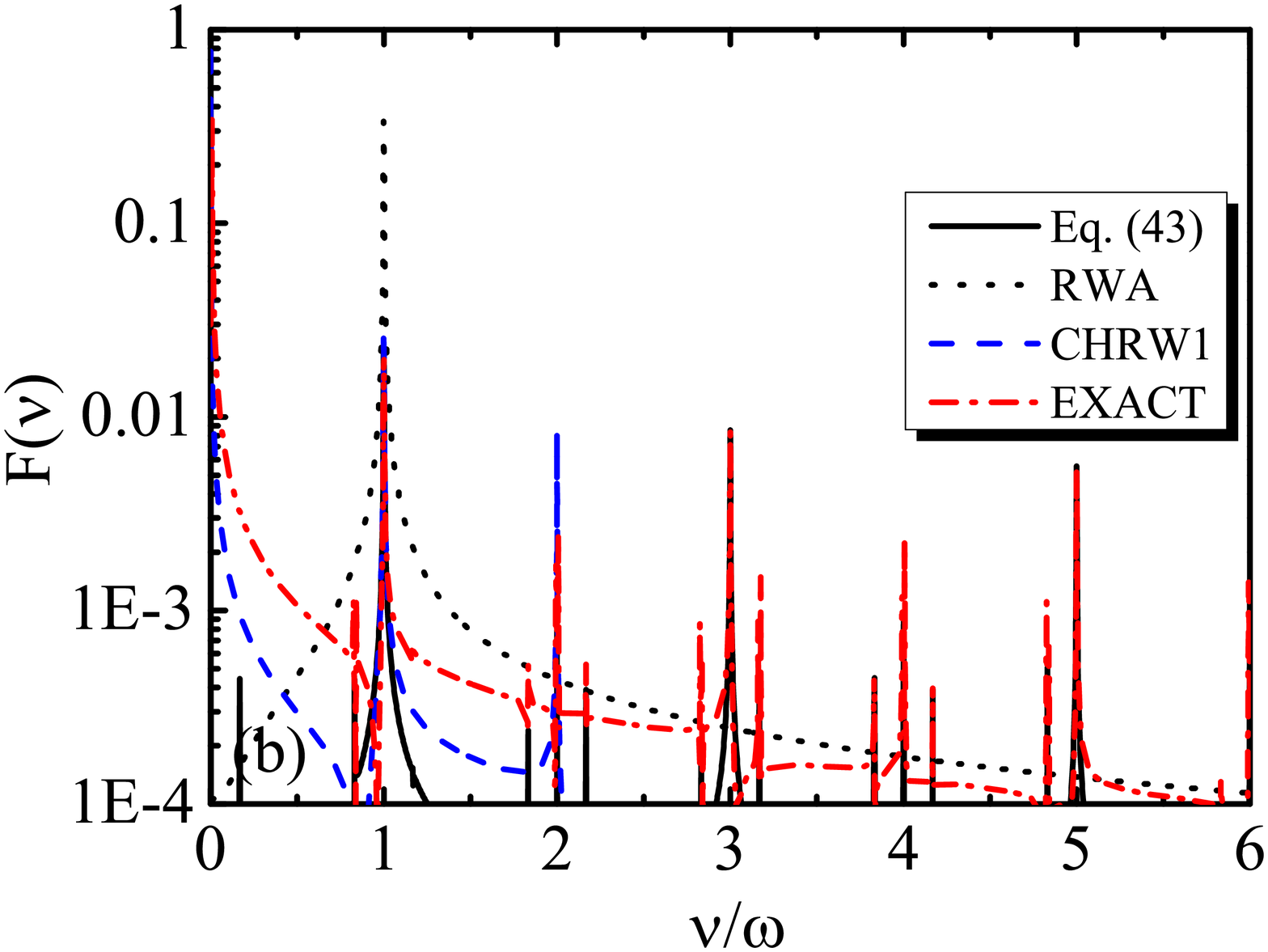}\\
  \caption{(color online) Fourier spectrum of $\langle \sigma_z \rangle$ with the same parameters as in Fig. \ref{fig4}. In each graph, the black line is the result of Eq. (\ref{cdt_z2}).}   \label{FigFFT4}
\end{figure}

It is interesting to demonstrate the Fourier transform of the CDT dynamics which is done by Eq. (\ref{fft}). Figures. \ref{FigFFT3} and \ref{FigFFT4} show the Fourier transform of the $\langle\sigma_z\rangle$ of Figs. \ref{fig3} and \ref{fig4}, respectively. It is obvious that
the RWA method shows only single frequency at $\nu=\omega$ and the CHRW1 approach depicts frequencies at $\nu=0, \omega,$ and $2\omega$, while Eq. (\ref{cdt_z2}) shows a series of peaks at each integer frequency $\nu= k\omega (k=0,1,2...)$. To understand the information in frequency domain, let us take Fig. \ref{FigFFT3}(a) as an example. It is clear to see that the peak near zero frequency predominates in the CDT dynamics, since the renormalized tunneling frequency $\tilde{\Delta}$ vanishes on the CDT condition. In contrast with the even-harmonic case in last section, it is surprising to underline that not only odd harmonic frequencies contribute to the CDT dynamics but also the even ones do. This fact could be explained by the series expansion of Eq. (\ref{cdt_z2}),
\begin{eqnarray}
  Z &\approx& 1-\frac{1}{2}\left\{2\Delta \sum_{k=1}^{\infty} \int_{0}^{t} J_{2k-1}\left(\frac{A}{\omega}\right)\sin\left[(2k-1)\omega\tau\right]d\tau\right\}^{2} \nonumber\\
   &=& 1-{\Delta}^{2}\sum_{k_1=1}^{\infty}\sum_{k_2=1}^{\infty} \frac{1}{(2k_1-1)(2k_2-1)} J_{2k_1-1}\left(\frac{A}{\omega}\right)J_{2k_2-1}\left(\frac{A}{\omega}\right)
   \left\{2-2\cos[(2k_1-1)\omega t] \right. \nonumber\\
   & &\left.-2\cos[(2k_2-1)\omega t]+\cos[(2k_1+2k_2-2)\omega t]+\cos[(2k_1-2k_2)\omega t] \right\}.
\end{eqnarray}
It is clear to see that both odd harmonic and even harmonic components appear in frequency domain. Therefore, we verify the effectiveness of Eq. (\ref{cdt_z2}) and restate the odd-harmonic effect is crucial to the armchair plateau.

From above, we can see that the different qubit response happens as the driving parameter changes. Especially, the temporal evolutions of Figs. \ref{fig3} and \ref{fig4} illustrate that the first harmonic plays an important role on the evolution in the CDT condition. In contrast, it is all of the even harmonic terms that lead to a nontrivial armchair plateau phenomenon. As we can see in Figs. \ref{fig3} and \ref{fig4}, the CHRW1 method, which takes the first harmonic into account, explicitly gives an oscillating structure in comparison with the numerically exact result. Under the CDT condition, the renormalized tunneling frequency $\tilde{\Delta}$ vanishes since $ J_0\left(\frac{A}{\omega}\right)=0$. As a result, the first harmonic determines the general tendency in armchair plateau, which is agreement with the numerically exact result. Furthermore, we discuss the collective effect of all the odd harmonic terms in Eq. (\ref{cdt_z2}), which eventually shows the exact plateau structure that the CHRW1 method cannot predict. Therefore, the higher-order odd-harmonic terms also play important roles in the construction of the armchair plateau phenomenon.

\section{Conclusion}
In summary, we have studied the dynamics of a strongly harmonically driven TLS in a systematic way by the improved CHRW method. To provide a comprehensive solution in strong driving case, we divide the situation into two parts: the even-harmonic situation and odd-harmonic one (CDT condition). Both of them exhibit the specific phenomenon of plateau with quantized oscillation resulting from the collective effects of the multiple harmonics. We demonstrate two kinds of plateau, one is the zigzag plateau corresponding to the even-harmonic situation, the other is armchair plateau corresponding to the odd-harmonic situation. It turns out that the plateau seems to be the universal dynamical character of a strongly driven TLS.

In the present work, after the first unitary transformation with the generator $S_1$, we obtain the transformed Hamiltonian Eq. (5). It is composed of different harmonics: zeroth harmonic, even harmonics, and odd harmonics ($n\omega$ terms, $n=0,1,2,...$). Therefore, the even and odd harmonic effects results from the even and odd harmonic terms, respectively. In the paper, the CHRW2 Hamiltonian is realized by the second unitary transformation with the novel generator $S_2$ to explore the strongly driven dynamics, especially, large-amplitude oscillatory case. It turns out that the collective effects of all even harmonics contribute to the generation of zigzag plateau with quantized oscillation and the general tendency of evolution coincides with the result of the CHRW1 method, which exhibits the even-harmonic effect in this situation. On the other hand, in CDT condition, we find that the even-harmonic terms vanishes as proved in Sec. IV, which indicates a simplification of the Hamiltonian in Eq. (5), only the odd-harmonic terms determine the CDT dynamics of the strongly driven system. In this case, we reveal an armchair plateau structure which is  not a complete destruction in the dynamical evolution. As a result, we obtain different patterns of evolution in these two situations. All in all, the effects of different multiple harmonics are treated by various methods, and under different conditions, analytical simple formula are derived for even harmonic effects and odd harmonic ones which are illustrated by the large amplitude and the small amplitude (CDT) oscillatory dynamics, respectively. Eq. (\ref{z_simplest}) and Eq. (\ref{cdt_z2}) are two important simplified analytical results for discussing zigzag and armchair plateaus, respectively. It is clear that the valid condition of Eq. (\ref{cdt_z2}) for the driving strength A is indicated as $(m-1/4)\pi$ and $J_0(A/\omega)=0$, while that of Eq. (\ref{z_simplest}) is away from $(m-1/4)\pi$ and at $\xi\rightarrow 1$.

Although the CHRW2 hamiltonian has taken into account the effects of even-harmonic terms which are not considered in the CHRW1 Hamiltonian \cite{CHRW:origin}, both the CHRW Hamiltonians hold a RWA mathematical form. As a result, the analytical expressions of the physical quantities given by the CHRW2 method, such as $\langle \sigma_x(t) \rangle$, possesses the renormalized framework similar to those of the CHRW1 method because of their mutual form of the effective Hamiltonian, which implies the simplicity for potential experimental applications. Of greater significance is the transformation matrix due to the representation transformation shown in Eq. (\ref{transform_matrix_for_even}), explicitly suggesting the multi-harmonic effect by the trigonometric functions. Therefore the CHRW2 method could work well in a much broader parameter regime.

The superiority of the CHRW2 method is embodied with the strong driving cases of the TLS. Frankly speaking, an extremely strong driving amplitude $A$ would lead to possible multi-harmonic processes in the procedure of tunneling phenomenon. By the comparisons of several analytical methods and numerical calculation about $\langle \sigma_z(t) \rangle$, we prove that the intriguing phenomenon, plateau with quantized oscillation, predicted by the numerically exact calculation, can be well illustrated and explicitly explained by the CHRW2 method. Significant dynamical features of $\langle \sigma_z(t) \rangle$ are found as follows: (i)this stunning plateau phenomenon has a periodical structure with the frequency $2\omega$; (ii) $\langle \sigma_z(t) \rangle$ oscillates around a certain mean value with a relatively tiny amplitude in each plateau except near the boundary where dramatic jumps of the evolution occur; (iii) by Eq. (\ref{z_simplest}), we demonstrate the detailed description about the oscillatory plateau, such as the quasi-periodical structure, and the quantized property of the oscillation whose number of periods is $N=\left \lfloor \frac{2A}{\omega\pi}\right \rfloor-\left\lfloor \frac{A}{\omega\pi} \right\rfloor$. Interestingly, the number of periodic oscillation is independent of the tunneling frequency $\Delta$. However, the amplitude of quantized oscillation is determined by $\Delta/A$, which has been shown in Eq. (\ref{inequality3}). Moveover, the mean values of these plateaus globally exhibit a harmonic oscillation with the amplitude equal to 1 which is corresponding to constructive tunneling. It can be derived directly from generally linear evolution of the phase function $f(t)=\tilde{\Delta}t$, which coincides with the result of the RWA-RF method \cite{RWA-RF}, a useful method handling the strong driving dynamics of the TLS. Therefore, the overall characteristics of the dynamics, both the general evolution and the detailed quantized oscillation on each plateau, can be elucidated by the CHRW2 method.

On the other hand, the CDT phenomenon, which is described as the situation that totally frozen tunneling occurs, also presents another quantized plateau phenomenon (armchair plateau) under the strong driving. In comparison with the even-harmonic cases, the CDT phenomenon means the diminishment of the linear evolution of the phase function because of $J_{0}(A/\omega)=0$, which is widely considered as the necessary and sufficient condition of this phenomenon. Nevertheless, the CDT possesses a novel two-stair structure beyond the general understanding, especially in the cases of extremely strong driving and near resonance. In order to give a comprehensive demonstration of this phenomenon, we apply the CHRW1 method to calculate the CDT dynamics. Note that the CDT condition signifies the vanishing of not only the zero-harmonic term $J_0\sigma_x$, but also all the even-harmonic terms $J_{2n} \sigma_x (n\geq 1)$  in the CHRW1 Hamiltonian (Eq. (5)). Therefore, when the even-harmonic effect disappears, the odd-harmonic effect dominates. Subsequently, we obtain the analytic expressions of the quantities of interest, $\langle \sigma_i(t) \rangle ~(i=x,y,z)$. $\langle \sigma_z(t) \rangle$  exhibits the armchair structure with quantized oscillations rather than the totally destructive tunneling.  When $A/\omega =m\pi-\frac{\pi}{4}$ for a large $m$ and $\Delta/\omega \rightarrow 0$, the dynamical pattern is undistinguishable for the diminishing difference between the two stairs. However, this kind of plateau results from the multi-harmonic effect involving only the odd-harmonic terms. As a result, there are also some significant differences between the CDT and the even-harmonic cases.

This paper sheds some light on the dynamical pattern of a TLS under the strong harmonic driving, where the multi-harmonic effect reveals a novel physical phenomenon, plateau with quantized oscillation. By the analysis of zigzag and armchair plateaus, we prove the profound merit of the CHRW method and believe that the dynamical patterns may illuminate the experiment to understand the stunning dynamics in different kinds of strongly driven systems.

\begin{acknowledgments}
Z.L. is supported by National Natural Science Foundation of China (Grants No. 11774226 and No. 61927822).
\end{acknowledgments}

\appendix
\section{the even-harmonic and odd-harmonic effects \label{appen1}}

In this Appendix, we give a quantitative analysis for the plateau structures: zigzag plateau and armchair plateau resulting from the even-harmonic terms and the odd-harmonic terms, respectively. We demonstrate these dynamical structures are determined by their intrinsic phase functions consisting of multiple harmonics.

In term of the phase function $\phi_1(t)$ in Sec. \ref{calc_even}, we define a function $
  g_1(t)=\phi_1\left(t+\frac{n\pi}{\omega}\right)-\phi_1\left(\frac{n\pi}{\omega}\right),
$
($t \in \left[-\frac{\pi}{2\omega},\frac{\pi}{2\omega}\right]$). Then we get
\begin{equation}\label{phase_rev}
 g_1(t)=\Delta \int_{0}^{t}\cos\left[\frac{A}{\omega}\xi\sin(\omega \tau)\right]d\tau,
\end{equation}
Substituting $\sin(\omega \tau)$ with $x$, we rewrite $g_1(t)$ as
\begin{equation}\label{phase_rev2}
  g_1(t)=h_1(X)=\frac{\Delta}{\omega}\int_{0}^{X} \frac{\cos\left(\frac{A}{\omega}\xi x\right)}{\sqrt{1-x^2}}dx,
\end{equation}
where $X=\sin(\omega t)$.

\begin{figure}
  \centering
  \includegraphics[width=8cm]{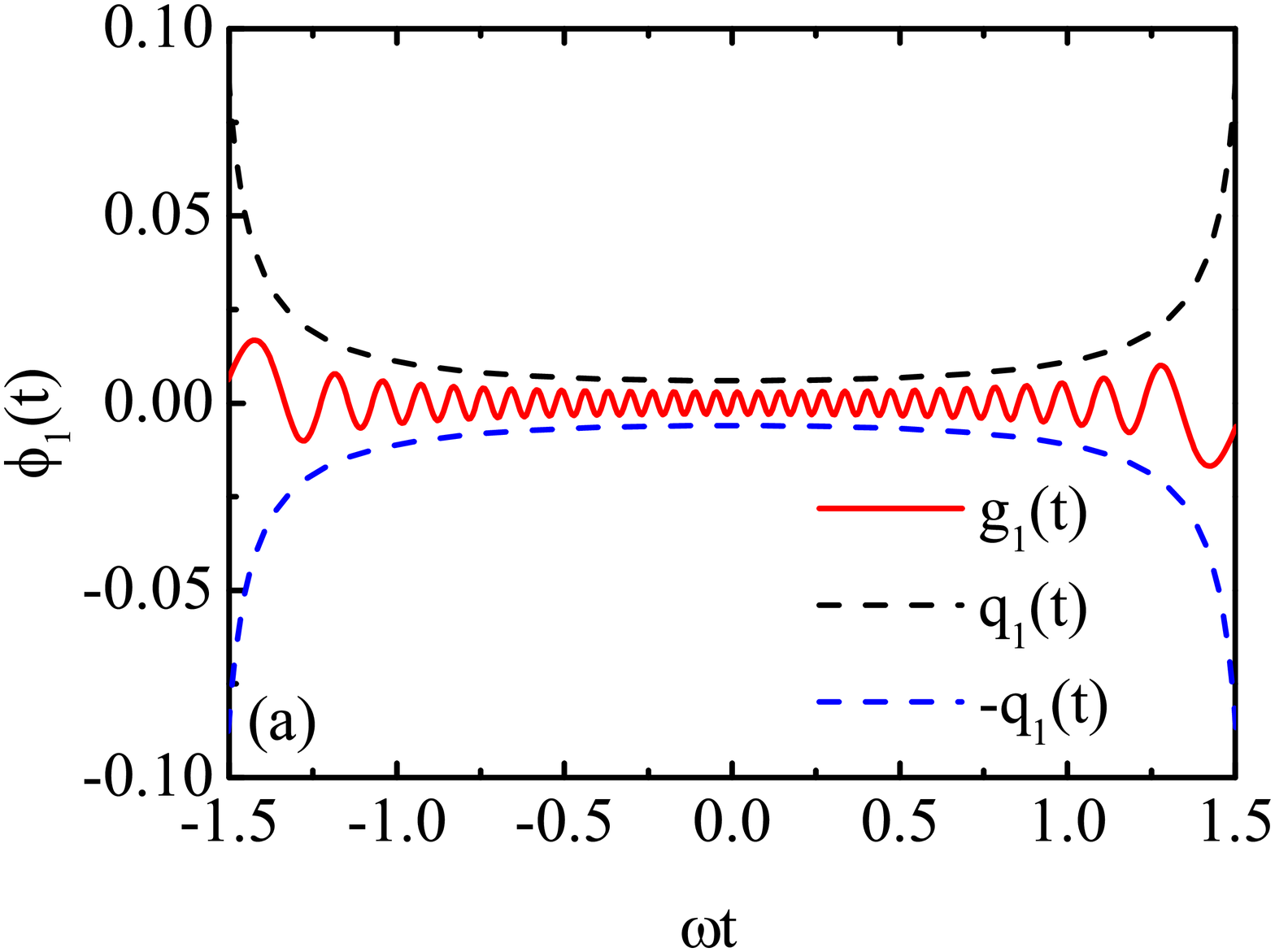}
  \includegraphics[width=8cm]{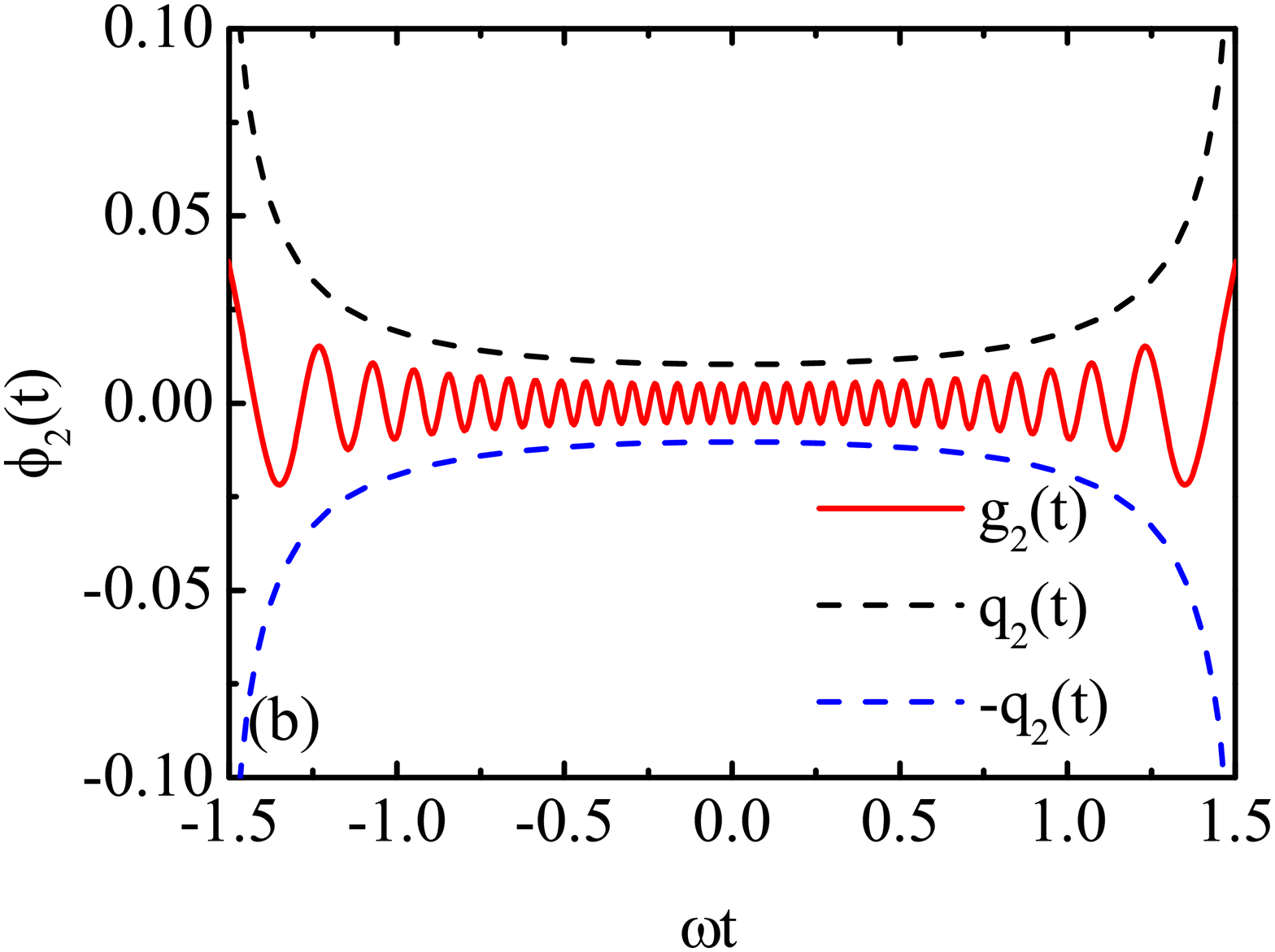}\\
  \caption{(Color online) (a) $g_1(t)$ shown in Eq. (\ref{phase_rev}), as a function of dimensionless time $\omega t$ for $A=100\omega$ and $\Delta=0.3\omega$.
  Also, $q_1(t)$ and $-q_1(t)$ are depicted as the upper bound and lower bound of $g_1(t)$.  (b) $g_2(t)$ shown in Eq. (\ref{phase_rev2}), as a function of dimensionless time $\omega t$, where $A=30.75\pi\omega$ and $\Delta=0.5\omega$. Also, $q_2(t)$ and $-q_2(t)$ are depicted as the upper bound and lower bound of $g_2(t)$. }\label{Figadd1}
\end{figure}

We analyze the distribution of extrema of $h_1(X)$, in which X satisfies $\frac{d(h_1(X))}{dX}=0$, or $\frac{A}{\omega}\xi x=n\pi+\frac{\pi}{2},~(n \in Z)$. Considering the sign of $\frac{d^2h_1(X)}{dX^2}$ we find that $g_1(t)$ reaches a maximum for $\frac{A}{\omega} \xi \sin(\omega t)=\left(2k+\frac{1}{2}\right)\pi$, and obtains a minimum for $\frac{A}{\omega} \xi \sin(\omega t)=\left(2k+\frac{3}{2}\right)\pi$. We denote these extrema by $x_n=\frac{\omega \pi}{A \xi}\left(n+\frac{3}{2}\right)$, and divide the integral in Eq. (\ref{phase_rev2}) into pieces as
\begin{eqnarray}
  A_n &=& h_1\left(\frac{\omega (n+1) \pi}{2 A \xi}\right)-h_1\left(\frac{\omega n \pi}{2 A \xi}\right) \nonumber\\
 &=& \frac{\Delta}{\omega}\int_{\frac{\omega n\pi}{2 A \xi}}^{\frac{\omega (n+1) \pi}{2 A \xi}} \frac{\cos\left(\frac{A}{\omega}\xi x\right)}{\sqrt{1-x^2}}dx.
\end{eqnarray}
It is easy to obtain that $\left|A_{n+1}\right|>\left|A_{n}\right|$ and the signs of $A_n$ are
\begin{equation}
{\rm sgn}(A_n)=  \left\{ \begin{array}{cc}
            + & (n ~~{\rm mod}~~ 4=0, 3) \\
            - & (n ~~{\rm mod}~~ 4=1, 2)
          \end{array} \right. \nonumber
\end{equation}
Because $h_1(X)$ is an odd function, we just take $h_1(x_n) (n \geq 0)$ into consideration. After some mathematical calculations, we get
\begin{eqnarray}
A_{2n}+A_{2n-1}-A_0>h(x_n)=\sum_{i=0}^{2n}A_i>A_0>0, (n ~{\rm mod}~ 2=0),\label{inequality1}
  \\
A_{2n}+A_{2n-1}+A_0<h(x_n)=\sum_{i=0}^{2n}A_i<A_0+A_1+A_2<0, (n ~{\rm mod}~ 2=1).\label{inequality2}
\end{eqnarray}
Thus, we find that all the maxima of $h_1(X)$ are positive while all the minima are negative. Returning to the phase function $\phi_1(t)$, we obtain that the value of the phase function oscillates around the average value $\phi_1\left(t=\frac{n\pi}{\omega}\right)=\frac{\tilde{\Delta}n\pi}{\omega}$, which the maxima and minima are distributed above and below the mean value alternately during time evolution. It is remarkable that the oscillating structure of $\phi_1(t)$ is like a harmonic function. Even if the distribution of the extremum does not satisfy a strictly periodical relationship, $\phi_1(t)$ is quasiperiodic expressed as $\frac{A\xi}{\omega}\sin(\omega t)=n\pi+\frac{\pi}{2}$. An oscillation with a quasi-period is defined as the process during which the phase function evolves from a maximum to a minimum and back to a maximum again. Accordingly, the number of the oscillation in a single plateau is calculated by $N=\left\lfloor \frac{2A}{\omega\pi}\right\rfloor-\left\lfloor \frac{A}{\omega\pi}\right\rfloor$. To estimate the amplitude of the oscillation, from Eqs. (\ref{inequality1}) and (\ref{inequality2}) we can find the bounds of these extrema

\begin{eqnarray}
  \left|h_1(x_n)\right| &<& \left|\frac{\Delta}{\omega}\int_{\frac{(2n-1)\omega\pi}{2A\xi}}^{\frac{(2n+1)\omega\pi}{2A\xi}} \frac{\cos\left(\frac{A}{\omega}\xi x\right)}{\sqrt{1-x^2}}dx\right|\nonumber \\
    &<& \frac{2\Delta}{A\xi}\left[1-\left(\frac{(2n+1)\omega\pi}{2A\xi}\right)^2\right]^{-\frac{1}{2}}\nonumber \\
    &<& \frac{2\Delta}{A\xi}\left(1-x_n^2\right)^{-\frac{1}{2}}. \label{bounded1}
\end{eqnarray}
The inequality Eq. (\ref{bounded1}) is also valid when $x_n$ is substituted with any $x\in [-1,1]$. Subsequently, $g_1(t)$($t \in [-\frac{\pi}{2\omega},\frac{\pi}{2\omega}]$) has
\begin{equation}\label{inequality3}
  |g_1(t)|<q_1(t)=\frac{2\Delta}{A\xi\cos(\omega t)}.
\end{equation}
To validate Eq. ({\ref{inequality3}}), we illustrate the results of $q_1(t)$, $-q_1(t)$ and $g_1(t)$ in Fig. \ref{Figadd1}. In the majority part of a plateau, the results of $q_1(t)$ and $-q_1(t)$ show the best estimation of the upper bound and lower bound, respectively. Moreover, we find that the amplitude of $g(t)$ fulfills a function proportional to $1/\cos(\omega t)$ in the main part of the plateau. Since $q_1(t)$ is proportional to $\Delta/A\xi$, the amplitude of the oscillation of the plateau could be suppressed with the increase of $A$ or the decrease of $\Delta$. Therefore, the plateau structure coming from even-harmonic effects can be attributed to the properties of $\phi_1(t)$.

Similarly, we would analyze the stair structure resulting from the odd-harmonic effects by the phase function $\phi_2(t)$ in Sec. \ref{calc_odd}, which oscillates around the average value $0$. We define a function
$
  g_2(t)=\phi_2\left(t+\frac{n\pi}{\omega}\right)-\phi_2\left(\frac{n\pi}{\omega}\right)
$
($t \in \left[-\frac{\pi}{2\omega},\frac{\pi}{2\omega}\right]$), and get
\begin{equation}\label{phase2_rev}
 g_2(t)=\Delta \int_{0}^{t}\sin\left[\frac{A}{\omega}\sin(\omega \tau)\right]d\tau,
\end{equation}
Substituting $\sin(\omega \tau)$ with $y$ we rewrite $g_2(t)$ as
\begin{equation}\label{phase2_rev2}
  g_2(t)\equiv h_2(Y)=\frac{\Delta}{\omega}\int_{0}^{Y} \frac{\sin\left(\frac{A}{\omega} y\right)}{\sqrt{1-y^2}}dy,
\end{equation}
where $Y=\sin(\omega t)$.

Analogously, we first solve the distribution of maxima and minima of $Y$ satisfying $\frac{d(h_2(Y))}{dY}=0$, or $\frac{A}{\omega} Y=n\pi (n \in Z)$.  Considering the sign of $\frac{d^2 h_2(Y)}{dY^2}$ we know that $g_2(t)$ gets a minimum for $\frac{A}{\omega} \sin(\omega t)=2k\pi$ and $g_2(t)$ reaches a maximum for $\frac{A}{\omega} \sin(\omega t)=\left(2k+1\right)\pi$. We denote these extrema by $y_n=\frac{n\omega \pi}{A}$, and divide the integral in Eq.(\ref{phase2_rev2}) into pieces as
\begin{eqnarray}
  B_n &=& h\left(\frac{\omega (n+1) \pi}{2 A }\right)-h\left(\frac{\omega n \pi}{2 A }\right) \nonumber\\
 &=& \frac{\Delta}{\omega}\int_{\frac{\omega n\pi}{2 A}}^{\frac{\omega (n+1) \pi}{2 A}} \frac{\sin\left(\frac{A}{\omega} x\right)}{\sqrt{1-x^2}}dx.
\end{eqnarray}
It is easy to obtain that $\left|B_{n+1}\right|>\left|B_{n}\right|$ and the sign of $B_n$ is
\begin{equation}
 {\rm sgn}(B_n)= \left\{ \begin{array}{cc}
            + & (n ~{\rm mod}~ 4=0, 3) \\
            - & (n ~{\rm mod}~ 4=1, 2)
          \end{array} \right. \nonumber
\end{equation}
Since $h_2(Y)$ is an even function, we just take $h_2(y_n)~ (n \geq 0)$ into consideration.  We get
\begin{eqnarray}
B_{2n-1}+B_{2n-2}+\frac{B_0+B_1}{2} &\leq& h_2(y_n)-\frac{B_0+B_1}{2}=\sum_{i=0}^{2n-1} B_i-\frac{B_0+B_1}{2}\leq -\frac{B_0+B_1}{2}<0  \nonumber
\\&~&(n ~{\rm mod}~ 2=0),\label{2inequality1}
  \\
B_{2n-1}+B_{2n-2}-\frac{B_0+B_1}{2} &\geq& h_2(y_n)-\frac{B_0+B_1}{2}=\sum_{i=0}^{2n-1} B_i-\frac{B_0+B_1}{2}\geq \frac{B_0+B_1}{2}>0 \nonumber
\\&~&(n ~{\rm mod}~ 2=1).\label{2inequality2}
\end{eqnarray}

\begin{figure}
  \centering
  \includegraphics[width=8cm]{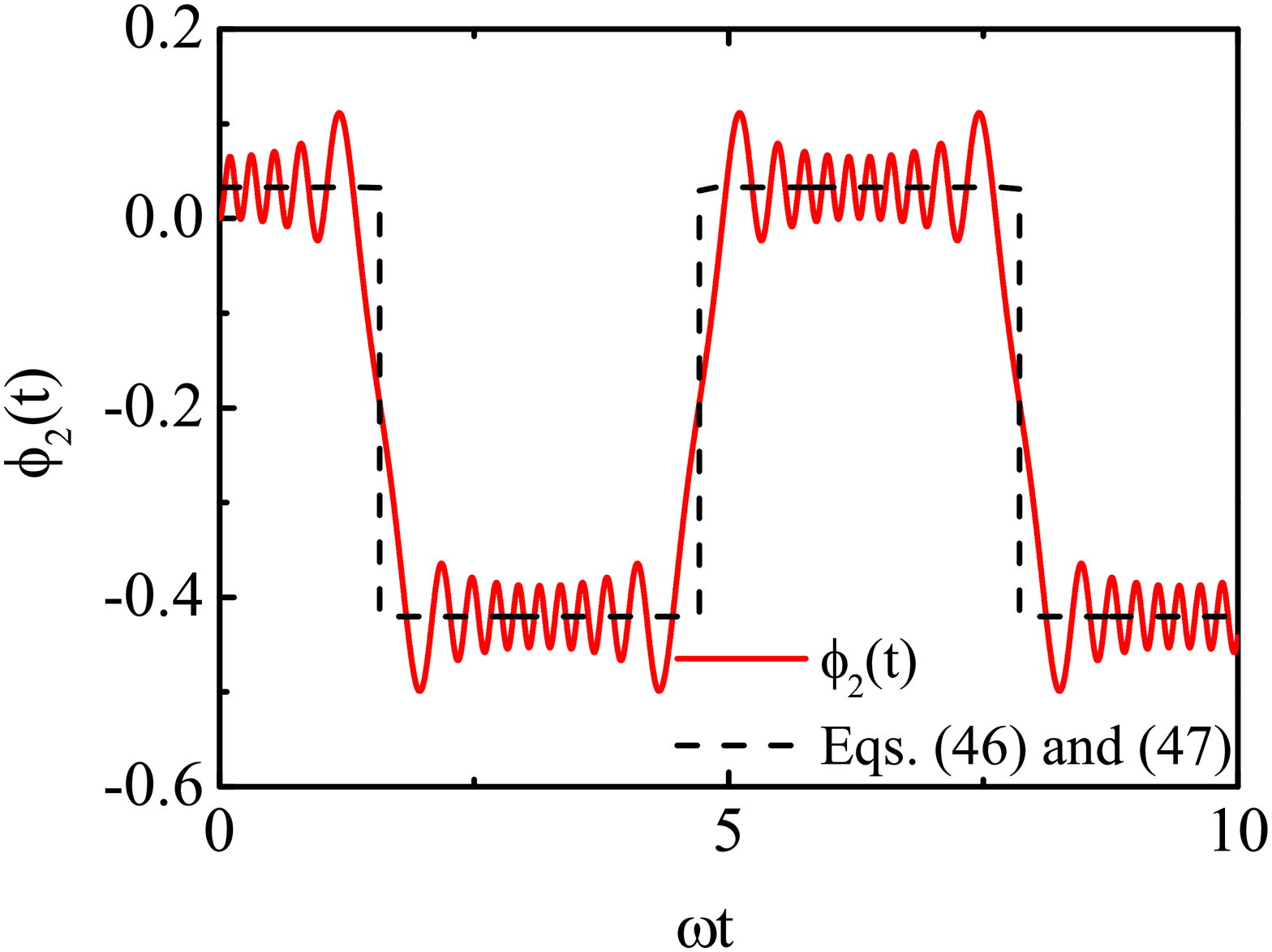}\\
  \caption{(Color online) The phase function, $\phi_2(t)=\Delta \int_{0}^{t} \sin\left[\frac{A}{\omega} \sin(\omega \tau)\right] d\tau$ shown in Eq. (\ref{cdt_z2}), as a function of dimensionless time $\omega t$ for $A=9.75\pi\omega$ and
  $\Delta=\omega$. The phase function is plotted by the red line with a two-stair plateau structure. The mean values of the plateaus are shown as a two-value
  function consisting of $L'_1$ and $L'_2$ given by Eq. (\ref{newL1}) and Eq. (\ref{newL2}), respectively, which is plotted by the black dashed line.}\label{fignewL}
\end{figure}

It is clear to see that all the maxima are positive while all the minima are negative in terms of $h_2(y_n)$. Returning to the phase function $\phi_2(t)$, we obtain that the value of $\phi_2(t)$ oscillates around a certain nonzero value, which is modified from $\phi_2\left(t=\frac{n\pi}{\omega}\right)$ to $\phi_2\left(t=\frac{n\pi}{\omega}\right)+\frac{B_0+B_1}{2}$. By $L_1$ and $L_2$ in Eq. (\ref{meanvalue1}), the average value of the plateaus can be obtained,
\begin{eqnarray}
  L'_1 &=& L_1+\frac{B_0+B_1}{2} \approx \frac{\Delta}{2}\int_{0}^{\frac{\pi}{A}} \sin\left[\frac{A}{\omega}\sin(\omega\tau)\right]d\tau \nonumber \\   \label{newL1}
       &=& \Delta\sum_{k=1}^{\infty}\frac{1}{(2k-1)\omega}J_{2k-1}\left(\frac{A}{\omega}\right)\left\{1-\cos\left[(2k-1)\frac{\pi\omega}{A}\right]\right\},\\
  L'_2 &=& L_2+\frac{B_0+B_1}{2} \nonumber\\
  & \approx & \Delta\sum_{k=1}^{\infty}\frac{1}{(2k-1)\omega}J_{2k-1}\left(\frac{A}{\omega}\right)\left\{3+\cos\left[(2k-1)\frac{\pi\omega}{A}\right]\right\}. \label{newL2}
\end{eqnarray}
The maxima and minima are distributed above and below the average value alternately as a function of time (see Fig. \ref{fignewL}). It is remarkable that the oscillating structure of $\phi_2(t)$ also satisfies a quasiperiodic condition, which can be written as $\frac{A\xi}{\omega}\sin(\omega t)=n\pi$. Accordingly, the number of the oscillation in a single plateau is calculated as $2 \left\lfloor\frac{A}{\omega\pi}\right\rfloor-\left\lfloor\frac{A}{\omega\pi}\right\rfloor$. To estimate the amplitude of the oscillation, from Eqs. (\ref{2inequality1}) and (\ref{2inequality2}), we can find reasonable bounds of these extrema as
\begin{eqnarray}
  \left|h_2(y_n)-\frac{B_0+B_1}{2}\right| &<& \left|\frac{\Delta}{\omega}\int_{\frac{n\omega\pi}{A}}^{\frac{(n-1)\omega\pi}{A}} \frac{\sin\left(\frac{A}{\omega} x\right)}{\sqrt{1-x^2}}dx\right|\nonumber \\
    &<& \frac{2\Delta}{A}\left[1-\left(\frac{n\omega\pi}{A}\right)^2\right]^{-\frac{1}{2}}\nonumber \\
    &<& \frac{2\Delta}{A}\left(1-y_n^2\right)^{-\frac{1}{2}}. \label{2bounded1}
\end{eqnarray}
The inequality Eq. (\ref{2bounded1}) is also valid when $y_n$ is substituted with any $y\in [-1,1]$. Subsequently, $g_2(t)$ ($t \in [-\frac{\pi}{2\omega},\frac{\pi}{2\omega}]$) holds
\begin{equation}\label{2inequality3}
  |g_2(t)|<q_2\equiv\frac{2\Delta}{A\cos(\omega t)}.
\end{equation}
To illustrate the validity of Eq. ({\ref{2inequality3}}), we show the results of $g_2(t)$, $q_2(t)$ and $-q_2(t)$ in Fig. \ref{Figadd1}. It is obvious to see that the $q_2(t)$ and $-q_2(t)$ are the best estimation of the upper bound and lower bound, respectively. Moreover, the envelope of $g_2(t)$ fulfills a function proportional to $1/\cos(\omega t)$ over time in most part of the plateau. Since the coefficient of $q_2(t)$ is proportional to $\Delta/A$, it indicates that the increase of $A$ or the decrease of $\Delta$ can suppress the amplitude of the oscillation on the plateau.


\begin{thebibliography}{56}%
\makeatletter
\providecommand \@ifxundefined [1]{%
 \@ifx{#1\undefined}
}%
\providecommand \@ifnum [1]{%
 \ifnum #1\expandafter \@firstoftwo
 \else \expandafter \@secondoftwo
 \fi
}%
\providecommand \@ifx [1]{%
 \ifx #1\expandafter \@firstoftwo
 \else \expandafter \@secondoftwo
 \fi
}%
\providecommand \natexlab [1]{#1}%
\providecommand \enquote  [1]{``#1''}%
\providecommand \bibnamefont  [1]{#1}%
\providecommand \bibfnamefont [1]{#1}%
\providecommand \citenamefont [1]{#1}%
\providecommand \href@noop [0]{\@secondoftwo}%
\providecommand \href [0]{\begingroup \@sanitize@url \@href}%
\providecommand \@href[1]{\@@startlink{#1}\@@href}%
\providecommand \@@href[1]{\endgroup#1\@@endlink}%
\providecommand \@sanitize@url [0]{\catcode `\\12\catcode `\$12\catcode
  `\&12\catcode `\#12\catcode `\^12\catcode `\_12\catcode `\%12\relax}%
\providecommand \@@startlink[1]{}%
\providecommand \@@endlink[0]{}%
\providecommand \url  [0]{\begingroup\@sanitize@url \@url }%
\providecommand \@url [1]{\endgroup\@href {#1}{\urlprefix }}%
\providecommand \urlprefix  [0]{URL }%
\providecommand \Eprint [0]{\href }%
\providecommand \doibase [0]{https://doi.org/}%
\providecommand \selectlanguage [0]{\@gobble}%
\providecommand \bibinfo  [0]{\@secondoftwo}%
\providecommand \bibfield  [0]{\@secondoftwo}%
\providecommand \translation [1]{[#1]}%
\providecommand \BibitemOpen [0]{}%
\providecommand \bibitemStop [0]{}%
\providecommand \bibitemNoStop [0]{.\EOS\space}%
\providecommand \EOS [0]{\spacefactor3000\relax}%
\providecommand \BibitemShut  [1]{\csname bibitem#1\endcsname}%
\let\auto@bib@innerbib\@empty
\bibitem [{\citenamefont {Grifoni}\ and\ \citenamefont
  {H\"anggi}(1998)}]{Hanggi}%
  \BibitemOpen
  \bibfield  {author} {\bibinfo {author} {\bibfnamefont {M.}~\bibnamefont
  {Grifoni}}\ and\ \bibinfo {author} {\bibfnamefont {P.}~\bibnamefont
  {H\"anggi}},\ }\bibfield  {title} {\bibinfo {title} {Driven quantum
  tunneling},\ }\href
  {https://doi.org/https://doi.org/10.1016/S0370-1573(98)00022-2} {\bibfield
  {journal} {\bibinfo  {journal} {Physics Reports}\ }\textbf {\bibinfo {volume}
  {304}},\ \bibinfo {pages} {229 } (\bibinfo {year} {1998})}\BibitemShut
  {NoStop}%
\bibitem [{\citenamefont {Leggett}\ \emph {et~al.}(1987)\citenamefont
  {Leggett}, \citenamefont {Chakravarty}, \citenamefont {Dorsey}, \citenamefont
  {Fisher}, \citenamefont {Garg},\ and\ \citenamefont
  {Zwerger}}]{Leggett-book_of_dissipative_TLS}%
  \BibitemOpen
  \bibfield  {author} {\bibinfo {author} {\bibfnamefont {A.~J.}\ \bibnamefont
  {Leggett}}, \bibinfo {author} {\bibfnamefont {S.}~\bibnamefont
  {Chakravarty}}, \bibinfo {author} {\bibfnamefont {A.~T.}\ \bibnamefont
  {Dorsey}}, \bibinfo {author} {\bibfnamefont {M.~P.~A.}\ \bibnamefont
  {Fisher}}, \bibinfo {author} {\bibfnamefont {A.}~\bibnamefont {Garg}},\ and\
  \bibinfo {author} {\bibfnamefont {W.}~\bibnamefont {Zwerger}},\ }\bibfield
  {title} {\bibinfo {title} {Dynamics of the dissipative two-state system},\
  }\href {https://doi.org/10.1103/RevModPhys.59.1} {\bibfield  {journal}
  {\bibinfo  {journal} {Rev. Mod. Phys.}\ }\textbf {\bibinfo {volume} {59}},\
  \bibinfo {pages} {1} (\bibinfo {year} {1987})}\BibitemShut {NoStop}%
\bibitem [{\citenamefont {Frisk~Kockum}\ \emph {et~al.}(2019)\citenamefont
  {Frisk~Kockum}, \citenamefont {Miranowicz}, \citenamefont {De~Liberato},
  \citenamefont {Savasta},\ and\ \citenamefont {Nori}}]{Kockum}%
  \BibitemOpen
  \bibfield  {author} {\bibinfo {author} {\bibfnamefont {A.}~\bibnamefont
  {Frisk~Kockum}}, \bibinfo {author} {\bibfnamefont {A.}~\bibnamefont
  {Miranowicz}}, \bibinfo {author} {\bibfnamefont {S.}~\bibnamefont
  {De~Liberato}}, \bibinfo {author} {\bibfnamefont {S.}~\bibnamefont
  {Savasta}},\ and\ \bibinfo {author} {\bibfnamefont {F.}~\bibnamefont
  {Nori}},\ }\bibfield  {title} {\bibinfo {title} {Ultrastrong coupling between
  light and matter},\ }\href {https://doi.org/10.1038/s42254-018-0006-2}
  {\bibfield  {journal} {\bibinfo  {journal} {Nature Reviews Physics}\ }\textbf
  {\bibinfo {volume} {1}},\ \bibinfo {pages} {19} (\bibinfo {year}
  {2019})}\BibitemShut {NoStop}%
\bibitem [{\citenamefont {Bloch}(1946)}]{TLS_relax:origin}%
  \BibitemOpen
  \bibfield  {author} {\bibinfo {author} {\bibfnamefont {F.}~\bibnamefont
  {Bloch}},\ }\bibfield  {title} {\bibinfo {title} {Nuclear induction},\ }\href
  {https://doi.org/10.1103/PhysRev.70.460} {\bibfield  {journal} {\bibinfo
  {journal} {Phys. Rev.}\ }\textbf {\bibinfo {volume} {70}},\ \bibinfo {pages}
  {460} (\bibinfo {year} {1946})}\BibitemShut {NoStop}%
\bibitem [{\citenamefont {Rourke}\ \emph {et~al.}(2004)\citenamefont {Rourke},
  \citenamefont {Khodarinova},\ and\ \citenamefont
  {Karabanov}}]{TLS_relax:David_PRL}%
  \BibitemOpen
  \bibfield  {author} {\bibinfo {author} {\bibfnamefont {D.~E.}\ \bibnamefont
  {Rourke}}, \bibinfo {author} {\bibfnamefont {L.}~\bibnamefont
  {Khodarinova}},\ and\ \bibinfo {author} {\bibfnamefont {A.~A.}\ \bibnamefont
  {Karabanov}},\ }\bibfield  {title} {\bibinfo {title} {Two-level systems with
  relaxation},\ }\href {https://doi.org/10.1103/PhysRevLett.92.163003}
  {\bibfield  {journal} {\bibinfo  {journal} {Phys. Rev. Lett.}\ }\textbf
  {\bibinfo {volume} {92}},\ \bibinfo {pages} {163003} (\bibinfo {year}
  {2004})}\BibitemShut {NoStop}%
\bibitem [{\citenamefont {Gorlicki}(1994)}]{TLS_relax:Gorlicki_adiabatic}%
  \BibitemOpen
  \bibfield  {author} {\bibinfo {author} {\bibfnamefont {M.}~\bibnamefont
  {Gorlicki}},\ }\bibfield  {title} {\bibinfo {title} {Evolution of a two-level
  system with relaxation driven by a time-dependent hamiltonian: A simple
  model},\ }\href {https://doi.org/10.1103/PhysRevA.49.4339} {\bibfield
  {journal} {\bibinfo  {journal} {Phys. Rev. A}\ }\textbf {\bibinfo {volume}
  {49}},\ \bibinfo {pages} {4339} (\bibinfo {year} {1994})}\BibitemShut
  {NoStop}%
\bibitem [{\citenamefont {Karabanov}\ and\ \citenamefont
  {Rourke}(2007)}]{TLS_relax:Karabanov_math}%
  \BibitemOpen
  \bibfield  {author} {\bibinfo {author} {\bibfnamefont {A.~A.}\ \bibnamefont
  {Karabanov}}\ and\ \bibinfo {author} {\bibfnamefont {D.~E.}\ \bibnamefont
  {Rourke}},\ }\bibfield  {title} {\bibinfo {title} {Spectral resolution and
  inversion of the Bloch equations with relaxation},\ }\href
  {https://doi.org/10.1007/s11005-007-0180-0} {\bibfield  {journal} {\bibinfo
  {journal} {Lett. Math. Phys.}\ }\textbf {\bibinfo {volume} {81}},\ \bibinfo
  {pages} {197} (\bibinfo {year} {2007})}\BibitemShut {NoStop}%
\bibitem [{\citenamefont {Lapert}\ \emph {et~al.}(2013)\citenamefont {Lapert},
  \citenamefont {Ass\'emat}, \citenamefont {Glaser},\ and\ \citenamefont
  {Sugny}}]{TLS_relax:Lapert}%
  \BibitemOpen
  \bibfield  {author} {\bibinfo {author} {\bibfnamefont {M.}~\bibnamefont
  {Lapert}}, \bibinfo {author} {\bibfnamefont {E.}~\bibnamefont {Ass\'emat}},
  \bibinfo {author} {\bibfnamefont {S.~J.}\ \bibnamefont {Glaser}},\ and\
  \bibinfo {author} {\bibfnamefont {D.}~\bibnamefont {Sugny}},\ }\bibfield
  {title} {\bibinfo {title} {Understanding the global structure of two-level
  quantum systems with relaxation: Vector fields organized through the magic
  plane and the steady-state ellipsoid},\ }\href
  {https://doi.org/10.1103/PhysRevA.88.033407} {\bibfield  {journal} {\bibinfo
  {journal} {Phys. Rev. A}\ }\textbf {\bibinfo {volume} {88}},\ \bibinfo
  {pages} {033407} (\bibinfo {year} {2013})}\BibitemShut {NoStop}%
\bibitem [{\citenamefont {Gershenzon}\ \emph {et~al.}(2009)\citenamefont
  {Gershenzon}, \citenamefont {Miller},\ and\ \citenamefont
  {Skinner}}]{TLS_relax:Naum}%
  \BibitemOpen
  \bibfield  {author} {\bibinfo {author} {\bibfnamefont {N.~I.}\ \bibnamefont
  {Gershenzon}}, \bibinfo {author} {\bibfnamefont {D.~F.}\ \bibnamefont
  {Miller}},\ and\ \bibinfo {author} {\bibfnamefont {T.~E.}\ \bibnamefont
  {Skinner}},\ }\bibfield  {title} {\bibinfo {title} {The design of excitation
  pulses for spin systems using optimal control theory: With application to nmr
  spectroscopy},\ }\href {https://doi.org/10.1002/oca.867} {\bibfield
  {journal} {\bibinfo  {journal} {Optimal Control Applications and Methods}\
  }\textbf {\bibinfo {volume} {30}},\ \bibinfo {pages} {463} (\bibinfo {year}
  {2009})}\BibitemShut {NoStop}%
\bibitem [{\citenamefont {Asban}\ \emph {et~al.}(2017)\citenamefont {Asban},
  \citenamefont {Amir}, \citenamefont {Imry},\ and\ \citenamefont
  {Schechter}}]{TLS_relax:Ofek_disorder}%
  \BibitemOpen
  \bibfield  {author} {\bibinfo {author} {\bibfnamefont {O.}~\bibnamefont
  {Asban}}, \bibinfo {author} {\bibfnamefont {A.}~\bibnamefont {Amir}},
  \bibinfo {author} {\bibfnamefont {Y.}~\bibnamefont {Imry}},\ and\ \bibinfo
  {author} {\bibfnamefont {M.}~\bibnamefont {Schechter}},\ }\bibfield  {title}
  {\bibinfo {title} {Effect of interactions and disorder on the relaxation of
  two-level systems in amorphous solids},\ }\href
  {https://doi.org/10.1103/PhysRevB.95.144207} {\bibfield  {journal} {\bibinfo
  {journal} {Phys. Rev. B}\ }\textbf {\bibinfo {volume} {95}},\ \bibinfo
  {pages} {144207} (\bibinfo {year} {2017})}\BibitemShut {NoStop}%
\bibitem [{\citenamefont {Castella}\ and\ \citenamefont
  {Zimmermann}(1999)}]{TLS_boson:Castella_coherent_ctrl}%
  \BibitemOpen
  \bibfield  {author} {\bibinfo {author} {\bibfnamefont {H.}~\bibnamefont
  {Castella}}\ and\ \bibinfo {author} {\bibfnamefont {R.}~\bibnamefont
  {Zimmermann}},\ }\bibfield  {title} {\bibinfo {title} {Coherent control for a
  two-level system coupled to phonons},\ }\href
  {https://doi.org/10.1103/PhysRevB.59.R7801} {\bibfield  {journal} {\bibinfo
  {journal} {Phys. Rev. B}\ }\textbf {\bibinfo {volume} {59}},\ \bibinfo
  {pages} {R7801} (\bibinfo {year} {1999})}\BibitemShut {NoStop}%
\bibitem [{\citenamefont {Chenu}\ \emph {et~al.}(2019)\citenamefont {Chenu},
  \citenamefont {Shiau},\ and\ \citenamefont {Combescot}}]{TLS_boson:Chenu}%
  \BibitemOpen
  \bibfield  {author} {\bibinfo {author} {\bibfnamefont {A.}~\bibnamefont
  {Chenu}}, \bibinfo {author} {\bibfnamefont {S.-Y.}\ \bibnamefont {Shiau}},\
  and\ \bibinfo {author} {\bibfnamefont {M.}~\bibnamefont {Combescot}},\
  }\bibfield  {title} {\bibinfo {title} {Two-level system coupled to phonons:
  Full analytical solution},\ }\href
  {https://doi.org/10.1103/PhysRevB.99.014302} {\bibfield  {journal} {\bibinfo
  {journal} {Phys. Rev. B}\ }\textbf {\bibinfo {volume} {99}},\ \bibinfo
  {pages} {014302} (\bibinfo {year} {2019})}\BibitemShut {NoStop}%
\bibitem [{\citenamefont {Mitzner}\ and\ \citenamefont
  {Campbell}(1995)}]{TLS_boson:Plohn_path-integral}%
  \BibitemOpen
  \bibfield  {author} {\bibinfo {author} {\bibfnamefont {R.}~\bibnamefont
  {Mitzner}}\ and\ \bibinfo {author} {\bibfnamefont {E.~E.~B.}\ \bibnamefont
  {Campbell}},\ }\bibfield  {title} {\bibinfo {title} {Optical emission studies
  of laser desorbed C60},\ }\href {https://doi.org/10.1063/1.469667} {\bibfield
   {journal} {\bibinfo  {journal} {The Journal of Chemical Physics}\ }\textbf
  {\bibinfo {volume} {103}},\ \bibinfo {pages} {2445} (\bibinfo {year}
  {1995})}\BibitemShut {NoStop}%
\bibitem [{\citenamefont {De~Raedt}\ and\ \citenamefont
  {De~Raedt}(1984)}]{TLS_boson:Raedt_thermo}%
  \BibitemOpen
  \bibfield  {author} {\bibinfo {author} {\bibfnamefont {B.}~\bibnamefont
  {De~Raedt}}\ and\ \bibinfo {author} {\bibfnamefont {H.}~\bibnamefont
  {De~Raedt}},\ }\bibfield  {title} {\bibinfo {title} {Thermodynamics of a
  two-level system coupled to bosons},\ }\href
  {https://doi.org/10.1103/PhysRevB.29.5325} {\bibfield  {journal} {\bibinfo
  {journal} {Phys. Rev. B}\ }\textbf {\bibinfo {volume} {29}},\ \bibinfo
  {pages} {5325} (\bibinfo {year} {1984})}\BibitemShut {NoStop}%
\bibitem [{\citenamefont {Shen}\ \emph {et~al.}(2016)\citenamefont {Shen},
  \citenamefont {Shao}, \citenamefont {Wang}, \citenamefont {Zhao},\ and\
  \citenamefont {Yi}}]{TLS_boson:Shen_QPT}%
  \BibitemOpen
  \bibfield  {author} {\bibinfo {author} {\bibfnamefont {H.~Z.}\ \bibnamefont
  {Shen}}, \bibinfo {author} {\bibfnamefont {X.~Q.}\ \bibnamefont {Shao}},
  \bibinfo {author} {\bibfnamefont {G.~C.}\ \bibnamefont {Wang}}, \bibinfo
  {author} {\bibfnamefont {X.~L.}\ \bibnamefont {Zhao}},\ and\ \bibinfo
  {author} {\bibfnamefont {X.~X.}\ \bibnamefont {Yi}},\ }\bibfield  {title}
  {\bibinfo {title} {Quantum phase transition in a coupled two-level system
  embedded in anisotropic three-dimensional photonic crystals},\ }\href
  {https://doi.org/10.1103/PhysRevE.93.012107} {\bibfield  {journal} {\bibinfo
  {journal} {Phys. Rev. E}\ }\textbf {\bibinfo {volume} {93}},\ \bibinfo
  {pages} {012107} (\bibinfo {year} {2016})}\BibitemShut {NoStop}%
\bibitem [{\citenamefont {Dakhnovskii}(1994)}]{EF:Dakhnovskii}%
  \BibitemOpen
  \bibfield  {author} {\bibinfo {author} {\bibfnamefont {Y.}~\bibnamefont
  {Dakhnovskii}},\ }\bibfield  {title} {\bibinfo {title} {Dynamics of a
  two-level system with Ohmic dissipation in a time-dependent field},\ }\href
  {https://doi.org/10.1103/PhysRevB.49.4649} {\bibfield  {journal} {\bibinfo
  {journal} {Phys. Rev. B}\ }\textbf {\bibinfo {volume} {49}},\ \bibinfo
  {pages} {4649} (\bibinfo {year} {1994})}\BibitemShut {NoStop}%
\bibitem [{\citenamefont {Wang}\ \emph
  {et~al.}(1998{\natexlab{a}})\citenamefont {Wang}, \citenamefont {Freire},\
  and\ \citenamefont {Zhao}}]{EF:Wang}%
  \BibitemOpen
  \bibfield  {author} {\bibinfo {author} {\bibfnamefont {H.}~\bibnamefont
  {Wang}}, \bibinfo {author} {\bibfnamefont {V.~N.}\ \bibnamefont {Freire}},\
  and\ \bibinfo {author} {\bibfnamefont {X.-G.}\ \bibnamefont {Zhao}},\
  }\bibfield  {title} {\bibinfo {title} {Dissipative quantum tunneling of
  two-level systems driven by dc-ac fields},\ }\href
  {https://doi.org/10.1103/PhysRevE.58.2632} {\bibfield  {journal} {\bibinfo
  {journal} {Phys. Rev. E}\ }\textbf {\bibinfo {volume} {58}},\ \bibinfo
  {pages} {2632} (\bibinfo {year} {1998}{\natexlab{a}})}\BibitemShut {NoStop}%
\bibitem [{\citenamefont {Grimaudo}\ \emph
  {et~al.}(2019)\citenamefont {Grimaudo}, \citenamefont {Vitanov},\
  and\ \citenamefont {Messina}}]{Qutrit:Grimaudo_twoQ}%
  \BibitemOpen
  \bibfield  {author} {\bibinfo {author} {\bibfnamefont {R.}~\bibnamefont
  {Grimaudo}}, \bibinfo {author} {\bibfnamefont {N.~V.}\ \bibnamefont {Vitanov}},\
  and\ \bibinfo {author} {\bibfnamefont {A.}\ \bibnamefont {Messina}},\
  }\bibfield  {title} {\bibinfo {title} {Landau-Majorana-St\"uckelberg-Zener dynamics driven by coupling for two interacting qutrit systems},\ }\href
  {https://doi.org/10.1103/PhysRevB.99.214406} {\bibfield  {journal} {\bibinfo
  {journal} {Phys. Rev. B}\ }\textbf {\bibinfo {volume} {99}},\ \bibinfo
  {pages} {214406} (\bibinfo {year} {2019})}\BibitemShut {NoStop}%
\bibitem [{\citenamefont {Han}\ \emph {et~al.}(2019)\citenamefont {Han},
  \citenamefont {Luo}, \citenamefont {Li}, \citenamefont {Zhang}, \citenamefont
  {Wang}, \citenamefont {Tsai}, \citenamefont {Nori},\ and\ \citenamefont
  {You}}]{Qutrit:Han}%
  \BibitemOpen
  \bibfield  {author} {\bibinfo {author} {\bibfnamefont {Y.}~\bibnamefont
  {Han}}, \bibinfo {author} {\bibfnamefont {X.-Q.}\ \bibnamefont {Luo}},
  \bibinfo {author} {\bibfnamefont {T.-F.}\ \bibnamefont {Li}}, \bibinfo
  {author} {\bibfnamefont {W.}~\bibnamefont {Zhang}}, \bibinfo {author}
  {\bibfnamefont {S.-P.}\ \bibnamefont {Wang}}, \bibinfo {author}
  {\bibfnamefont {J.}~\bibnamefont {Tsai}}, \bibinfo {author} {\bibfnamefont
  {F.}~\bibnamefont {Nori}},\ and\ \bibinfo {author} {\bibfnamefont
  {J.}~\bibnamefont {You}},\ }\bibfield  {title} {\bibinfo {title} {Time-domain
  grating with a periodically driven qutrit},\ }\href
  {https://doi.org/10.1103/PhysRevApplied.11.014053} {\bibfield  {journal}
  {\bibinfo  {journal} {Phys. Rev. Applied}\ }\textbf {\bibinfo {volume}
  {11}},\ \bibinfo {pages} {014053} (\bibinfo {year} {2019})}\BibitemShut
  {NoStop}%
\bibitem [{\citenamefont {Wang}\ \emph {et~al.}(2015)\citenamefont {Wang},
  \citenamefont {Sun},\ and\ \citenamefont {Li}}]{Qutrit:QED}%
  \BibitemOpen
  \bibfield  {author} {\bibinfo {author} {\bibfnamefont {Z.~H.}\ \bibnamefont
  {Wang}}, \bibinfo {author} {\bibfnamefont {C.~P.}\ \bibnamefont {Sun}},\ and\
  \bibinfo {author} {\bibfnamefont {Y.}~\bibnamefont {Li}},\ }\bibfield
  {title} {\bibinfo {title} {Microwave degenerate parametric down-conversion
  with a single cyclic three-level system in a circuit-QED setup},\ }\href
  {https://doi.org/10.1103/PhysRevA.91.043801} {\bibfield  {journal} {\bibinfo
  {journal} {Phys. Rev. A}\ }\textbf {\bibinfo {volume} {91}},\ \bibinfo
  {pages} {043801} (\bibinfo {year} {2015})}\BibitemShut {NoStop}%
\bibitem [{\citenamefont {Vitanov}(2012)}]{Qutrit:Vitanov_Arbitrary_SU3}%
  \BibitemOpen
  \bibfield  {author} {\bibinfo {author} {\bibfnamefont {N.~V.}\ \bibnamefont
  {Vitanov}},\ }\bibfield  {title} {\bibinfo {title} {Synthesis of arbitrary
  SU(3) transformations of atomic qutrits},\ }\href
  {https://doi.org/10.1103/PhysRevA.85.032331} {\bibfield  {journal} {\bibinfo
  {journal} {Phys. Rev. A}\ }\textbf {\bibinfo {volume} {85}},\ \bibinfo
  {pages} {032331} (\bibinfo {year} {2012})}\BibitemShut {NoStop}%
\bibitem [{\citenamefont {Rabi}(1936)}]{Rabi:origin}%
  \BibitemOpen
  \bibfield  {author} {\bibinfo {author} {\bibfnamefont {I.~I.}\ \bibnamefont
  {Rabi}},\ }\bibfield  {title} {\bibinfo {title} {On the process of space
  quantization},\ }\href {https://doi.org/10.1103/PhysRev.49.324} {\bibfield
  {journal} {\bibinfo  {journal} {Phys. Rev.}\ }\textbf {\bibinfo {volume}
  {49}},\ \bibinfo {pages} {324} (\bibinfo {year} {1936})}\BibitemShut
  {NoStop}%
\bibitem [{\citenamefont {Berezovsky}\ \emph {et~al.}(2008)\citenamefont
  {Berezovsky}, \citenamefont {Mikkelsen}, \citenamefont {Stoltz},
  \citenamefont {Coldren},\ and\ \citenamefont {Awschalom}}]{EXP:Berezovsky}%
  \BibitemOpen
  \bibfield  {author} {\bibinfo {author} {\bibfnamefont {J.}~\bibnamefont
  {Berezovsky}}, \bibinfo {author} {\bibfnamefont {M.~H.}\ \bibnamefont
  {Mikkelsen}}, \bibinfo {author} {\bibfnamefont {N.~G.}\ \bibnamefont
  {Stoltz}}, \bibinfo {author} {\bibfnamefont {L.~A.}\ \bibnamefont
  {Coldren}},\ and\ \bibinfo {author} {\bibfnamefont {D.~D.}\ \bibnamefont
  {Awschalom}},\ }\bibfield  {title} {\bibinfo {title} {Picosecond coherent
  optical manipulation of a single electron spin in a quantum dot},\ }\href
  {https://doi.org/10.1126/science.1154798} {\bibfield  {journal} {\bibinfo
  {journal} {Science}\ }\textbf {\bibinfo {volume} {320}},\ \bibinfo {pages}
  {349} (\bibinfo {year} {2008})}\BibitemShut {NoStop}%
\bibitem [{\citenamefont {Deblock}\ \emph {et~al.}(2003)\citenamefont
  {Deblock}, \citenamefont {Onac}, \citenamefont {Gurevich},\ and\
  \citenamefont {Kouwenhoven}}]{EXP:Deblock}%
  \BibitemOpen
  \bibfield  {author} {\bibinfo {author} {\bibfnamefont {R.}~\bibnamefont
  {Deblock}}, \bibinfo {author} {\bibfnamefont {E.}~\bibnamefont {Onac}},
  \bibinfo {author} {\bibfnamefont {L.}~\bibnamefont {Gurevich}},\ and\
  \bibinfo {author} {\bibfnamefont {L.~P.}\ \bibnamefont {Kouwenhoven}},\
  }\bibfield  {title} {\bibinfo {title} {Detection of quantum noise from an
  electrically driven two-level system},\ }\href
  {https://doi.org/10.1126/science.1084175} {\bibfield  {journal} {\bibinfo
  {journal} {Science}\ }\textbf {\bibinfo {volume} {301}},\ \bibinfo {pages}
  {203} (\bibinfo {year} {2003})}\BibitemShut {NoStop}%
\bibitem [{\citenamefont {Nakamura}\ \emph {et~al.}(1999)\citenamefont
  {Nakamura}, \citenamefont {Pashkin},\ and\ \citenamefont
  {Tsai}}]{EXP:Nakamura}%
  \BibitemOpen
  \bibfield  {author} {\bibinfo {author} {\bibfnamefont {Y.}~\bibnamefont
  {Nakamura}}, \bibinfo {author} {\bibfnamefont {Y.}~\bibnamefont {Pashkin}},\
  and\ \bibinfo {author} {\bibfnamefont {J.}~\bibnamefont {Tsai}},\ }\bibfield
  {title} {\bibinfo {title} {Coherent control of macroscopic quantum states in
  a single-Cooper-pair box},\ }\href {https://doi.org/10.1038/19718} {\bibfield
   {journal} {\bibinfo  {journal} {Nature}\ }\textbf {\bibinfo {volume}
  {398}},\ \bibinfo {pages} {786} (\bibinfo {year} {1999})}\BibitemShut
  {NoStop}%
\bibitem [{\citenamefont {Laucht}\ \emph {et~al.}(2016)\citenamefont {Laucht},
  \citenamefont {Simmons}, \citenamefont {Kalra}, \citenamefont {Tosi},
  \citenamefont {Dehollain}, \citenamefont {Muhonen}, \citenamefont {Freer},
  \citenamefont {Hudson}, \citenamefont {Itoh}, \citenamefont {Jamieson},
  \citenamefont {McCallum}, \citenamefont {Dzurak},\ and\ \citenamefont
  {Morello}}]{EXP:spin-qubit}%
  \BibitemOpen
  \bibfield  {author} {\bibinfo {author} {\bibfnamefont {A.}~\bibnamefont
  {Laucht}}, \bibinfo {author} {\bibfnamefont {S.}~\bibnamefont {Simmons}},
  \bibinfo {author} {\bibfnamefont {R.}~\bibnamefont {Kalra}}, \bibinfo
  {author} {\bibfnamefont {G.}~\bibnamefont {Tosi}}, \bibinfo {author}
  {\bibfnamefont {J.~P.}\ \bibnamefont {Dehollain}}, \bibinfo {author}
  {\bibfnamefont {J.~T.}\ \bibnamefont {Muhonen}}, \bibinfo {author}
  {\bibfnamefont {S.}~\bibnamefont {Freer}}, \bibinfo {author} {\bibfnamefont
  {F.~E.}\ \bibnamefont {Hudson}}, \bibinfo {author} {\bibfnamefont {K.~M.}\
  \bibnamefont {Itoh}}, \bibinfo {author} {\bibfnamefont {D.~N.}\ \bibnamefont
  {Jamieson}}, \bibinfo {author} {\bibfnamefont {J.~C.}\ \bibnamefont
  {McCallum}}, \bibinfo {author} {\bibfnamefont {A.~S.}\ \bibnamefont
  {Dzurak}},\ and\ \bibinfo {author} {\bibfnamefont {A.}~\bibnamefont
  {Morello}},\ }\bibfield  {title} {\bibinfo {title} {Breaking the rotating
  wave approximation for a strongly driven dressed single-electron spin},\
  }\href {https://doi.org/10.1103/PhysRevB.94.161302} {\bibfield  {journal}
  {\bibinfo  {journal} {Phys. Rev. B}\ }\textbf {\bibinfo {volume} {94}},\
  \bibinfo {pages} {161302} (\bibinfo {year} {2016})}\BibitemShut {NoStop}%
\bibitem [{\citenamefont {Yoshihara}\ \emph {et~al.}(2014)\citenamefont
  {Yoshihara}, \citenamefont {Nakamura}, \citenamefont {Yan}, \citenamefont
  {Gustavsson}, \citenamefont {Bylander}, \citenamefont {Oliver},\ and\
  \citenamefont {Tsai}}]{fluxstrongdriving}%
  \BibitemOpen
  \bibfield  {author} {\bibinfo {author} {\bibfnamefont {F.}~\bibnamefont
  {Yoshihara}}, \bibinfo {author} {\bibfnamefont {Y.}~\bibnamefont {Nakamura}},
  \bibinfo {author} {\bibfnamefont {F.}~\bibnamefont {Yan}}, \bibinfo {author}
  {\bibfnamefont {S.}~\bibnamefont {Gustavsson}}, \bibinfo {author}
  {\bibfnamefont {J.}~\bibnamefont {Bylander}}, \bibinfo {author}
  {\bibfnamefont {W.~D.}\ \bibnamefont {Oliver}},\ and\ \bibinfo {author}
  {\bibfnamefont {J.-S.}\ \bibnamefont {Tsai}},\ }\bibfield  {title} {\bibinfo
  {title} {Flux qubit noise spectroscopy using Rabi oscillations under strong
  driving conditions},\ }\href {https://doi.org/10.1103/PhysRevB.89.020503}
  {\bibfield  {journal} {\bibinfo  {journal} {Phys. Rev. B}\ }\textbf {\bibinfo
  {volume} {89}},\ \bibinfo {pages} {020503} (\bibinfo {year}
  {2014})}\BibitemShut {NoStop}%
\bibitem [{\citenamefont {Deng}\ \emph {et~al.}(2016)\citenamefont {Deng},
  \citenamefont {Shen}, \citenamefont {Ashhab},\ and\ \citenamefont
  {Lupascu}}]{SCstrongdriving}%
  \BibitemOpen
  \bibfield  {author} {\bibinfo {author} {\bibfnamefont {C.}~\bibnamefont
  {Deng}}, \bibinfo {author} {\bibfnamefont {F.}~\bibnamefont {Shen}}, \bibinfo
  {author} {\bibfnamefont {S.}~\bibnamefont {Ashhab}},\ and\ \bibinfo {author}
  {\bibfnamefont {A.}~\bibnamefont {Lupascu}},\ }\bibfield  {title} {\bibinfo
  {title} {Dynamics of a two-level system under strong driving: Quantum-gate
  optimization based on Floquet theory},\ }\href
  {https://doi.org/10.1103/PhysRevA.94.032323} {\bibfield  {journal} {\bibinfo
  {journal} {Phys. Rev. A}\ }\textbf {\bibinfo {volume} {94}},\ \bibinfo
  {pages} {032323} (\bibinfo {year} {2016})}\BibitemShut {NoStop}%
\bibitem [{\citenamefont {Lobanov}\ \emph {et~al.}(2017)\citenamefont
  {Lobanov}, \citenamefont {Gippius}, \citenamefont {Tikhodeev},\ and\
  \citenamefont {Butov}}]{Yu-strongdriving}%
  \BibitemOpen
  \bibfield  {author} {\bibinfo {author} {\bibfnamefont {S.~V.}\ \bibnamefont
  {Lobanov}}, \bibinfo {author} {\bibfnamefont {N.~A.}\ \bibnamefont
  {Gippius}}, \bibinfo {author} {\bibfnamefont {S.~G.}\ \bibnamefont
  {Tikhodeev}},\ and\ \bibinfo {author} {\bibfnamefont {L.~V.}\ \bibnamefont
  {Butov}},\ }\bibfield  {title} {\bibinfo {title} {Control of light
  polarization by voltage in excitonic metasurface devices},\ }\href
  {https://doi.org/10.1063/1.5005827} {\bibfield  {journal} {\bibinfo
  {journal} {Applied Physics Letters}\ }\textbf {\bibinfo {volume} {111}},\
  \bibinfo {pages} {241101} (\bibinfo {year} {2017})}\BibitemShut {NoStop}%
\bibitem [{\citenamefont {Yu}\ \emph {et~al.}(2012)\citenamefont {Yu},
  \citenamefont {Zhu}, \citenamefont {Liang}, \citenamefont {Chen},\ and\
  \citenamefont {Jia}}]{Rabi:analytic_solution1}%
  \BibitemOpen
  \bibfield  {author} {\bibinfo {author} {\bibfnamefont {L.}~\bibnamefont
  {Yu}}, \bibinfo {author} {\bibfnamefont {S.}~\bibnamefont {Zhu}}, \bibinfo
  {author} {\bibfnamefont {Q.}~\bibnamefont {Liang}}, \bibinfo {author}
  {\bibfnamefont {G.}~\bibnamefont {Chen}},\ and\ \bibinfo {author}
  {\bibfnamefont {S.}~\bibnamefont {Jia}},\ }\bibfield  {title} {\bibinfo
  {title} {Analytical solutions for the Rabi model},\ }\href
  {https://doi.org/10.1103/PhysRevA.86.015803} {\bibfield  {journal} {\bibinfo
  {journal} {Phys. Rev. A}\ }\textbf {\bibinfo {volume} {86}},\ \bibinfo
  {pages} {015803} (\bibinfo {year} {2012})}\BibitemShut {NoStop}%
\bibitem [{\citenamefont {C\'ardenas}\ \emph {et~al.}(2017)\citenamefont
  {C\'ardenas}, \citenamefont {Teixeira},\ and\ \citenamefont
  {Semi\~ao}}]{Rabi:coupled_qubit}%
  \BibitemOpen
  \bibfield  {author} {\bibinfo {author} {\bibfnamefont {P.~C.}\ \bibnamefont
  {C\'ardenas}}, \bibinfo {author} {\bibfnamefont {W.~S.}\ \bibnamefont
  {Teixeira}},\ and\ \bibinfo {author} {\bibfnamefont {F.~L.}\ \bibnamefont
  {Semi\~ao}},\ }\bibfield  {title} {\bibinfo {title} {Coupled modes locally
  interacting with qubits: Critical assessment of the rotating-wave
  approximation},\ }\href {https://doi.org/10.1103/PhysRevA.95.042116}
  {\bibfield  {journal} {\bibinfo  {journal} {Phys. Rev. A}\ }\textbf {\bibinfo
  {volume} {95}},\ \bibinfo {pages} {042116} (\bibinfo {year}
  {2017})}\BibitemShut {NoStop}%
\bibitem [{\citenamefont {Yan}\ \emph {et~al.}(2017{\natexlab{a}})\citenamefont
  {Yan}, \citenamefont {L\"u}, \citenamefont {Luo},\ and\ \citenamefont
  {Zheng}}]{Rabi:CR_term}%
  \BibitemOpen
  \bibfield  {author} {\bibinfo {author} {\bibfnamefont {Y.}~\bibnamefont
  {Yan}}, \bibinfo {author} {\bibfnamefont {Z.}~\bibnamefont {L\"u}}, \bibinfo
  {author} {\bibfnamefont {J.}~\bibnamefont {Luo}},\ and\ \bibinfo {author}
  {\bibfnamefont {H.}~\bibnamefont {Zheng}},\ }\bibfield  {title} {\bibinfo
  {title} {Effects of counter-rotating couplings of the Rabi model with
  frequency modulation},\ }\href {https://doi.org/10.1103/PhysRevA.96.033802}
  {\bibfield  {journal} {\bibinfo  {journal} {Phys. Rev. A}\ }\textbf {\bibinfo
  {volume} {96}},\ \bibinfo {pages} {033802} (\bibinfo {year}
  {2017}{\natexlab{a}})}\BibitemShut {NoStop}%
\bibitem [{\citenamefont {Braak}(2011)}]{Rabi:integrability}%
  \BibitemOpen
  \bibfield  {author} {\bibinfo {author} {\bibfnamefont {D.}~\bibnamefont
  {Braak}},\ }\bibfield  {title} {\bibinfo {title} {Integrability of the Rabi
  model},\ }\href {https://doi.org/10.1103/PhysRevLett.107.100401} {\bibfield
  {journal} {\bibinfo  {journal} {Phys. Rev. Lett.}\ }\textbf {\bibinfo
  {volume} {107}},\ \bibinfo {pages} {100401} (\bibinfo {year}
  {2011})}\BibitemShut {NoStop}%
\bibitem [{\citenamefont {Hausinger}\ and\ \citenamefont
  {Grifoni}(2010)}]{Rabi:untrastrong_regime}%
  \BibitemOpen
  \bibfield  {author} {\bibinfo {author} {\bibfnamefont {J.}~\bibnamefont
  {Hausinger}}\ and\ \bibinfo {author} {\bibfnamefont {M.}~\bibnamefont
  {Grifoni}},\ }\bibfield  {title} {\bibinfo {title} {Qubit-oscillator system:
  An analytical treatment of the ultrastrong coupling regime},\ }\href
  {https://doi.org/10.1103/PhysRevA.82.062320} {\bibfield  {journal} {\bibinfo
  {journal} {Phys. Rev. A}\ }\textbf {\bibinfo {volume} {82}},\ \bibinfo
  {pages} {062320} (\bibinfo {year} {2010})}\BibitemShut {NoStop}%
\bibitem [{\citenamefont {Grossmann}\ \emph {et~al.}(1991)\citenamefont
  {Grossmann}, \citenamefont {Dittrich}, \citenamefont {Jung},\ and\
  \citenamefont {H\"anggi}}]{CDT:origin}%
  \BibitemOpen
  \bibfield  {author} {\bibinfo {author} {\bibfnamefont {F.}~\bibnamefont
  {Grossmann}}, \bibinfo {author} {\bibfnamefont {T.}~\bibnamefont {Dittrich}},
  \bibinfo {author} {\bibfnamefont {P.}~\bibnamefont {Jung}},\ and\ \bibinfo
  {author} {\bibfnamefont {P.}~\bibnamefont {H\"anggi}},\ }\bibfield  {title}
  {\bibinfo {title} {Coherent destruction of tunneling},\ }\href
  {https://doi.org/10.1103/PhysRevLett.67.516} {\bibfield  {journal} {\bibinfo
  {journal} {Phys. Rev. Lett.}\ }\textbf {\bibinfo {volume} {67}},\ \bibinfo
  {pages} {516} (\bibinfo {year} {1991})}\BibitemShut {NoStop}%
\bibitem [{\citenamefont {Stockburger}(1999)}]{CDT:stabilizing_CDT}%
  \BibitemOpen
  \bibfield  {author} {\bibinfo {author} {\bibfnamefont {J.~T.}\ \bibnamefont
  {Stockburger}},\ }\bibfield  {title} {\bibinfo {title} {Stabilizing coherent
  destruction of tunneling},\ }\href
  {https://doi.org/10.1103/PhysRevE.59.R4709} {\bibfield  {journal} {\bibinfo
  {journal} {Phys. Rev. E}\ }\textbf {\bibinfo {volume} {59}},\ \bibinfo
  {pages} {R4709} (\bibinfo {year} {1999})}\BibitemShut {NoStop}%
\bibitem [{\citenamefont {Luo}\ \emph {et~al.}(2011)\citenamefont {Luo},
  \citenamefont {Huang},\ and\ \citenamefont {Lee}}]{CDT:experiment}%
  \BibitemOpen
  \bibfield  {author} {\bibinfo {author} {\bibfnamefont {X.}~\bibnamefont
  {Luo}}, \bibinfo {author} {\bibfnamefont {J.}~\bibnamefont {Huang}},\ and\
  \bibinfo {author} {\bibfnamefont {C.}~\bibnamefont {Lee}},\ }\bibfield
  {title} {\bibinfo {title} {Coherent destruction of tunneling in a lattice
  array under selective in-phase modulations},\ }\href
  {https://doi.org/10.1103/PhysRevA.84.053847} {\bibfield  {journal} {\bibinfo
  {journal} {Phys. Rev. A}\ }\textbf {\bibinfo {volume} {84}},\ \bibinfo
  {pages} {053847} (\bibinfo {year} {2011})}\BibitemShut {NoStop}%
\bibitem [{\citenamefont {Bloch}\ and\ \citenamefont
  {Siegert}(1940)}]{BS:Bloch}%
  \BibitemOpen
  \bibfield  {author} {\bibinfo {author} {\bibfnamefont {F.}~\bibnamefont
  {Bloch}}\ and\ \bibinfo {author} {\bibfnamefont {A.}~\bibnamefont
  {Siegert}},\ }\bibfield  {title} {\bibinfo {title} {Magnetic resonance for
  nonrotating fields},\ }\href {https://doi.org/10.1103/PhysRev.57.522}
  {\bibfield  {journal} {\bibinfo  {journal} {Phys. Rev.}\ }\textbf {\bibinfo
  {volume} {57}},\ \bibinfo {pages} {522} (\bibinfo {year} {1940})}\BibitemShut
  {NoStop}%
\bibitem [{\citenamefont {Yan}\ \emph {et~al.}(2015)\citenamefont {Yan},
  \citenamefont {L\"u},\ and\ \citenamefont {Zheng}}]{BS:Yan}%
  \BibitemOpen
  \bibfield  {author} {\bibinfo {author} {\bibfnamefont {Y.}~\bibnamefont
  {Yan}}, \bibinfo {author} {\bibfnamefont {Z.}~\bibnamefont {L\"u}},\ and\
  \bibinfo {author} {\bibfnamefont {H.}~\bibnamefont {Zheng}},\ }\bibfield
  {title} {\bibinfo {title} {Bloch-Siegert shift of the rabi model},\ }\href
  {https://doi.org/10.1103/PhysRevA.91.053834} {\bibfield  {journal} {\bibinfo
  {journal} {Phys. Rev. A}\ }\textbf {\bibinfo {volume} {91}},\ \bibinfo
  {pages} {053834} (\bibinfo {year} {2015})}\BibitemShut {NoStop}%
\bibitem [{\citenamefont {Zhang}\ \emph {et~al.}(2018)\citenamefont {Zhang},
  \citenamefont {Saha},\ and\ \citenamefont {Suter}}]{BS:Zhang}%
  \BibitemOpen
  \bibfield  {author} {\bibinfo {author} {\bibfnamefont {J.}~\bibnamefont
  {Zhang}}, \bibinfo {author} {\bibfnamefont {S.}~\bibnamefont {Saha}},\ and\
  \bibinfo {author} {\bibfnamefont {D.}~\bibnamefont {Suter}},\ }\bibfield
  {title} {\bibinfo {title} {Bloch-Siegert shift in a hybrid quantum register:
  Quantification and compensation},\ }\href
  {https://doi.org/10.1103/PhysRevA.98.052354} {\bibfield  {journal} {\bibinfo
  {journal} {Phys. Rev. A}\ }\textbf {\bibinfo {volume} {98}},\ \bibinfo
  {pages} {052354} (\bibinfo {year} {2018})}\BibitemShut {NoStop}%
\bibitem [{\citenamefont {Zheng}\ \emph {et~al.}(2008)\citenamefont {Zheng},
  \citenamefont {Zhu},\ and\ \citenamefont {Zubairy}}]{QZE}%
  \BibitemOpen
  \bibfield  {author} {\bibinfo {author} {\bibfnamefont {H.}~\bibnamefont
  {Zheng}}, \bibinfo {author} {\bibfnamefont {S.~Y.}\ \bibnamefont {Zhu}},\
  and\ \bibinfo {author} {\bibfnamefont {M.~S.}\ \bibnamefont {Zubairy}},\
  }\bibfield  {title} {\bibinfo {title} {Quantum Zeno and anti-Zeno effects:
  Without the rotating-wave approximation},\ }\href
  {https://doi.org/10.1103/PhysRevLett.101.200404} {\bibfield  {journal}
  {\bibinfo  {journal} {Phys. Rev. Lett.}\ }\textbf {\bibinfo {volume} {101}},\
  \bibinfo {pages} {200404} (\bibinfo {year} {2008})}\BibitemShut {NoStop}%
\bibitem [{\citenamefont {Ashhab}\ \emph {et~al.}(2007)\citenamefont {Ashhab},
  \citenamefont {Johansson}, \citenamefont {Zagoskin},\ and\ \citenamefont
  {Nori}}]{RWA-RF}%
  \BibitemOpen
  \bibfield  {author} {\bibinfo {author} {\bibfnamefont {S.}~\bibnamefont
  {Ashhab}}, \bibinfo {author} {\bibfnamefont {J.~R.}\ \bibnamefont
  {Johansson}}, \bibinfo {author} {\bibfnamefont {A.~M.}\ \bibnamefont
  {Zagoskin}},\ and\ \bibinfo {author} {\bibfnamefont {F.}~\bibnamefont
  {Nori}},\ }\bibfield  {title} {\bibinfo {title} {Two-level systems driven by
  large-amplitude fields},\ }\href {https://doi.org/10.1103/PhysRevA.75.063414}
  {\bibfield  {journal} {\bibinfo  {journal} {Phys. Rev. A}\ }\textbf {\bibinfo
  {volume} {75}},\ \bibinfo {pages} {063414} (\bibinfo {year}
  {2007})}\BibitemShut {NoStop}%
\bibitem [{\citenamefont {Shirley}(1965)}]{Shirley}%
  \BibitemOpen
  \bibfield  {author} {\bibinfo {author} {\bibfnamefont {J.~H.}\ \bibnamefont
  {Shirley}},\ }\bibfield  {title} {\bibinfo {title} {Solution of the
  Schr\"odinger equation with a Hamiltonian periodic in time},\ }\href
  {https://doi.org/10.1103/PhysRev.138.B979} {\bibfield  {journal} {\bibinfo
  {journal} {Phys. Rev.}\ }\textbf {\bibinfo {volume} {138}},\ \bibinfo {pages}
  {B979} (\bibinfo {year} {1965})}\BibitemShut {NoStop}%
\bibitem [{\citenamefont {Dai}\ \emph {et~al.}(2016)\citenamefont {Dai},
  \citenamefont {Shi},\ and\ \citenamefont {Yi}}]{Yi}%
  \BibitemOpen
  \bibfield  {author} {\bibinfo {author} {\bibfnamefont {C.~M.}\ \bibnamefont
  {Dai}}, \bibinfo {author} {\bibfnamefont {Z.~C.}\ \bibnamefont {Shi}},\ and\
  \bibinfo {author} {\bibfnamefont {X.~X.}\ \bibnamefont {Yi}},\ }\bibfield
  {title} {\bibinfo {title} {Floquet theorem with open systems and its
  applications},\ }\href {https://doi.org/10.1103/PhysRevA.93.032121}
  {\bibfield  {journal} {\bibinfo  {journal} {Phys. Rev. A}\ }\textbf {\bibinfo
  {volume} {93}},\ \bibinfo {pages} {032121} (\bibinfo {year}
  {2016})}\BibitemShut {NoStop}%
\bibitem [{\citenamefont {Garraway}\ and\ \citenamefont
  {Vitanov}(1997)}]{Garraway}%
  \BibitemOpen
  \bibfield  {author} {\bibinfo {author} {\bibfnamefont {B.~M.}\ \bibnamefont
  {Garraway}}\ and\ \bibinfo {author} {\bibfnamefont {N.~V.}\ \bibnamefont
  {Vitanov}},\ }\bibfield  {title} {\bibinfo {title} {Population dynamics and
  phase effects in periodic level crossings},\ }\href
  {https://doi.org/10.1103/PhysRevA.55.4418} {\bibfield  {journal} {\bibinfo
  {journal} {Phys. Rev. A}\ }\textbf {\bibinfo {volume} {55}},\ \bibinfo
  {pages} {4418} (\bibinfo {year} {1997})}\BibitemShut {NoStop}%
\bibitem [{\citenamefont {L\"u}\ and\ \citenamefont
  {Zheng}(2012)}]{CHRW:origin}%
  \BibitemOpen
  \bibfield  {author} {\bibinfo {author} {\bibfnamefont {Z.}~\bibnamefont
  {L\"u}}\ and\ \bibinfo {author} {\bibfnamefont {H.}~\bibnamefont {Zheng}},\
  }\bibfield  {title} {\bibinfo {title} {Effects of counter-rotating
  interaction on driven tunneling dynamics: Coherent destruction of tunneling
  and Bloch-Siegert shift},\ }\href
  {https://doi.org/10.1103/PhysRevA.86.023831} {\bibfield  {journal} {\bibinfo
  {journal} {Phys. Rev. A}\ }\textbf {\bibinfo {volume} {86}},\ \bibinfo
  {pages} {023831} (\bibinfo {year} {2012})}\BibitemShut {NoStop}%
\bibitem [{\citenamefont {L\"u}\ \emph {et~al.}(2016)\citenamefont {L\"u},
  \citenamefont {Yan}, \citenamefont {Goan},\ and\ \citenamefont
  {Zheng}}]{CHRW:bias}%
  \BibitemOpen
  \bibfield  {author} {\bibinfo {author} {\bibfnamefont {Z.}~\bibnamefont
  {L\"u}}, \bibinfo {author} {\bibfnamefont {Y.}~\bibnamefont {Yan}}, \bibinfo
  {author} {\bibfnamefont {H.-S.}\ \bibnamefont {Goan}},\ and\ \bibinfo
  {author} {\bibfnamefont {H.}~\bibnamefont {Zheng}},\ }\bibfield  {title}
  {\bibinfo {title} {Bias-modulated dynamics of a strongly driven two-level
  system},\ }\href {https://doi.org/10.1103/PhysRevA.93.033803} {\bibfield
  {journal} {\bibinfo  {journal} {Phys. Rev. A}\ }\textbf {\bibinfo {volume}
  {93}},\ \bibinfo {pages} {033803} (\bibinfo {year} {2016})}\BibitemShut
  {NoStop}%
\bibitem [{\citenamefont {Zhang}\ \emph {et~al.}(2019)\citenamefont {Zhang},
  \citenamefont {Yang}, \citenamefont {Fu},\ and\ \citenamefont
  {Wang}}]{CHRW:battery}%
  \BibitemOpen
  \bibfield  {author} {\bibinfo {author} {\bibfnamefont {Y.-Y.}\ \bibnamefont
  {Zhang}}, \bibinfo {author} {\bibfnamefont {T.-R.}\ \bibnamefont {Yang}},
  \bibinfo {author} {\bibfnamefont {L.}~\bibnamefont {Fu}},\ and\ \bibinfo
  {author} {\bibfnamefont {X.}~\bibnamefont {Wang}},\ }\bibfield  {title}
  {\bibinfo {title} {Powerful harmonic charging in a quantum battery},\ }\href
  {https://doi.org/10.1103/PhysRevE.99.052106} {\bibfield  {journal} {\bibinfo
  {journal} {Phys. Rev. E}\ }\textbf {\bibinfo {volume} {99}},\ \bibinfo
  {pages} {052106} (\bibinfo {year} {2019})}\BibitemShut {NoStop}%
\bibitem [{\citenamefont {Yan}\ \emph {et~al.}(2017{\natexlab{b}})\citenamefont
  {Yan}, \citenamefont {L\"u}, \citenamefont {Luo},\ and\ \citenamefont
  {Zheng}}]{CHRW:Effects_of_CR}%
  \BibitemOpen
  \bibfield  {author} {\bibinfo {author} {\bibfnamefont {Y.}~\bibnamefont
  {Yan}}, \bibinfo {author} {\bibfnamefont {Z.}~\bibnamefont {L\"u}}, \bibinfo
  {author} {\bibfnamefont {J.}~\bibnamefont {Luo}},\ and\ \bibinfo {author}
  {\bibfnamefont {H.}~\bibnamefont {Zheng}},\ }\bibfield  {title} {\bibinfo
  {title} {Effects of counter-rotating couplings of the Rabi model with
  frequency modulation},\ }\href {https://doi.org/10.1103/PhysRevA.96.033802}
  {\bibfield  {journal} {\bibinfo  {journal} {Phys. Rev. A}\ }\textbf {\bibinfo
  {volume} {96}},\ \bibinfo {pages} {033802} (\bibinfo {year}
  {2017}{\natexlab{b}})}\BibitemShut {NoStop}%
\bibitem [{\citenamefont {L\"u}\ \emph {et~al.}(2017)\citenamefont {L\"u},
  \citenamefont {Zhao},\ and\ \citenamefont {Zheng}}]{CHRW:2photon-Rabimodel}%
  \BibitemOpen
  \bibfield  {author} {\bibinfo {author} {\bibfnamefont {Z.}~\bibnamefont
  {L\"u}}, \bibinfo {author} {\bibfnamefont {C.}~\bibnamefont {Zhao}},\ and\
  \bibinfo {author} {\bibfnamefont {H.}~\bibnamefont {Zheng}},\ }\bibfield
  {title} {\bibinfo {title} {Quantum dynamics of two-photon quantum Rabi
  model},\ }\href {https://doi.org/10.1088/1751-8121/aa5537} {\bibfield
  {journal} {\bibinfo  {journal} {Journal of Physics A: Mathematical and
  Theoretical}\ }\textbf {\bibinfo {volume} {50}},\ \bibinfo {pages} {074002}
  (\bibinfo {year} {2017})}\BibitemShut {NoStop}%
\bibitem [{\citenamefont {Pietik\"ainen}\ \emph {et~al.}(2017)\citenamefont
  {Pietik\"ainen}, \citenamefont {Danilin}, \citenamefont {Kumar},
  \citenamefont {Veps\"al\"ainen}, \citenamefont {Golubev}, \citenamefont
  {Tuorila},\ and\ \citenamefont
  {Paraoanu}}]{citeCHRW:BS_shift_q_to_c_transition}%
  \BibitemOpen
  \bibfield  {author} {\bibinfo {author} {\bibfnamefont {I.}~\bibnamefont
  {Pietik\"ainen}}, \bibinfo {author} {\bibfnamefont {S.}~\bibnamefont
  {Danilin}}, \bibinfo {author} {\bibfnamefont {K.~S.}\ \bibnamefont {Kumar}},
  \bibinfo {author} {\bibfnamefont {A.}~\bibnamefont {Veps\"al\"ainen}},
  \bibinfo {author} {\bibfnamefont {D.~S.}\ \bibnamefont {Golubev}}, \bibinfo
  {author} {\bibfnamefont {J.}~\bibnamefont {Tuorila}},\ and\ \bibinfo {author}
  {\bibfnamefont {G.~S.}\ \bibnamefont {Paraoanu}},\ }\bibfield  {title}
  {\bibinfo {title} {Observation of the Bloch-Siegert shift in a driven
  quantum-to-classical transition},\ }\href
  {https://doi.org/10.1103/PhysRevB.96.020501} {\bibfield  {journal} {\bibinfo
  {journal} {Phys. Rev. B}\ }\textbf {\bibinfo {volume} {96}},\ \bibinfo
  {pages} {020501} (\bibinfo {year} {2017})}\BibitemShut {NoStop}%
\bibitem [{\citenamefont {Shan}\ \emph {et~al.}(2018)\citenamefont {Shan},
  \citenamefont {Dai}, \citenamefont {Shen},\ and\ \citenamefont
  {Yi}}]{citeCHRW:controlled_state_transfer}%
  \BibitemOpen
  \bibfield  {author} {\bibinfo {author} {\bibfnamefont {H.~J.}\ \bibnamefont
  {Shan}}, \bibinfo {author} {\bibfnamefont {C.~M.}\ \bibnamefont {Dai}},
  \bibinfo {author} {\bibfnamefont {H.~Z.}\ \bibnamefont {Shen}},\ and\
  \bibinfo {author} {\bibfnamefont {X.~X.}\ \bibnamefont {Yi}},\ }\bibfield
  {title} {\bibinfo {title} {Controlled state transfer in a Heisenberg spin
  chain by periodic drives},\ }\href
  {https://doi.org/10.1038/s41598-018-31552-w} {\bibfield  {journal} {\bibinfo
  {journal} {Scientific Reports}\ }\textbf {\bibinfo {volume} {8}},\ \bibinfo
  {pages} {13565} (\bibinfo {year} {2018})}\BibitemShut {NoStop}%
\bibitem [{\citenamefont {Guo}\ \emph {et~al.}(2018)\citenamefont {Guo},
  \citenamefont {Deng}, \citenamefont {Li}, \citenamefont {Song}, \citenamefont
  {Wang}, \citenamefont {Su}, \citenamefont {Li}, \citenamefont {Jin},\ and\
  \citenamefont {Zheng}}]{citeCHRW:LFM}%
  \BibitemOpen
  \bibfield  {author} {\bibinfo {author} {\bibfnamefont {X.}~\bibnamefont
  {Guo}}, \bibinfo {author} {\bibfnamefont {H.}~\bibnamefont {Deng}}, \bibinfo
  {author} {\bibfnamefont {H.}~\bibnamefont {Li}}, \bibinfo {author}
  {\bibfnamefont {P.}~\bibnamefont {Song}}, \bibinfo {author} {\bibfnamefont
  {Z.}~\bibnamefont {Wang}}, \bibinfo {author} {\bibfnamefont {L.}~\bibnamefont
  {Su}}, \bibinfo {author} {\bibfnamefont {J.}~\bibnamefont {Li}}, \bibinfo
  {author} {\bibfnamefont {Y.}~\bibnamefont {Jin}},\ and\ \bibinfo {author}
  {\bibfnamefont {D.}~\bibnamefont {Zheng}},\ }\bibfield  {title} {\bibinfo
  {title} {Modulation of energy spectrum and control of coherent microwave
  transmission at single-photon level by longitudinal field in a
  superconducting quantum circuit},\ }\href
  {https://doi.org/10.1088/1674-1056/27/7/074206} {\bibfield  {journal}
  {\bibinfo  {journal} {Chinese Physics B}\ }\textbf {\bibinfo {volume} {27}},\
  \bibinfo {pages} {074206} (\bibinfo {year} {2018})}\BibitemShut {NoStop}%
\bibitem [{\citenamefont {Chen}\ \emph {et~al.}(2018)\citenamefont {Chen},
  \citenamefont {Wang}, \citenamefont {Liang},\ and\ \citenamefont
  {Wang}}]{citeCHRW:maximal_quantum_Fisher_information_semi_Rabi_model}%
  \BibitemOpen
  \bibfield  {author} {\bibinfo {author} {\bibfnamefont {J.}~\bibnamefont
  {Chen}}, \bibinfo {author} {\bibfnamefont {Z.}~\bibnamefont {Wang}}, \bibinfo
  {author} {\bibfnamefont {H.}~\bibnamefont {Liang}},\ and\ \bibinfo {author}
  {\bibfnamefont {X.}~\bibnamefont {Wang}},\ }\bibfield  {title} {\bibinfo
  {title} {Maximal quantum fisher information in the semi-classical Rabi
  model},\ }\href {https://doi.org/10.1140/epjd/e2018-90024-0} {\bibfield
  {journal} {\bibinfo  {journal} {The European Physical Journal D}\ }\textbf
  {\bibinfo {volume} {72}},\ \bibinfo {pages} {145} (\bibinfo {year}
  {2018})}\BibitemShut {NoStop}%
\bibitem [{\citenamefont {Ferr\'on}\ \emph {et~al.}(2010)\citenamefont
  {Ferr\'on}, \citenamefont {Dom\'{\i}nguez},\ and\ \citenamefont
  {S\'anchez}}]{plateau:magnetic_flux}%
  \BibitemOpen
  \bibfield  {author} {\bibinfo {author} {\bibfnamefont {A.}~\bibnamefont
  {Ferr\'on}}, \bibinfo {author} {\bibfnamefont {D.}~\bibnamefont
  {Dom\'{\i}nguez}},\ and\ \bibinfo {author} {\bibfnamefont {M.~J.}\
  \bibnamefont {S\'anchez}},\ }\bibfield  {title} {\bibinfo {title}
  {Large-amplitude harmonic driving of highly coherent flux qubits},\ }\href
  {https://doi.org/10.1103/PhysRevB.82.134522} {\bibfield  {journal} {\bibinfo
  {journal} {Phys. Rev. B}\ }\textbf {\bibinfo {volume} {82}},\ \bibinfo
  {pages} {134522} (\bibinfo {year} {2010})}\BibitemShut {NoStop}%
\bibitem [{\citenamefont {Poggi}\ \emph {et~al.}(2014)\citenamefont {Poggi},
  \citenamefont {Arranz}, \citenamefont {Benito}, \citenamefont {Borondo},\
  and\ \citenamefont {Wisniacki}}]{plateau:dc-ac-LiCN}%
  \BibitemOpen
  \bibfield  {author} {\bibinfo {author} {\bibfnamefont {P.~M.}\ \bibnamefont
  {Poggi}}, \bibinfo {author} {\bibfnamefont {F.~J.}\ \bibnamefont {Arranz}},
  \bibinfo {author} {\bibfnamefont {R.~M.}\ \bibnamefont {Benito}}, \bibinfo
  {author} {\bibfnamefont {F.}~\bibnamefont {Borondo}},\ and\ \bibinfo {author}
  {\bibfnamefont {D.~A.}\ \bibnamefont {Wisniacki}},\ }\bibfield  {title}
  {\bibinfo {title} {Maximum population transfer in a periodically driven
  quantum system},\ }\href {https://doi.org/10.1103/PhysRevA.90.062108}
  {\bibfield  {journal} {\bibinfo  {journal} {Phys. Rev. A}\ }\textbf {\bibinfo
  {volume} {90}},\ \bibinfo {pages} {062108} (\bibinfo {year}
  {2014})}\BibitemShut {NoStop}%
 \bibitem {note1}
We have used the relations $\sum_{n=-\infty}^{\infty} J_n(\Xi) \cos(n\omega t)= J_0(\Xi)+2\sum_{n=1}^{\infty} J_{2n}(\Xi) \cos(2n\omega t)$, and
$\sum_{n=-\infty}^{\infty} J_n(\Xi) \sin(n\omega t)= 2\sum_{n=1}^{\infty} J_{2n-1}(\Xi) \sin[(2n-1)\omega t]$.
\end{thebibliography}

\newpage

\end{document}